\documentclass[aip,reprint]{revtex4-1}

\usepackage{graphicx}
\usepackage{amsmath, amssymb, amsthm}
\usepackage{framed}
\usepackage{mathrsfs}
\usepackage{esint}

\begin{document}

\title{Quantum electrodynamic theory of the cardiac excitation propagation I: construction of quantum electrodynamics in the bidomain}

\author{Sehun Chun}
\email{schun@aims.ac.za}
\affiliation{African Institute for Mathematical Sciences and Stellenbosch University, 5 Melrose road, Muizenberg, Cape Town, South Africa 7945}

\begin{abstract}
To provide a unified theoretical framework ranging from a cellular-level excitation mechanism to organic-level geometric propagation, a new theory inspired by quantum electrodynamic theory for light propagation is proposed by describing the cardiac excitation propagation as the continuation of absorption and emission of charged ions by myocardial cells. By the choice of gauge and the membrane current density, a set of Maxwell's equations with a charge density and a current density is constructed in macroscopic bidomain and is shown to be equivalent to the diffusion-reaction system with the B. van der Pol oscillator. The derived Maxwell's equations for the excitation propagation obeys the conservational laws of the number of the cations, energy and momentum, but the total charge is not conserved. The Lagrangian is derived to reveal that the trajectory and wavefront of the excitation propagation are the same as the electrodynamic wave if ion channels work uniformly. From the second quantization, the Hamiltonian is also derived to explain the excitation mechanism of the myocardial cell by Feynman's diagram and the mechanism of the refractory period in the perspective of positron. The effects of the external electromagnetic field are explained both from the action of the Lagrangian and the interaction by the Hamiltonian.
\end{abstract}


\maketitle


\section{Introduction: Why QED theory for electrophysiology?}

Quantum electrodynamics (QED) is the theory of interaction between light and matter, mostly about the absorption and emission of photons. QED lies in the hearth of quantum physics on the scale of one thousandth of picometer $(10^{-15} \sim 10^{-14} m)$ and seems to be not on the right scale to be correlated with the mechanism of the cardiac excitation propagation on the tissue scale of micrometer $(10^{-5} \sim 10^{-4} m)$. This irrelevance seems to reflect more than the axiomatic schisms between physics and biology. An intriguing nomenclature such as \textit{quantum biology} has been used by some scholars to explain the photosynthesis of leaf and the bird's eye campus \cite{Ball2011}, but what is proposed in this paper is not the direct applications of quantum theory such as the effect of the electromagnetic field on the ions in the heart. Instead, by adapting some analogies on the implicit changes of scales, we propose that QED also provides the fundamental mechanism of the cardiac excitation propagation, saying that QED is also the theory of interaction between the \textit{electrical signal} and the \textit{cardiac tissue} concerning the absorption and emission of the \textit{propagating cations}.

As most of the new theories are proposed to explain the unexplainable phenomena in view of classical theories, the introduction of QED to cardiac electrophysiology is caused by the same motivation. On a macroscopic scale, the propagation of the cardiac excitation, or the cardiac action potential or membrane potential, has been often modeled as \textit{waves} by a system of diffusion-reaction equations. But this wave model often fails to provide theoretical explanations on fundamental mechanisms, for example, the necessary conditions of conduction failure. Some disruptions of the excitation propagation have been explained by the curvature of the wavefront or the subsequent changes of propagational velocities \cite{Zykov1} \cite {Zykov2}, but these explanations are only applied to the simplest cases and often fail to be applied to actual geometry with anisotropy and complex curvature. Moreover, as soon as the excitation propagation looses the properties of the wave or meets the discontinuity of the excitable media such as myocardial farction or the discontinuities of myocardial fibre, the mathematical modeling or its computational simulation fails to reflect the real electrophysiological phenomena. The most significant and required mathematical studies for electrophysiological pathologies seem to be in the realm of the non-wavelike properties of the cardiac excitation propagation such as partial propagation through damaged myocardial tissue or, the movement of ions after the collision of two propagations, thus we should be prepared to accept the \textit{particle aspect} of the excitation propagation on a macroscopic scale. Contrary to the popular ion models on a cellular scale such as the Luo-Rudy model \cite{LuoRudy} providing no theoretical explanation on an organic scale, the incorporation of the particle aspect on an organic scale from the original wave model will provide both explanations for phenomena ranging from the cellular level to the organic level.

The quanta aspect of the cardiac excitation propagation is more naturally accepted than that of light in the early 1900's by Plank \cite{Planck}, Dirac \cite{Dirac1927} and Einstein \cite{Einstein1909}. Maybe the wave aspect of the excitation propagation was never accepted by cardiologists since they understand, by observations or textbooks, that myocardial cells are excited by the influx of charged ions. Then, we may ask ourselves on whether the quanta of the excitation propagation travels continuously following the law of classical mechanics possibly under the influence of an electromagnetic field. Looking at the molecular propagation of the cations such as potassium $(K^+)$ reveals that they propagate in the continuous procedure of being \textit{absorbed and emitted} by myocardial  cells. The classical concept of trajectory can be mathematically constructed from the orthogonal direction (in the Riemannian sense) to the wavefront, but clinical observations have never confirmed the existence of such solid object traveling continuously from the initial time to the final time. Substituting \textit{matter in the aether} with the \textit{myocardial cell}, we find a close similarity between the interaction between light and matter and its equivalence in the excitation of the myocardial cells. Briefly stated, as light propagates through the space in a continuous absorption and emission of photons by electron in the aether, we claim in this paper that the electric signal propagates through the heart tissue in a continuous absorption and emission of the propagational cations by myocardial cells. The only noticeable difference is that the electron moves in space, but the myocardial cell is stationary. This analogy may be no surprise if light is classified as a (the fastest) \textit{signal in an aether}, the same as the cardiac excitation propagation.

All the motivations for the quantization theory of the excitation propagation, however, have stemmed from undeniable demands to express the diffusion-reaction equations of the propagation in the language of Maxwell's equations. Then, the quanta theory of the excitation propagation is a direct consequence of the second quantization of Maxwell's equations, not to mention the geometric theory of the propagation with a very high frequency. The use of Maxwell's equations as the governing equations for the propagation has the following advantages: (i) The first is its versatile expressions in terms of the field or the potential. Thus, we can directly incorporate the external electromagnetic field or the external potential. (ii) The second advantage is the convenient derivation of Lagrangians and Hamiltonians possibly in the simplest form because the governing equations can be expressed under conservational laws. The Lagrangian can be used to trace the trajectory of the propagation of the excitation and the simple expression of the Lagrangian or, the comparison of it to that of the classical Maxwell's equations will shed more light on the behavior of the excitation propagation even in complex curved anisotropic space.

(iii) The third is the geometric expression of the governing equations. The eikonal equation for the diffusion equation has been derived by Keener \cite{Keener1991} to trace the wavefront of the excitation propagation, but the expression is too complicated to be practically useful for clinical studies, not to mention that it is derived from the FitzHugh-Nagumo (FHN) equations, or just the diffusion operator, thus, subsequently, inherits the restriction of the wave-like properties. As popularly used in geometric optics, Maxwell's equations for the excitation propagation may yield the simplest form of the eikonal equation with a very high frequency. (iv) The last, but the most important advantage is the quantization of the field induced by the excitation propagation. This means that the excitation propagation is considered as the movement of corpuscular positive ions of spin 1 (or bosons) which satisfy Bose-Einstein statistics, such as photons. This is necessary in view of cellular-level dynamics because it is well known that the movement of electrically charged ions such as $K^+$, $Ca^{2+}$, $Na^+$, $Cl^-$ across the intercellular space or membrane induces the excitation propagation. In spite of subsequent simplifications of ion-pumping processes, the benefit of quantization can serve as a powerful tool for unanswered phenomena. Well-known examples could be the interaction of the magnetic field with the propagating ions, the influence of the magnetic field on the resting state which we may call a \textit{vacuum}, and the collisions of multiple wavelets of the excitation propagation, all of which cannot be explained by the wave theory of the propagation.

Nevertheless, the QED theory for the cardiac excitation propagation opens up many fundamental questions in the perspective classical or quantum electrodynamics. (i) The first is whether the (electric) excitation can be viewed as an \textit{electrodynamic wave}. The main difference from classical electrodynamic waves lies in the fact that the excitation propagation always requires the media for propagation, for example, the cardiac or nerve tissue. This may be analogous to \textit{aether} which was abstractly used to explain the propagation of light by the early 1900s' \cite{Whittacker}. As the concept of aether has become redundant by the field theory of classical electromagnetic waves, can a similar procedure be also legitimately applied to the excitation propagation to eliminate the biological media by adapting the field theory?

(ii) Secondly, by the second quantization, we obtain the creation and annihilation operator representing the interaction between photon and matter. In the biological system, matter can be naturally replaced by the basic unit of the media such as the cardiac cell or nerve cell, and the photon can be replaced by the cation or the positively-charged ion in the tissue. The first replacement requires us to change only the notion of \textit{matter}, but the second replacement requires us to additionally change the size of the propagating particle. Consequently, the absorption or emission of photons by matter should be translated as \textit{entrance or exit of the cation thorough a biological cell} representing the similar mechanisms of light, but on a enormously large scale compared to it. A question rises whether this replacement or translation is legitimate in Maxwell's equations and its subsequent quantization such that the measure of \textit{quantum} is not absolute, but may depend on the type of media.

(iii) The last question is on the use of the \textit{bidomain} space to represent the biological space, instead of the classical \textit{mono-domain} space to represent the physical world. The bidomain space means that one point on a macroscopic scale always represents two separate points in different microscopic spaces. The introduction of the bidomain is inevitable for the construction of a conservational system for the excitation propagation. For example, consider the forest fire: Energy increases in the domain consisting of the trees only, but energy is preserved in the domain consisting of the trees \textit{and} the air around it. Interpreting this bidomain into the languages of the modo-domain of the physical world seems to be valid and produces an unconventional concept of \textit{time-varying point charge} with respect to time which will be elaborated in the latter part of this paper, but does this existence of the two domains at every point of the world violate any axiom of physical laws?

In the remainder of the paper, we will not prove or justify these axiomatic questions and leave them for later discussions and publications. This is a reasonable excuse because answering these rather philosophical questions seems not to be required for the analysis and explanation shown in this paper. In the next section, for readers who are not familiar with the classical diffusion-reaction model, for example, the FHN equations for the excitation propagation, the diffusion-reaction equations and its limitations will be explained in brief.

\begin{figure}[h]
\centering
\vbox{
\includegraphics[height=4cm, width=4cm] {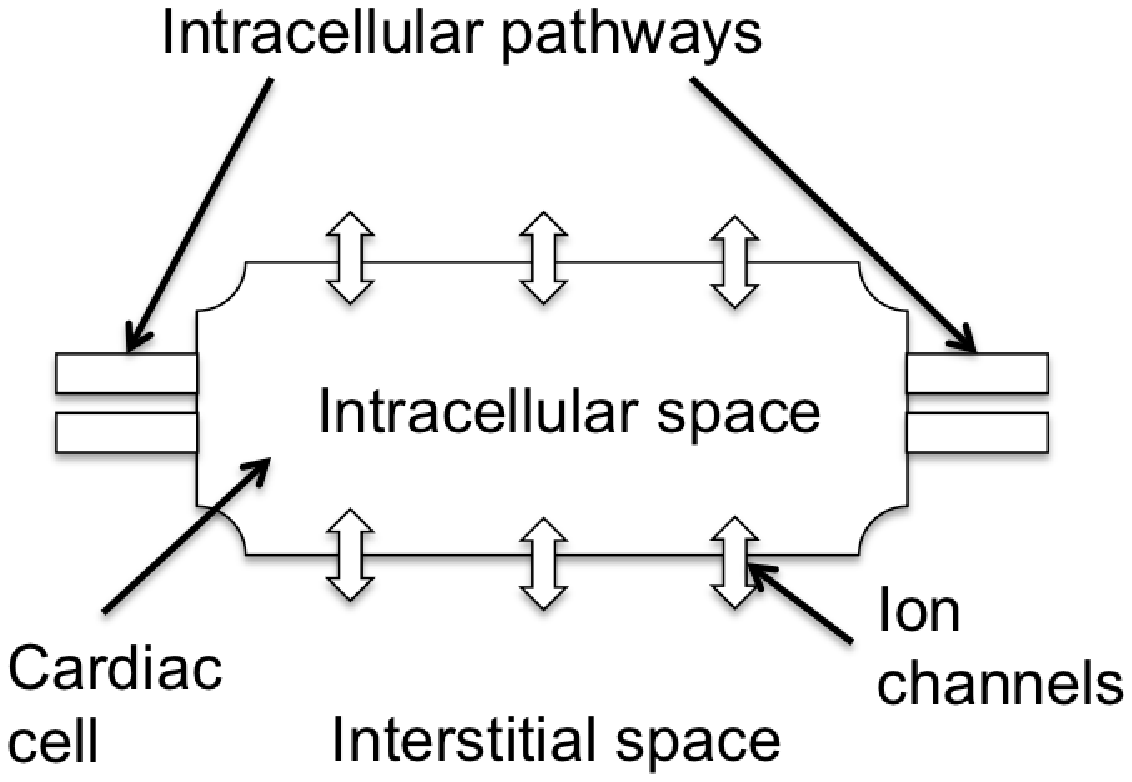}  \includegraphics[height=4cm, width=4cm] {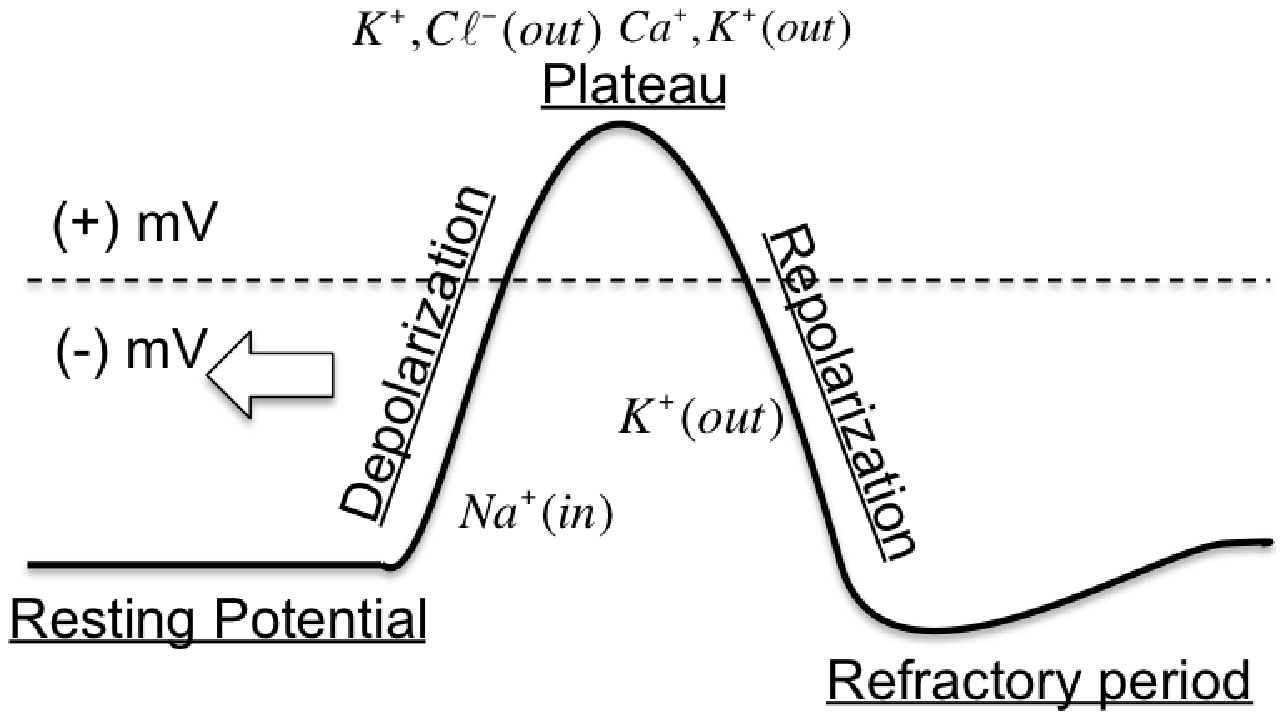}   }
\caption{Illustrations of the cardiac cell (left) and the action potential (right).}
\label {fig:cellandap}
\end{figure}

\subsection{Brief review and restrictions of the diffusion-reaction model}
Inspired by the FitzHugh's model \cite{FitzHugh1} \cite{FitzHugh2}, adapting the design of the Nagumo's electric circuit \cite{Nagumo} for excitation in nerves, the excitation propagation has been most widely characterized by the Bonhoeffer van der Pol (BvP) model for a relaxation oscillator \cite {vdP} which can be expressed for an oscillating quantity $x$ such as \cite{FitzHugh1}
\begin{equation}
\ddot{x} + a_1 (x^2 - 1) \dot{x} + x = 0 ,  \label{BvPorig0}
\end{equation}
where $a_1 \in \mathbb{R}^+$ and the damping coefficients depend on $x$ quadratically. By introducing the variable $y$ from the Li\'{e}nard's transformation \cite{Lienard}, this oscillator is alternatively expressed as
\begin{align}
\dot{x} & = a_1 \left ( y + x - \frac{x^3}{3}  \right )  ,~~~~~ \label{BvPorg1}   \\
\dot{y} & = - \frac{1}{c} (x - a_2 + a_3 y)  ,~~~~~~ \label{BvPorg2}
\end{align} 
where $a_2,~a_3 \in \mathbb{R}^+$. The biological tissue consists of two kinds of media: One is the \textit{intracellular space} such as the myocardial cell in the heart and the other is the \textit{interstitial space} such as the ambient medium surrounding the cell known as the bath (left of Figure \ref{fig:cellandap}).

Measured by the electric potential difference between the intracellular space and the interstitial space, the action potential, as illustrated in the right of Figure \ref{fig:cellandap}, is diffused to the neighboring cells. This propagation may or may not excite all of them because the mechanism of excitation is well characterized by the BvP oscillator of equation (\ref{BvPorig0}) which is only activated by a certain magnitude of the membrane potential or a pre-determined voltage threshold. The FHN model is derived as an one-dimensional oscillator, but has been widely used as the reaction function of equations (\ref{BvPorg1}) and (\ref{BvPorg2}) in the diffusion-reaction model which was first proposed as a similar class by Hodgkin and Huxley \cite{Hodgkin} and inspired by Turing's monumental work for animal coats \cite{Turing}. Tung \cite {Tung} extended this mechanism to divide the domain into two separate, but communicating domains and included the diffusion process in the interstitial space while maintaining the BvP oscillator for the intracellular space. Nevertheless, when the conductivity ratio between the two media remains relatively constant, the Tung's model, also known as the \textit{bidomain} model, reduces to a simpler model in order to depict the mechanism of the BvP oscillator in one space, known as the \textit{mono-domain} model \cite{Keenerbook}. There are several variations of the FHN equations which fit better with the real shape of the action potential, for example, the Rogers-McCulloch model \cite{Rogers} or the Aliev-Panfilov model \cite{AlievPanfilov} for the cardiac action potential, but all of them naturally share the same critical properties as the FHN model with the BvP oscillator.

In the multi-dimensional space, the FHN model is often expressed as the system of the diffusion-reaction equations as
\begin{align}
 \frac{\partial \phi}{\partial t} & = \nabla \cdot \left ( \mathcal{D} \nabla \phi \right ) + F (\phi, \phi^3, \psi),~~   \label{FHN1}\\ 
 \frac{\partial \psi}{\partial t} &= G(\phi, \psi),~~~~~~ \label{FHN2}
\end{align}
where $\phi$ is the membrane potential as an activator and $\psi$ is the refractoriness as an inhibitor. $F (\phi, \phi^3, \psi)$ and $G(\phi, \psi)$ are reaction functions such that $F,~G: \mathbf{R} \times \mathbf{R} \rightarrow \mathbf{R}$. The diffusivity tensor $\mathcal{D}$ represents the conductivity and directionality of myocardial fibre. With the emphasis on the wave-like property of the excitation propagation, the diffusion-reaction (DR) equations have enjoyed unprecedented success in the modeling of electrophysiological phenomena in nerves and in the heart, but at the same time, they have also revealed some restrictions in analyzing diverse and complex electrophysiological phenomena.

The first restriction comes from the fact that (i) the DR model does not obey conservational laws for energy and momentum. The variables $\phi$ and $\psi$ only indicate the difference between two variables measured at the different spaces, thus energy or momentum is generally not conserved in a physical domain as intuitively being recognized from the equivalent phenomena of forest fire. Consequently, many useful physical concepts and mathematical devices remain out of reach for the analysis of the excitation propagation due to the non-conservational properties of the DR model. (ii) Moreover, the analysis of the DR model is restricted with the given scale, thus the mathematical analysis of the different scale cannot be analyzed. This happens because the microscopic DR model shares the  same diffusion operator with the macroscopic DR model, but its reaction functions are significantly different. Computationally, a large sum of the microscopic DR model can be an approximation of the macroscopic DR model, but mathematically, they are not equivalent. This inconsistency between different scales prevents the understanding of the phenomena occurring on the different scale.

The third restriction arises because (iii) the DR equations have actually one physical variable, the membrane potential denoted as $\phi$ in equations (\ref{FHN1}) and (\ref{FHN2}). The FHN equations are written with two variables $\phi$ and $\psi$, but the second variable $\psi$, expressed as a function of the membrane potential and its time derivative as $\psi = f ( \phi, \dot{\phi} ) $, works as the inhibitor of the membrane potential and does not represent a substantially different field in the perspective of classical electrodynamics. If the excitation is regarded as the electrodynamic field in three-dimensional space and the membrane potential as the scalar potential, then the above DR equations contain only the scalar potential $\phi$ without three components of the vector potential. This restriction results in the undetermined electric and magnetic field even with the time-dependent solution of the DR equations. The underdetermined electromagnetic field induced by the excitation propagation from the governing equations directly means the lack of important tools in the perspective of \textit{field} in the study of complex cardiac electrophysiology as well as the ignorance of the coupling effect of the external electromagnetic field.

Clinical problems related to the external electromagnetic field can be briefly described as follows: The internal electric current in the heart has been widely studied \textit{in vivo} or \textit{in vitro}. The clinical studies of the external electric current are not as active as those of the internal electric current, but the original research of the former may date back to the 1930s \cite {Ferris}. After the seminal papers demonstrating the effects of the external electric currents for producing effective cardiac beats \cite {Zoll1}, mainly for the termination of ventricular tachycardia or fibrillation \cite {Beck} \cite {Zoll2}, this procedure has become one of the most effective and popular treatments for cardiac patients. However, its mechanism remains largely unknown on both microscopic and macroscopic scales. The biggest difficulty arises when we try to answer how the exterior electric current is coupled with the membrane potential that does not completely determine the internal electric current up to a constant. If we assume that the time variation of the vector potential $\mathbf{A}$ is approximately zero, then the gradient of the membrane potential is the same as that of the electric current, but this assumption may not represent reality if varying magnetic fields are present in the heart internally or externally.

Similar arguments can be applied to the magnetic field. Since Baule and McFee  \cite{Baule} first reported the magnetic field of the heart by magneto-cardiogram in 1962, the magnetic field of frog-heart muscle \cite{Burstein} and a single axon \cite {Wikswo} \cite {Roth1985} \textit{in vitro} seemed to validate the intrinsic magnetic field induced by the excitation. To date, no clinical implementations have been devised for the use of the magnetic field. Moreover, the effect of the external magnetic field, which has never gained substantial attraction in either of the neurology or cardiology communities, remains largely unknown as well. One may argue that this ignorance is due to the dependency of the magnetic field on the electric field such that an independent consideration of the magnetic field is negligible. In 1982's publication, Plonsey mathematically showed the similar claim that the magnetic field is completely determined by the electric field \cite {Plonsey}, but Plonesy implicitly used the aforementioned assumption on vector potential. Roth and Wikswo  \cite{Roth1986} provided a counter-example to this claim by showing that the magnetic field induced by the excitation may exist without the presence of the electric field. 

\subsection{Goals, notations and order of this paper}
The goals of the paper can be summarized according to two different perspectives. The first is focused on the practical aspect of this study for clinical applications: (i) The derivation of a mathematical expression to show that the functionality of ion channels reflecting the shape of the action potential can change the direction and velocity of the propagation. (ii) The derivation of a mathematical expression on the effect of the external electromagnetic fields for the propagation. (iii) The derivation of the geometrical action potential propagation and its eikonal equation.

The second perspective is on the theoretical aspect of this study: (i) The diffusion-reaction system with the BvP oscillator can be equivalently expressed by Maxwell's equations in the bidomain with an appropriate choice of gauge and the membrane current density. (ii) The one-dimensional BvP oscillator in reciprocal space directly contributes to the reaction function of the excitation in multi-dimensional space. (iii) The Maxwell's equations for the excitation propagation conserve the total number of the cations, the total energy and the momentum. (iv) The Lagrangian of the Maxwell's equations for the excitation propagation is the same as the Lagrangian of the classical Maxwell's equations if ion changes work uniformly in all the media. (v) The Hamiltonian of the excitation can be expressed with quantum operators and the refractory region can be described as an analogy of a positron.

The derivation and quantization of Maxwell's equations do not go beyond the level of textbooks, especially following the book by C. Cohen-Tannoudji et. al. \cite{CohenPA} \cite{CohenAP} to compare its results from the classical Maxwell's equations with the Coulomb gauge. The description of the excitation and its subsequent derivation of a set of Maxwell's equations also can be applied to the nerve cell in neuroscience, but for the sake of convenience and consistency, we mainly consider cases from cardiac electrophysiology.

The most important terminology in this paper is the cation, but it could mean ambiguously multiple objects, probably the same as the ambiguity of the meaning of a photon. Mostly, the \textit{cation} means a positively-charged ion traveling in myocardial cells for excitation. The cations are of a single kind, identical and indistinguishable, obeying the laws of Bose-Einstein statistics, the same as the properties of a photon. Among several ions such as $K^+$, $Ca^{2+}$, $Na^+$, $Cl^-$ being involved in the excitation mechanism of myocardial cells, the potassium $K^+$ could have the closest properties to the cation. However, it may be more accurate to say that the cation means the \textit{corpuscular of energy and momentum delivered by the propagation}, not a specific type of charged ion. Thus, we may call it a photon as well, but to avoid confusion, we stick to its nomenclature as the cation.

Consequently, ion channels are only related to the influx and efflux of the cations and this means that we only pay our attention to the changes of the membrane potential induced by the membrane current of the cations. This could be an excessive simplification for complex ion channels with several ions, but may reveal the fundamental functions and goals of ion channels on a macroscopic scale. For example, the excitation of the cardiac cell is mainly aimed to induce calcium $Ca^{2+}$ for muscular contraction, thus it may be not relevant to include the ion channels for $Ca^{2+}$ to understand the effects of ion channels for the excitation propagation. On the other hand, the sodium $Na^+$ channels plays a critical role especially in the depolarization phase, but in this paper the sodium channels are regarded as the channels of the cations, which is likely to pose no serious problem because they are all positively charged. The chloride $C\ell^-$ is similarly substituted by the channels of the cations, but due to the different signs, the influx of $C \ell^-$ is considered as the efflux of the cation and \textit{vice versa}.

This paper is organized as follows: In Section II, Maxwell's equations are constructed from a microscopic bidomain to a macroscopic bidomain. In Section III, the choice of gauge and the membrane current density are described and the BvP oscillator is constructed in reciprocal space. Section IV shows that the diffusion-reaction system for the excitation propagation is equivalent to the derived Maxwell's equations and its meaning is explained in the perspective of the semiclassical theory of radiation. In Section V, the conservation of the total number of the cations is proved, but the total charge is shown not to be conserved. Also, for the system of the particle and the field, the total energy and the total momentum are shown to be conserved. Section VI proposes the Lagrangian for the derived Maxwell's equations and shows that the Lagrangian is the same as the electromagnetic waves in homogeneous and isotropic media. The effects of the external electromagnetic field on the trajectory of the propagation are also shown. In Section VII, the Hamiltonian of the Maxwell's equations is derived. Moreover, the excitation mechanism and the refractory period are described by Feynman's diagram and the transition amplitude. The effects of the external electromagnetic field on the excitation mechanism are also shown. Appendix is organized as follows: Appendix I provides the proof of proposition 4 (A), proposition 5 (B). Appendix II provides the proof of lemma 2 (A) and Appendix III provides the proof of lemma 4 (A) and proposition 8 (B).

\begin{table}[htdp]
\caption{List of notations}
\begin{center}
\begin{tabular}{|c|l|}
\hline
$\pi^i$ & Microscopic intercellular domain \\
$\pi^o$ & Microscopic interstitial domain \\
$\pi^i \cap \pi^o$ & Membrane in the microscopic domain \\
$\Pi$ & Macroscopic domain \\
$\mathbf{V}^i$ & Field or variable in $\pi^i$ \\
$\mathbf{V}^o$ & Field or variable in $\pi^o$ \\
$\mathbf{V}_k$ & Field or variable in reciprocal space \\
$\mathbf{V}^{\parallel} $ & Parallel component to the wave vector $\mathbf{k}$ \\
$\mathbf{V}^{\perp} $ & Perpendicular component to the wave vector $\mathbf{k}$\\
\hline
\end{tabular}
\end{center}
\label{notation}
\end{table}

\section{From microscopic to macroscopic bi-doman}

\subsection{Microscopic domain}

Let us begin with Maxwell's equations on the microscopic domain being described as follows: Suppose that the microscopic domain contains the collection of myocardial cells, each of which is typically $100~\mu m$ long and $15~\mu m$ in diameter as well as the surrounding bath \cite{Keenerbook}. The intracellular space is denoted as $\pi^i$ representing myocardial cells, while the interstitial space is denoted as $\pi^o$ representing the bath. For simplicity, we assume that each domain is homogeneous. By the microscopic scale for the excitation propagation, we mean that $\pi^i$ and $\pi^o$ are microscopically separable with a clear boundary as $\pi^i \cap \pi^o = 0$. Thus, one point in the microscopic domain belongs to either of $\pi^i$ or $\pi^o$ while disregarding the thin membrane of $\pi^i \cap \pi^o$. Suppose that, in each microscopic domain, the dynamics of electromagnetic field induced by the presence or the movement of point charges are well expressed by Maxwell's equations as follows. In SI units, for the intracellular space $\pi^i$,
\begin{align}
\nabla \cdot \mathbf{e}^i &= \frac{\varrho^i}{\varepsilon_i},~~\nabla \cdot \mathbf{b}^i = 0,  \label{Maxwell1}\\  
\nabla \times \mathbf{e}^i &= - \frac{\partial \mathbf{b}^i }{\partial t},~~\frac{1}{\mu_i} \nabla \times \mathbf{b}^i  = \varepsilon_i \frac{\partial \mathbf{e}^i }{\partial t} + \mathbf{j}^i, \label{Maxwell2}
\end{align}
and for the interstitial space $\pi^o$,
\begin{align}
\nabla \cdot \mathbf{e}^o &= \frac{\varrho^o}{\varepsilon_o},~~\nabla \cdot \mathbf{b}^o = 0,  \label{Maxwell3}\\  
\nabla \times \mathbf{e}^o &= - \frac{\partial \mathbf{b}^o}{\partial t},~~\frac{1}{\mu_o} \nabla \times \mathbf{b}^o  = \varepsilon_o \frac{\partial \mathbf{e}^o }{\partial t} + \mathbf{j}^o, \label{Maxwell4}
\end{align}
where the superscript $i$ and $o$ indicate the variables and fields belonging to the intercellular domain $\pi^i$ and interstitial domain $\pi^o$, respectively. The use of Maxwell's equations for the electric signal propagation in the heart or the brain has been widely accepted theoretically and experimentally \cite{Plonsey} \cite{Ragan} \cite{Schwan1957}, not to mention light scattering and birefringence \cite{Cohen1969} and light absorption \cite{Ross1974} by the action potential. Thus we will not discuss the further justification of the biological electrodynamic field. Representing the propagation of the electric signal in the resting state, equations (\ref{Maxwell1}) - (\ref{Maxwell4}) are written in the same expression as those of the classical electrodynamics that representing the propagation of light in the space devoid of matter. But they should be interpreted in the different context because we consider different kinds of signal propagation in different media.

Let $\varrho$ be the charge density and $\mathbf{j}$ be the current density in $\pi^i$ or $\pi^o$. The magnitude of $\varrho$ and $\mathbf{j}$ in each microscopic domain $\pi^i$ and $\pi^o$ are not trivial for almost everywhere because ion-pumps being attached to $\pi^i$ and $\pi^i$ generate \textit{sources} by transferring charged ions from the other domain. But they are only non-trivial in the duration of the excitation of the myocardial cell when ion pumps are activated. Then, $\mathbf{e}$ and $\mathbf{b}$ are the electric and magnetic fields induced by them in the \textit{cardiac tissue}, not necessarily meaning the same kind of the electromagnetic field of the classical electrodynamics in the \textit{physical space}. Similarly, the permittivity $\varepsilon$ and permeability $\mu$ should also be redefined corresponding to those of the classical electromagnetics. Let us define the \textit{resting state} as the condition of the myocardial cell where the membrane potential, or the difference of the scalar potential between $\pi^i$ and $\pi^o$, is stable and no macro-dynamics of the ions occurs. Let $c$ be the \textit{maximum speed of the cardiac excitation propagation} in the resting state of the cardiac tissue which is known to be approximately $1~ m/s$ \cite{Guyton}. Then, for permeability $\mu_0$ of the vacuum state, the permittivity $\varepsilon_0$ is defined as $c^{-2}/\mu_0$, equivalently, ${\varepsilon_0 \mu_0} = c^2$ and we suppose that they remain constant in each domain unless mentioned otherwise. Then we propose the following axiom which naturally holds in the classical electrodynamics:\\
\\
\textbf{Axiom 1}: In the vacuum state, the macroscopic phase velocity by the intercellular space $\pi^i$ is the same as that by the macroscopic interstitial space $\pi^o$ as $\varepsilon_i \mu_i = \varepsilon_o \mu_o $ and consequently the same at the membrane.\\
\\
If we consider the generation of the electromagnetic field as the consequences of the moving of charged ions, the equality $\varepsilon_i \mu_i = \varepsilon_o \mu_o $ means that the maximum speed of the electric signal in $\pi^i$ and $\pi^o$ is the same. Since this equality is well accepted in the classical electrodynamics since the maximum speed of light is constant everywhere in the relativistic sense, we may apply the same principle that the electric signal travels at the same speed in the resting state $\pi^i$ and $\pi^o$. In fact, this axiom is supported by more fundamental observations on the existence of the membrane potential, or the scalar potential difference between in $\pi^i$ and $\pi^o$ (otherwise the membrane potential would collapse) and its constant speed in homogeneous resting media. Then the time variable $t$ corresponding to this signal is accordingly defined as $t = \ell / c $ for any length $\ell$, but the remaining analysis is non-relativistic, thus for the sake of simplicity the time $t$ is just set as the physical time $t = \ell /$(speed of light). The charge density $\varrho$ and the current density $\mathbf{j}$ are considered \textit{discretely} with point charge $\iota^i_{\alpha}$ in $\pi^i$ or $\iota^o_{\alpha}$ in $\pi^o$ and are expressed as
\begin{align}
\varrho^{i,o} ( \mathbf{r}, t ) &= \sum_{\alpha} \iota_{\alpha}^{i,o} \delta [ \mathbf{r} - \mathbf{r}_{\alpha} ( t) ],  \label{discreterho}, \\
\mathbf{j}^{i,o}  ( \mathbf{r}, t ) &= \sum_{\alpha} \iota_{\alpha}^{i,o} \boldsymbol{\nu}_{\alpha}^{i,o} (t) \delta [ \mathbf{r} - \mathbf{r}_{\alpha} ( t) ], \label{discretej} ,
\end{align}
where $\alpha$ is the index of each point charge and $\delta$ is the Dirac-delta function. $\mathbf{r}_{\alpha}$ indicates the location of point charge indexed as $\alpha$, while $\boldsymbol{\nu}_{\alpha}^{i}$ and $\boldsymbol{\nu}_{\alpha}^{o}$ is the velocity of point charge $\alpha$ in each domain $\pi^i$ and $\pi^o$. In electrostatic conditions where there is no movement of charged particles and consequently no excitation occurs, the conservation of the electric charge and the electric current holds such that
\begin{align*}
&\int_{\cup \pi^i } \frac{\partial \varrho^i}{\partial t} dx + \int_{\cup \pi^o} \frac{\partial \rho^o}{\partial t} dx = 0 ,\\
 &\int_{\cup \partial \pi^i } \mathbf{j}^i \cdot \mathbf{n} ds + \int_{ \cup \partial \pi^i } \mathbf{j}^o \cdot \mathbf{n} ds = 0 .
\end{align*}
In words, the first equality means that the total charge is conserved in $\pi^i \cup \pi^o$. Charged ions can change the domain but always stays in $\pi^i \cup \pi^o$. The second equality means the net current is zero in $\pi^i \cup \pi^o$. The zero net current becomes more obvious by introducing the membrane current $\mathbf{j}^m$ which measures the electric current through the membrane where each electric current is expressed as $\mathbf{j}^i = \mathbf{j}^m$ and  $\mathbf{j}^o = - \mathbf{j}^m$ in electrostatic conditions.

\begin{figure}[h]
\centering
\vbox{
\includegraphics[height=3cm, width=8cm] {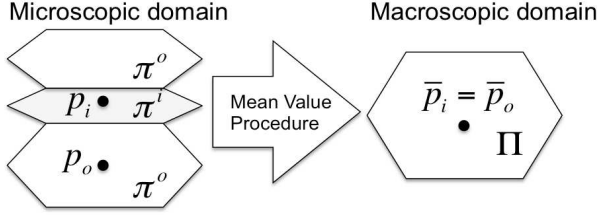}  }
\caption{From microscopic domain $\pi^i$ and $\pi^o$ to macroscopic domain $\Pi$. $p_i$ and $p_o$ are the \textit{microscopic} point in $\pi^i$ and $\pi^o$, respectively, and $\bar{p}_i$ and $\bar{p}_o$ are the \textit{macroscopic} point that are obtained as the mean value.}
\label {fig:micromacro}
\end{figure}

\subsection{Macroscopic domain}

Next, we integrate the two sets of Maxwell's equations (\ref{Maxwell1}) - (\ref{Maxwell4}) of a microscopic domain into the equivalent equations on a macroscopic scale. The macroscopic scale corresponds to the domain where one point always represents one point in $\pi^i$ and one point in $\pi^o$. This is possible because we assume that the size of the myocardial cell is much smaller than the unit of the macroscopic unit. Consequently, there is no spacial measurement in $\Pi$ to distinguish between $\pi^i$ and $\pi^o$. For example a macroscopic domain means a myocardial tissue on an organic scale consisting of hundreds and thousands of myocardial cells. The boundary between the macroscopic domain and the microscopic domain could be ambiguous, but we roughly regard the macroscopic scale as the equivalence of the organic scale by which the cell and the bath cannot be differentiated as shown in Figure (\ref{fig:micromacro}). Let us denote this macroscopic domain as $\Pi$. Shifting from the microscopic domain to the macroscopic domain follows the classical mean value approach first used by H. A. Lorentz for macroscopic Maxwell's equations \cite {Lorentz}; the macroscopic field component is obtained as the average of the microscopic field components. For example, the electric field $\mathbf{E}$ and the magnetic field $\mathbf{B}$ on the macroscopic scale are obtained as
\begin{equation*}
\mathbf{E} = \bar{\mathbf{e}} \equiv \frac{1}{V} \int \mathbf{e} d {V} ,~~~~~ \mathbf{B} = \bar{\mathbf{b}} \equiv  \frac{1}{V} \int \mathbf{b} d {V} ,
\end{equation*}
where $V$ is the volume of the sphere centered at each point in the microscopic domain and the bar notation indicate that the corresponding quantity is obtained from the mean value. Suppose that the sphere is sufficiently large so that the sphere does not divide point charge $\iota_{\alpha}^i$ in $\pi^i$. By the mean value procedure, $\iota_{\alpha}^i$ becomes macroscopic with the corresponding new index $\alpha$ for macroscopic point charges, but $\iota_{\alpha}^i$ in the intercellular space $\pi^i$ is not added with $\iota_{\alpha}^o$ in the interstitial space $\pi^o$. The macroscopic point charge $q_{\alpha}^i$ and $q_{\alpha}^o$ is the average of point charge $\iota_{\alpha}^i$ and $\iota_{\alpha}^o$ in each microscopic domain $\pi^i$ and $\pi^o$ such as $q_{\alpha}^i \equiv ({1}/ {V}) \int \iota_{\alpha}^i d V$ and $q_{\alpha}^o \equiv ({1}/{V}) \int \iota_{\alpha}^o d V $ where the sphere $V$ contains both $\pi^i$ and $\pi^o$, but $\iota_{\alpha}^i $ and $\iota_{\alpha}^o $ only exist in $\pi^i$ and $\pi^o$, respectively.

The macroscopic domain $\Pi$ can also be constructed from $\pi^i$ and $\pi^o$. One macroscopic point represents each microscopic point in the two different microscopic spaces and subsequently, all parameters may have two different values at each macroscopic point. Note that, as aforementioned in the Introduction, this is no more than the bidomain premise which is most popularly used in biological modeling  \cite{Tung}. As a result, Maxwell's equations for the intracellular space and the interstitial space are written in the same macroscopic domain $\Pi$: From Maxwell's equations in $\pi^i$ (\ref{Maxwell1}) and (\ref{Maxwell2}),
\begin{align}
\nabla \cdot \mathbf{E}^i &= \frac{\rho^i}{\varepsilon_i},~~\nabla \cdot \mathbf{B}^i = 0,  \label{Maxwell5}\\  
\nabla \times \mathbf{E}^i &= - \frac{\partial \mathbf{B}^i }{\partial t},~~\frac{1}{\mu_i} \nabla \times \mathbf{B}^i  = \varepsilon_i \frac{\partial \mathbf{E}^i }{\partial t} + \mathbf{J}^i, \label{Maxwell6}
\end{align}
and from Maxwell's equations in $\pi^o$ (\ref{Maxwell3}) and (\ref{Maxwell4}),
\begin{align}
\nabla \cdot \mathbf{E}^o &= \frac{\rho^o}{\varepsilon_o},~~\nabla \cdot \mathbf{B}^o = 0,  \label{Maxwell7}\\  
\nabla \times \mathbf{E}^o &= - \frac{\partial \mathbf{B}^o}{\partial t},~~\frac{1}{\mu_o} \nabla \times \mathbf{B}^o  = \varepsilon_o \frac{\partial \mathbf{E}^o }{\partial t} + \mathbf{J}^o, \label{Maxwell8}
\end{align}
where the capital letter of the fields such as $\mathbf{E}$, $\mathbf{B}$, and $\mathbf{J}$ indicates that the corresponding field is macroscopic and the permittivity and permeability in the macroscopic domain $\Pi$ are obtained similarly, but they are the same as those of the microscopic domain due to the homogeneous assumption of the media such that they are constant in each microscopic domain. With a point charge $q_{\alpha}^i$ and $q_{\alpha}^o$ being derived from $\iota_{\alpha}^i$ and $\iota_{\alpha}^o$, the charge density $\rho^i$ and $\rho^o$ and the current density $\mathbf{J}^i$ and $\mathbf{J}^o$ are expressed as 
\begin{align*}
\rho^{i,o} (\mathbf{r},t) &=  \sum_{\alpha} q_{\alpha}^{i,o} \delta [ \mathbf{r} - \mathbf{r}_{\alpha} ], \\
\mathbf{J}^{i,o} (\mathbf{r},t) &=  \sum_{\alpha} q_{\alpha}^{i,o} \mathbf{v}_{\alpha}^{i,o} \delta [ \mathbf{r} - \mathbf{r}_{\alpha} ] ,
\end{align*}
where the velocity of the macroscopic particles $\mathbf{v}_{\alpha}^i$ and $\mathbf{v}_{\alpha}^o$ are defined as the weighted average velocity of $ \iota^i_{\alpha}\boldsymbol{\nu}^i_{\alpha}$ and $ \iota^o_{\alpha}\boldsymbol{\nu}^o_{\alpha}$, respectively, such that
\begin{equation}
\mathbf{v}_{\alpha}^{i,o} \equiv \frac{ 1}{ q_{\alpha}^{i,o} } \left ( \frac{1}{V} \int \iota^{i,o}_{\alpha}\boldsymbol{\nu}^{i,o}_{\alpha} dV \right ). \label{macrovel}
\end{equation}
Both of sets of Maxwell's equations (\ref{Maxwell5}) - (\ref{Maxwell8}) are defined in the same macroscopic domain $\Pi$. This unusual co-existence does not mean that two electromagnetic fields interfere with each other in the near-field of the domain $\mathbf{r}_{\alpha}$. They appear to be at the same location macroscopically, but each field actually lies in the different space microscopically. Consequently, the only place they may interact is at the membrane, the boundaries of the intercellular space and the interstitial space, i.e., $\pi^o \cap \pi^i$. For every point of the domain $\Pi$, there are two distinct fields which do not interact with each other in the near field such as inside the myocardial tissue. However, the point charge in each field may work as a dipole moment, thus they are likely to interact in the far-field such as outside the heart. This can be summarized as the following axiom: \\
\\
\textbf{Axiom 2}: The field $( \mathbf{E}^i, \mathbf{B}^i )$ in $\pi^i$ does not interfere with the field $( \mathbf{E}^o, \mathbf{B}^o )$ in $\pi^o$ at the near-field of the macroscopic domain $\Pi$ except at the membrane $\pi^i \cap\pi^o$. \\
\\
The electromagnetic field is only generated by a point charge which only lies in either $\pi^i$ or $\pi^o$. Let us denote $q^i$ and $q^o$ as point charges staying in $\pi^i$and $\pi^o$, respectively. However, we suppose that a point charge never stays in the membrane $\pi^i \cap \pi^o$ because the membrane is relatively thin on the scale of nanometer $(10^{-9} m)$ \cite{Curtis} and the movement through ion pumps is relatively instantaneous. In other words, we do not consider the membrane as the space such that point charges only travel in $\pi^i$ and $\pi^o$.  \\
\\
\textbf{Axiom 3}:  The membrane is sufficiently thin everywhere relative to $\pi^i$ and $\pi^o$. Thus, point charges can only stay in and travel through either $\pi^i$ or $\pi^o$, but not in $\pi^i \cap\pi^o$.

\subsection{Weighted difference of the field and potential}

In order to retrieve the well-known observables such as the membrane potential, we will express the governing Maxwell's equations as the weighted difference between the field in $\pi_i$ and $\pi_o$. Since equations (\ref{Maxwell5}) - (\ref{Maxwell8}) are in the same domain, we can multiply equations (\ref{Maxwell7}) and (\ref{Maxwell8}) with $\sqrt{\varepsilon_o / \varepsilon_i}$ and subtract from equations (\ref{Maxwell5}) and (\ref{Maxwell6}). Using axiom 1, we obtain
\begin{align}
\nabla \cdot \mathbf{E} &= \frac{\rho}{\varepsilon_i},\label{MW1} \\
\nabla \cdot \mathbf{B} &= 0,  \label{MW2} \\  
\nabla \times \mathbf{E} &= - \frac{\partial \mathbf{B} }{\partial t}, \label{MW3} \\
\frac{1}{\mu_i} \nabla \times \mathbf{B}  &= \varepsilon_i \frac{\partial \mathbf{E} }{\partial t} + \mathbf{J}, \label{MW4}
\end{align}
where the new fields and parameters are defined as the weighted difference such as
\begin{align*}
&\mathbf{E} \equiv \mathbf{E}^i - \lambda \mathbf{E}^o,~~~ \mathbf{B}  \equiv \mathbf{B}^i -\lambda \mathbf{B}^o, \\
&\rho \equiv \rho^i - \lambda^{-1} \rho^o, ~~~\mathbf{J} \equiv \mathbf{J}^i - \lambda^{-1} \mathbf{J}^o, 
\end{align*}
where $\lambda$ is a scalar defined as $\lambda = \sqrt{\varepsilon_o / \varepsilon_i} = \sqrt{\mu_i/ \mu_o}$. Since the imaginary component of permittivity is the conductivity divided by the frequency, the ratio $\sqrt{\sigma_o / \sigma_i}$ is only proportional to the imaginary part of $\lambda$ without being related to the real part. But this does not imply that $\lambda = 1.0$. The experimental value of $\lambda$ is unknown, but the presumed value from the well-known phenomena will be discussed in section VI. This weight difference can be practically measured at the membrane, i.e. at the boundaries of the two spaces, but we prefer to maintain the intracellular permittivity and permeability constants $\mu_i$ and $\varepsilon_i$ instead of $\mu_i - \mu_o$ and $\varepsilon_i - \varepsilon_o$. Moreover, suppose that $q^i_{\alpha}$ and $q^o_{\alpha}$ are non-negligible at every location of $\mathbf{r}_{\alpha}$ as physiologically measured, as in ref \cite{Katz}. Then, $\rho$ and $\mathbf{J}$ are expressed as
\begin{equation}
\rho(\mathbf{r},t) =  \sum_{\alpha}  \chi_{\alpha} \delta [ \mathbf{r} - \mathbf{r}_{\alpha} ] ,~~\mathbf{J}(\mathbf{r},t) = \sum_{\alpha} \chi_{\alpha} \mathbf{v}_{\alpha}  \delta [ \mathbf{r} - \mathbf{r}_{\alpha} ] , \label{discreterhoandj}
\end{equation}
where the new point charge $\chi_{\alpha}$, namely \textit{point charge difference}, and the velocity $\mathbf{v}_{\alpha}$, namely \textit{velocity difference}, are defined in $\Pi$ as 
\begin{equation}
\chi_{\alpha} \equiv q^i_{\alpha} - \lambda^{-1} q^o_{\alpha}, ~~\mathbf{v}_{\alpha} \equiv \frac{1 }{ \chi_{\alpha} } \left (q^i_{\alpha} \mathbf{v}^i_{\alpha} - \lambda^{-1} q^o_{\alpha} \mathbf{v}^o_{\alpha} \right )  .  \label{defpcandvel}
\end{equation}
Contrary to $q_{\alpha}^i$ and $q_{\alpha}$, $\chi_{\alpha}$ is defined only in $\Pi$ due to the property that the magnitude of $\chi_{\alpha} (\mathbf{r})$ can be changed from the definition of $\chi_{\alpha}$. Consequently, $\chi_{\alpha}$ does not explicitly obey axiom 3, but its variation is closely related to it. It is important to note that ion pumps can significantly change $\chi_{\alpha}$. The operations of ion pumps to change $\chi_{\alpha}$ will be discussed in detail in the later part of this paper. Before proceeding further, we need to briefly mention that the Maxwell's equations (\ref{MW1}) - (\ref{MW4}) and the fields $\mathbf{E},~\mathbf{B},~\rho,~\mathbf{J}$ are well defined in $\Pi$.\\
\\
\textbf{Proposition 1}: The Maxwell's equations (\ref{MW1}) - (\ref{MW4}) with the weighted difference fields are well defined everywhere in $\Pi$.\\
\\
\textbf{Proof}:  By axiom 3, point charge $q$ lies either in $\pi^i$ or $\pi^o$, thus the Maxwell's equations (\ref{MW1}) - (\ref{MW4}) only represent the well-defined Maxwell's equations (\ref{Maxwell5}) - (\ref{Maxwell8}). For example, consider point charge $q_{\alpha}^i$ lies in $\pi^i$. Then, by axiom 2, the Maxwell's equations turn out to be equations (\ref{Maxwell5}) - (\ref{Maxwell6}) since $\mathbf{E}^o$ and $\mathbf{B}^o$ are zero in $\pi^i$. A similar argument exists for $q_{\alpha}^o$ lying in $\pi^o$ $\square$. \\
\\
An additional advantage of the expression of (\ref{MW1}) - (\ref{MW4}) is that it represents the field value at the membrane $\pi^i \cap \pi^o$ which is crucial for the initiation of the membrane current density $\mathbf{J}^m$. If we consider the vector potential $\mathbf{A}$ and the scalar potential $\phi$ being derived from equations (\ref{MW1}) - (\ref{MW4}) such as 
\begin{align}
\mathbf{B} &= \nabla \times \mathbf{A} , \label{MW5} \\
 \mathbf{E} &= - \nabla \phi - \frac{ \partial \mathbf{A}}{\partial t},  \label{MW6}
\end{align}
then we can verify that $\mathbf{A}$ and $\phi$ are defined as $\mathbf{A} \equiv \mathbf{A}^i -\lambda \mathbf{A}^o$ and $\phi \equiv \phi^i - \lambda \phi^o$. Note that the classical membrane potential is now generalized as $\phi$, the weighted difference by $\lambda$ to the potentials in $\pi^o$. Moreover, substituting equation (\ref{MW1}) into the divergence of equation (\ref{MW4}) yields the conservation of charge density difference as 
\begin{equation}
\frac{\partial \rho }{\partial t} + \nabla \cdot \mathbf{J}  = 0,~~\mbox{or}~~\frac{\partial \rho^i }{\partial t} + \nabla \cdot \mathbf{J}^i   = \frac{1}{\lambda} \left ( \frac{\partial \rho^o }{\partial t} + \nabla \cdot \mathbf{J}^o \right ) .   \label{conservcharge}
\end{equation}
The conservation of the first equality means that the time variation of the charge density difference $\rho$ is only caused by the current density difference $\mathbf{J}$. On the other hand, the second equality only implies that charge density is conserved in $\pi^i \cup \pi^o$. We notice that $\mathbf{J}$ is zero even with the significant current density $\mathbf{J}^i$ in the intracellular space if there is the same magnitude and direction of the current density $\mathbf{J}^o$ in the interstitial space. If we consider the membrane current density $\mathbf{J}^m$, then the current densities for each microscopic domain are expressed as $\mathbf{J}^i = \mathbf{J}^m$ and $\mathbf{J}^o = - \mathbf{J}^m$ and the current density difference $\mathbf{J}$ is expressed as $\mathbf{J} = \mathbf{J}^i - \lambda^{-1} \mathbf{J}^o =  ( 1 + 1/\lambda ) \mathbf{J}^m $. In the next section, the construction of the BvP oscillator from the Maxwell's equations (\ref{MW1}) - (\ref{MW4}) will show that the membrane current density $\mathbf{J}^m$ is a function of the scalar potential and its time derivative. Then, $\rho$ is also the function of a scalar potential such as $\mathbf{J}^m = \mathbf{J}^m ( \phi, \dot {\phi} ), ~\rho = \rho ( \phi )$. Each variable is naturally a function of permittivity $\varepsilon_i$ and conductivity $\sigma^i$, but we drop the notations for simplicity.

\begin{table}[htdp]
\caption{Fields and variables by weighted difference}
\begin{center}
\begin{tabular}{|c|c||c|c|}
\hline
Symbol & Definition & Symbol &  Definition \\
\hline
$\mathbf{E}$  & $ \mathbf{E}^i - \lambda \mathbf{E}^o$ & $\mathbf{B}$ & $\mathbf{B}^i - \lambda \mathbf{B}^o$ \\
$\rho$ & $ \rho^i - \lambda^{-1} \rho^o$ &  $\mathbf{J}$  & $\mathbf{J}^i - \lambda^{-1} \mathbf{J}^o$ \\
$\mathbf{A}$ & $  \mathbf{A}^i - \lambda \mathbf{A}^o$ & $\phi$ & $\phi^i - \lambda \phi^o$ \\
$\chi_{\alpha}$ & $q^i_{\alpha} - \lambda^{-1} q^o_{\alpha} $ & $\mathbf{v}_{\alpha}$ & $ (q^i_{\alpha} \mathbf{v}^i_{\alpha} - \lambda^{-1} q^o_{\alpha} \mathbf{v}^o_{\alpha}  )/\chi_{\alpha}$ \\
\hline
\end{tabular}
\end{center}
\label{weighted}
\end{table}

\section{Choice of gauge and membrane current density}
The FHN model, a diffusion-reaction model with the BvP oscillator, is popularly used for mathematical modeling of the excitation propagation, thus the derivation of the FHN model from the Maxwell's equations (\ref{MW1}) - (\ref{MW4}) mean that the two equations are actually equivalent or one system of equations are a subsystem of the other and may show that the Maxwell's equations (\ref{MW1}) - (\ref{MW4}) can also represent the dynamics of the cardiac excitation propagation. This derivation consists of two procedures: one is to derive the diffusion operator and the other is to derive the BvP oscillator for the reaction.

\subsection{Gauge choice}
Firstly, the diffusion operator is easily obtained by gauge choice. By applying the divergence operator to equation (\ref{MW6}) and using equation (\ref{MW1}), we obtain ${ \rho (\phi)} / {\varepsilon_i} = - \nabla^2 \phi - {\partial ( \nabla \cdot \mathbf{A} ) } / {\partial t}$. In Maxwell's equations, the choice of $\nabla \cdot \mathbf{A}$ is known as \textit{gauge} and remains redundant for the same fields $\mathbf{E}$ and $\mathbf{B}$, but is rather chosen according to the type of electromagnetic propagation \cite{Jackson1}. For example, with Coulomb gauge $\nabla \cdot \mathbf{A} = 0$, the above equation becomes $\nabla \phi^2 = {\rho (\phi) }/ {\varepsilon_i} $, which describes the instantaneous distribution of a scalar potential. Lorentz gauge defined as $ \nabla \cdot \mathbf{A} = (1/c^2) \partial \phi / \partial t$ transforms the above equation into $\nabla^2 \phi - (1/c^2) {\partial^2 \phi} / {\partial t^2} = - {\rho (\phi) } / {\varepsilon_i} $ which gives the special solution of the time dependent Poisson equation describing the retarded radiation \cite{Jackson2}. But, since neither of them seems to represent the dynamics of the propagation, we propose the new gauge to be defined as
\begin{equation}
\nabla \cdot \mathbf{A} = - \phi.  \label{DRgauge}
\end{equation}
By using this gauge, we obtain
\begin{equation}
 \frac{\partial \phi }{\partial t} =   \nabla^2 \phi  +  \frac{\rho(\phi)}{\varepsilon_i}  .    \label{DR1}
\end{equation} 
Note that the isotropic elliptic operator is obtained by the use of the new gauge. The physical meaning of the new gauge can be understood in several ways: If we integrate equation (\ref{DRgauge}) over a small region $\Omega \in \Pi$, then by the divergence theorem, we obtain $\int_{\partial \Omega} \mathbf{A} \cdot \mathbf{n} d S = - \int_{\Omega} \phi dV$. This equality means that the scalar potential $\phi$ is determined by the flux of the vector potential $\mathbf{A}$ across the boundaries. Moreover, if we decompose the vector potential $\mathbf{A}$ into the longitudinal component ($\mathbf{A}^{\parallel}$) and the transverse component ($\mathbf{A}^{\perp}$) such as $\mathbf{A} = \nabla A^{\parallel} + \nabla \times A^{\perp}$, then by substituting this expression into equation (\ref{DRgauge}), we obtain $ \nabla^2 A^{\parallel} = - \phi$ to imply that the longitudinal component of $\mathbf{A}$ determines $\phi$, but the transverse component still remains undetermined and independent of $\phi$. In general, we do not assume that the transverse component is zero because equation (\ref{MW2}) is $\mathbf{B} = \nabla \times ( \nabla \times A^{\perp} )$ implying that the magnetic field for the gauge choice (\ref{DRgauge}) is also zero. However, such a strong restriction is not required for the remaining analysis of this paper.

Nevertheless, the new gauge choice still does not determine the potentials because various potentials can produce the same electromagnetic field. For example, the following gauge transformation also yields the same electromagnetic field in equations (\ref{MW1}) - (\ref{MW4}); $\mathbf{A} \rightarrow \mathbf{A} + \nabla \Lambda$ and $\phi  \rightarrow \phi - {\partial \Lambda} / {\partial t}$ for a scalar function $\Lambda$ known as \textit{gauge function}. Substituting the above transformations into equation (\ref{DRgauge}) reveals that the gauge function for the gauge (\ref{DRgauge}) satisfies the simple diffusion equation ${\partial \Lambda} / {\partial t} - \nabla^2 \Lambda = 0$. Therefore, for any function $\Lambda$ satisfying the above equality, multiple scalar potentials being added by $- \nabla^2 \Lambda$ and its corresponding vector potentials produce the same electric field $\mathbf{E}$ and magnetic field $\mathbf{B}$. The existence of gauge function and subsequence gauge invariance are also well described in quantum mechanics for example from the invariance of the Pauli equation \cite{QEDFeynman}.

As aforementioned, this gauge choice (\ref{DRgauge}) does not change the electromagnetic field and its potential up to a constant, but reflects the different mechanism of the light oscillator for excitation, or signal oscillator for excitation in general. Consider the following wavelike equation being obtained by substituting the gauge choice (\ref{DRgauge}) into equation (\ref{MW4}):
\begin{equation*}
\nabla^2 \mathbf{A} - \frac{1}{c^2} \frac{\partial \mathbf{A}}{\partial t^2} = - \nabla \phi +  \frac{1}{c^2} \frac{\partial ( \nabla \phi) }{\partial t} + \mu_i \mathbf{J} (\phi, \dot{\phi} ) .
\end{equation*}
As a similar procedure done by Heitler \cite{Heitler}, let us consider the vector potential $\mathbf{A}$ as a series of orthogonal plane waves with the wave number $k_{\alpha}$ and frequency $\nu_{\alpha}$ such as $\mathbf{A} = \sum_{\alpha} a_{\alpha} (t) \mathbf{A}_{\alpha} (\mathbf{r})$ where $a_{\alpha}$ depends only on $t$, and $\mathbf{A}_{\alpha}$ depends only on $\mathbf{r}$. Expand $\phi$ similarly as $\phi= \sum_{\alpha} b_{\alpha} (t) \phi_{\alpha} (\mathbf{r})$. For simplicity, we only consider the transverse component of $\mathbf{A}_{\alpha}$ with the velocity of the electron as $\mathbf{v}_{\alpha} = \mathbf{v}_0 \cos \nu_0 t$, then the above equation turns out to be
\begin{equation}
\ddot{a}_{\alpha} + \nu_{\alpha}^2 a_{\alpha} = k_{\alpha} ( c^2 b_{\alpha} + \dot{b}_{\alpha} ) + f_{\alpha} \cos \nu_0 t ,  \label{Aosc}
\end{equation}
where $c$ is the velocity of electron $e$ equivalent to $4 \pi c^2 \rho$ such that $c = \nu_{\alpha} / k_{\alpha}$ and $f_{\alpha} = ({e}/{c}) v_0 \left | a_{\alpha} ( k)  \right | cos \Theta$ for the angle $\Theta$ between the polarization and the oscillator. For Lorentz gauge with no $b_{\alpha}$ and $\dot{b}_{\alpha}$ in equation (\ref{Aosc}), $a_{\alpha}$ is analytically given at time $t=0$ as $a_{\alpha} = ( f_{\alpha} /  ( \nu^2_{\alpha} - \nu^2_0 ) ) ( \cos \nu_0 t - \cos \nu_{\alpha} t )$, thus the oscillators are only excited at the wave with the same frequency as the electron. On the other hand, the new gauge (\ref{DRgauge}) generates additional terms involving the coefficients of $\phi$, $b_{\alpha}$ and $\dot{b}_{\alpha}$, as a source term added to $f_{\alpha} \cos \nu_0 t$. Thus, the excitation of the oscillators now significantly depends on the scalar potential $\phi_{\alpha}$ and its time variation $\dot{\phi}_{\alpha}$. Moreover, the energy of the oscillator $H_{\alpha}$ after the time $t$ subsequently depends on $b_{\alpha}$ and $\dot{b}_{\alpha}$, and subsequently $\phi_{\alpha}$ and $\dot{\phi}_{\alpha}$ as
\begin{equation*}
H_{\alpha} (k,t) = \int_0^t \dot{a}_{\alpha} \left ( k_{\alpha} ( c^2 b_{\alpha} + \dot{b}_{\alpha} ) + f_{\alpha} \cos \nu_0 t \right ) d t .
\end{equation*}
The presence of $\phi_{\alpha}$ and $\dot{\phi}_{\alpha}$ in the energy of the oscillator implies that (1) the energy of the oscillators having a frequency between $\nu$ and $\nu + d \nu$ is no more proportional to the time $t$ and (2) the amount of energy transferred to the oscillators is \textit{not the same} as that of the energy flow out of the oscillators.

\subsection{Choice of the membrane current density}

\begin{figure}[h]
\centering
\vbox{
\includegraphics[height=4cm, width=4cm] {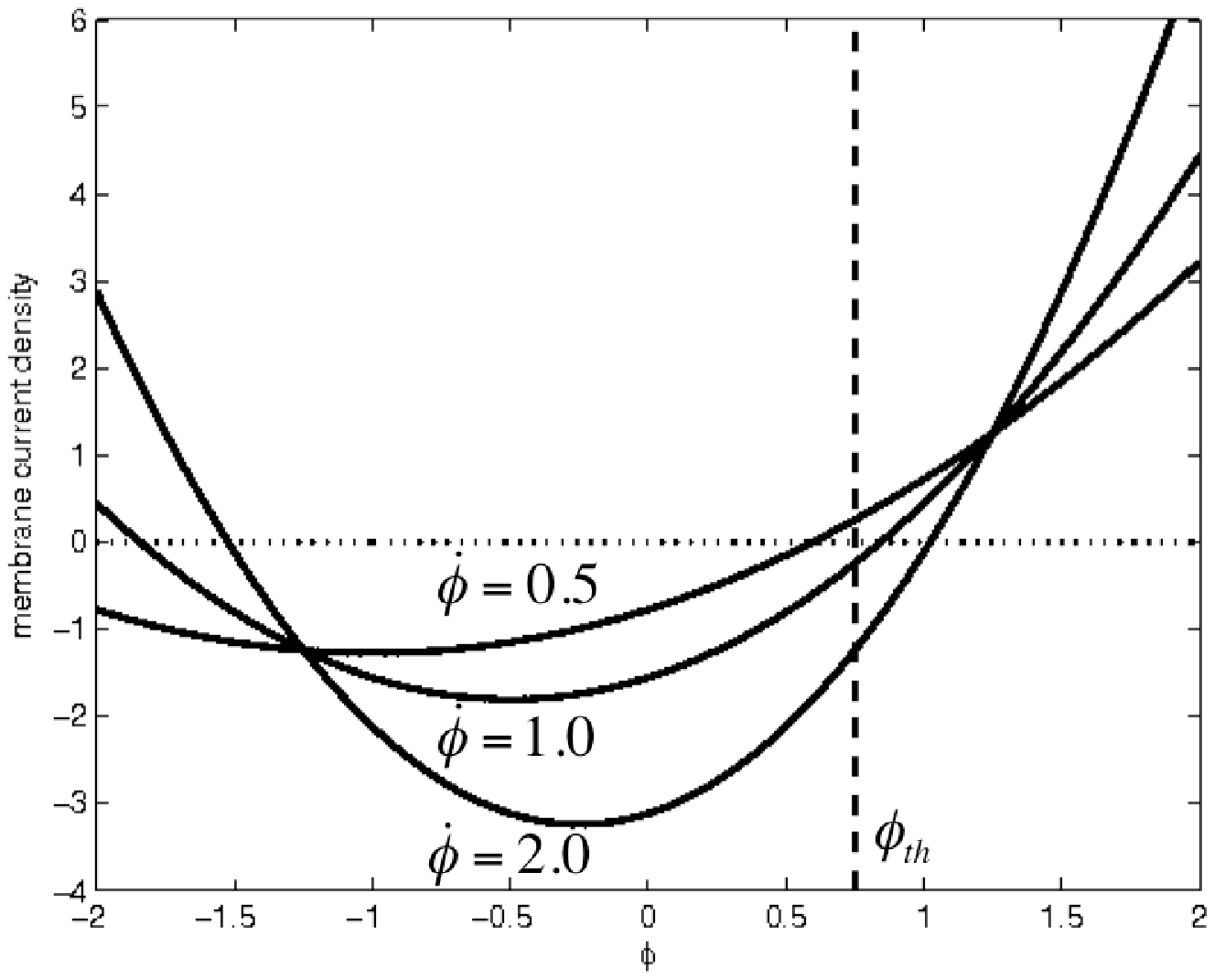}  \includegraphics[height=4cm, width=4cm] {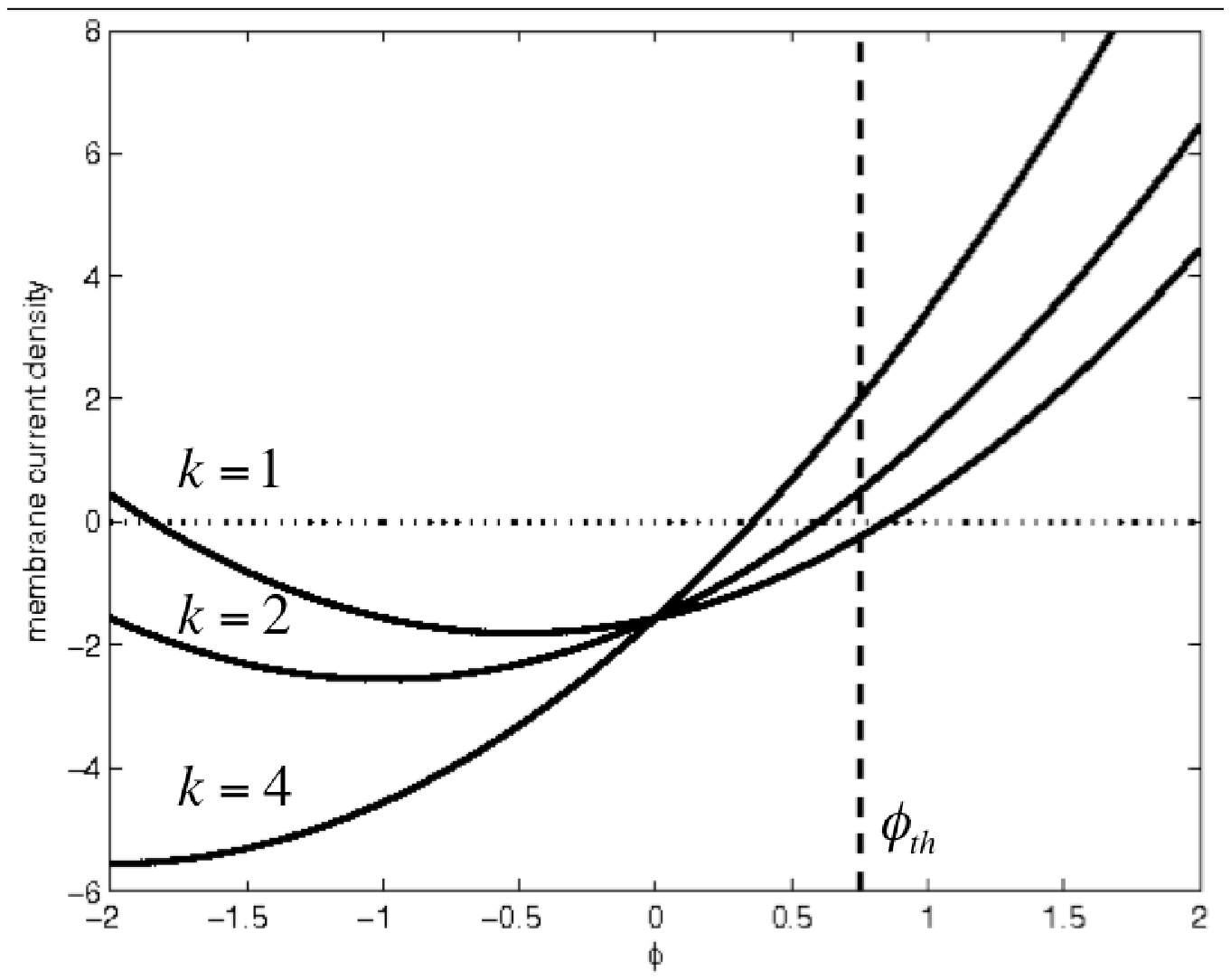}   }
\caption{The current density $\mathbf{j}_k$ versus the membrane potential $\phi$ for a constant wave number $k$ (left) and for a constant time variation of the scalar potential $\dot{\phi}$. $\sigma_i = \varepsilon_i =1.0$. $k=1$ for the left plot and $\dot{\phi} = 1.0$ for the right plot.}
\label {fig:mcd}
\end{figure}

Instead of assigning the charge density $\rho(\phi)$ for the BvP oscillator directly, an oscillator will be first constructed in reciprocal space for the membrane current density $\mathbf{J}^m (\phi,\dot{\phi})$ to retrieve the original form of the BvP oscillator. Then, the reaction function in real space will be subsequently determined. This procedure will yield similar results to those in the construction of $\rho(\phi)$ for the BvP model in real space, but will verify the connection between the oscillators in reciprocal space and the reaction function in real space. Let's consider all the fields and variables as running plain waves \cite{CohenPA} \cite{Pathint} such as
\begin{align*}
\mathbf{A} ( \mathbf{r},t) &= \frac{1}{ ( 2 \pi )^{3/2}}  \int \mathbf{a}_k ( \mathbf{k}, t ) e^{ i \mathbf{k} \cdot \mathbf{r} } d^3 k, \\
\phi ( \mathbf{r},t) &= \frac{1}{ ( 2 \pi )^{3/2}}  \int \phi_k ( \mathbf{k}, t ) e^{ i \mathbf{k} \cdot \mathbf{r} } d^3 k , \\
\mathbf{J} ( \mathbf{r},t) &= \frac{1}{ ( 2 \pi )^{3/2}}  \int \mathbf{j}_k ( \mathbf{k}, t ) e^{ i \mathbf{k} \cdot \mathbf{r} } d^3 k, \\
\rho ( \mathbf{r},t) &= \frac{1}{ ( 2 \pi )^{3/2}}  \int \rho_k ( \mathbf{k}, t ) e^{ i \mathbf{k} \cdot \mathbf{r} } d^3 k  ,\\
\mathbf{E} ( \mathbf{r},t) &= \frac{1}{ ( 2 \pi )^{3/2}}  \int \mathbf{e}_k ( \mathbf{k}, t ) e^{ i \mathbf{k} \cdot \mathbf{r} } d^3 k, \\
\mathbf{B} ( \mathbf{r},t) &= \frac{1}{ ( 2 \pi )^{3/2}}  \int \mathbf{b}_k ( \mathbf{k}, t ) e^{ i \mathbf{k} \cdot \mathbf{r} } d^3 k ,
\end{align*}
where $\mathbf{r}$ is the position vector, $\mathbf{k}$ is the wave vector and $t$ is the time variable. This is also known as the \textit{Fourier spatial transformation}. The fields $\mathbf{A}$, $\mathbf{J}$, $\mathbf{E}$, $\mathbf{B}$, $\phi$, and $\rho$ are in real space, while $\mathbf{a}_k$, $\mathbf{j}_k$, $\mathbf{e}_k$, $\mathbf{b}_k$, $\phi_k$ and $\rho_k$ are in the space, known as \textit{reciprocal space} or \textit{frequency domain}, where the subscript $k$ represents the coefficient of the plane wave with the wave number $k$. We only consider that all the fields are real such as
\begin{align*}
\mathbf{e}_k^* ( \mathbf{k}, t ) &= \mathbf{e}_k ( - \mathbf{k}, t),~~\phi_k^* ( \mathbf{k}, t ) = \phi_k ( - \mathbf{k}, t), \\
\mathbf{j}_k^* ( \mathbf{k}, t ) &= \mathbf{j}_k ( - \mathbf{k}, t) ,~~ \rho_k^* ( \mathbf{k}, t ) = \rho_k ( - \mathbf{k}, t), \\
\mathbf{e}_k^* ( \mathbf{k}, t ) &= \mathbf{e}_k ( - \mathbf{k}, t),~~\mathbf{b}_k^* ( \mathbf{k}, t ) = \mathbf{b}_k ( - \mathbf{k}, t) ,
\end{align*}
where the superscript $*$ means the complex conjugate. We often decompose vectors into the longitudinal vector fields and transverse vector fields: the longitudinal vector field $\mathbf{v}^{\parallel}$ is parallel to the wave vector $\mathbf{k}$ and is defined as $\mathbf{v}^{\parallel}_k ( \mathbf{k} ) \equiv \boldsymbol{\kappa} \left [ \boldsymbol{\kappa} \cdot \mathbf{v}_k ( \mathbf{k} )  \right ]$ for the normalized wave vector $\boldsymbol{\kappa} = \mathbf{k} / k $. The transverse vector field $\mathbf{v}^{\perp}$ is perpendicular to the wave vector $\mathbf{k}$ and is defined as $\mathbf{v}^{\perp}_k ( \mathbf{k} ) \equiv \mathbf{v}_k - \mathbf{v}^{\parallel}_k $. Then it can be easily shown that $i \mathbf{k} \cdot \mathbf{v}_k^{\perp} = 0 $ and $i \mathbf{k} \times \mathbf{v}_k^{\parallel}= 0$. In reciprocal space, the Maxwell's equations (\ref{MW1}) - (\ref{MW4}) are written as 
\begin{align}
i \mathbf{k} \cdot \mathbf{e}_k &= \frac{\rho_k ( \phi_k ) }{\varepsilon_i}  , \label{MWk1}   \\
i \mathbf{k} \cdot \mathbf{b}_k &= 0    , \label{MWk2}  \\
i \mathbf{k} \times \mathbf{e}_k &= - \dot {\mathbf{b}}_k   ,  \label{MWk3}  \\
\frac{1}{\mu_i} i \mathbf{k} \times \mathbf{b}_k &= \varepsilon_i \dot{\mathbf{e}}_k + \frac{1+\lambda}{\lambda} \mathbf{j}^m_k  ( \phi_k , \dot{\phi}_k )  , \label{MWk4} 
\end{align}
and equations (\ref{MW5}) - (\ref{MW6}) and the conservation of charge (\ref{conservcharge}) are given by
\begin{align}
 \mathbf{b}_k &= i \mathbf{k} \times \mathbf{a}_k , \label{MWk5} \\
 \mathbf{e}_k &= - i \mathbf{k} \phi_k - \dot{\mathbf{a}}_k, \label{MWk6}  \\ 
 \dot{\rho}_k &= - i \frac{1+\lambda}{\lambda} \mathbf{k} \cdot \mathbf{j}^m_k ,   \label{MWk7}
\end{align}
where the dot notation is used to represent the differentiation with respect to the time variable $t$ in reciprocal space. Then, the gauge choice (\ref{DRgauge}) is also expressed as $i \mathbf{k} \cdot \mathbf{a}_k = - \phi_k  , $ or, in the transverse direction, $\mathbf{a}^{\parallel}_k = -  ( {i \mathbf{k}}/{k^2 }) \phi_k $. Similarly, equation (\ref{DR1}) is well expressed in reciprocal space and by differentiating this equation with respect to $t$ and by substituting equation (\ref{MWk7}), we obtain
\begin{equation}
\ddot{\phi}_k + k^2 \dot{\phi}_k = - \frac{i}{\varepsilon_i} \frac{1 + \lambda}{\lambda} { \mathbf{k} \cdot  \mathbf{j}^m_k } .   \label{bvp0}
\end{equation}
The construction of $\mathbf{j}^m_k$ will be conveniently achieved by decomposing it into two components: One is the current density induced by the electric field $\mathbf{e}_k$ and the other is the current density induced by the BvP oscillator. Let the former component be denoted by $\mathbf{j}^c_k$, namely the \textit{conducting membrane current density} where $c$ stands for conducting, and the latter by $\mathbf{j}^r_k$, namely the \textit{reactive membrane current density}, where $r$ stands for reaction. Thus, $\mathbf{j}^m_k$ is given by
\begin{equation}
\mathbf{j}^m_k = \frac{\lambda}{1 + \lambda} \left (  \mathbf{j}^c_k + \mathbf{j}^r_k \right ) .  \label{membranecurrent}
\end{equation}
The conducting membrane current density $\mathbf{j}^c_k$ is simply caused by the membrane potential difference between $\pi^i$ and $\pi^o$. This is similar to the early model on ion channels based on electro-diffusion described by the Nernst-Planck equation \cite{Keenerbook}. $\mathbf{j}^c_k$ can be decomposed into two directions: One is in the parallel direction to $\mathbf{k}$ and the other is in the perpendicular direction such as $\mathbf{j}^c_k =  \mathbf{j}^{c \parallel}_k + \mathbf{j}^{c \perp}_k $ where $\mathbf{j}^{c \parallel}_k$ can be expressed as $\mathbf{j}^{c \parallel}_k = -\sigma_i \mathbf{e}^{\parallel}_k$ for the electric conductivity $\sigma_i$. Since $\mathbf{e}^{\parallel}_k$ can be expressed in terms of $\phi_k$ and $\dot{\phi}_k$ from equations (\ref{MWk5}), (\ref{MWk6}), and the gauge choice, the conducting current difference $\mathbf{j}^c_k$ is given by
\begin{equation}
\mathbf{j}^c_k =  - i \sigma_i \mathbf{k} (   \phi_k - \frac{1}{k^2} \dot{\phi}_k ) + \mathbf{j}^{c \perp}_k .  \label{cdchoice1} 
\end{equation}
On the other hand, the reactive membrane current density $\mathbf{j}^r_k$ is controlled by the macroscopic mechanism of the ion channels featured as a resilient oscillator. The choice of $\mathbf{j}^r_k$ is obviously not defined in the classical electrodynamics because the physical domain is not a bidomain. Thus, we resort to the previous modeling of the excitation mechanism in the biological tissue. For example, in order to reflect the biological mechanism of the membrane current flow as first modeled by FitzHugh \cite{FitzHugh1} \cite{FitzHugh2}, we adapt the BvP oscillator as shown in equation (\ref{BvPorig0}). Various ways of constructing BvP oscillators are possible, but for the sake of simplicity, we construct the simple $\mathbf{j}^r_k$ as
\begin{align}
\mathbf{j}^r_k &= i \varepsilon_i \frac{\mathbf{k}}{k^2} ( \phi_{th}^2  - \phi_k^2 ) \dot{\phi}_k,  \label{cdchoice2} 
\end{align}
where $\phi_{th}$ is called the \textit{threshold potential} as the lowest level of the electric potential for excitation. The membrane current density $\mathbf{j}^r_k$ is not defined as the weighted difference, but we let the positive sign of $\mathbf{j}^r_k$ be the influx into $\pi^i$ (or efflux of $\pi^o$) and the negative sign of $\mathbf{j}^r_k$ be the influx into $\pi^o$ (or efflux of $\pi^i$). If  $\phi_k$ is larger than the threshold potential $\phi_{th}$, then the membrane current occurs in the negative direction of the wave vector $\mathbf{k}$ and is added to the magnitude of $\mathbf{j}^{c \parallel}_k$ for the rapid increase of the potential difference $\phi_k$. On the other hand, if $\phi_k$ is less than the threshold potential $\phi_{th}$, then it flows in the direction of the wave vector $\mathbf{k}$ and it is likely to cancel out $\mathbf{j}^{c \parallel}_k$ which normally occurs in the opposite direction. Consequently, the cardiac cell is only excited when $\phi_k$ is sufficiently larger than $\phi_{th}$. In reciprocal space, the membrane current density $\mathbf{j}^r_k$ is constructed in the direction of the wave vector $\mathbf{k}$ such that $\mathbf{j}^r_k$ is proportional to $\dot{\phi}$ and $\phi^2_{ph} - \phi^2_k$. In real space, by the inverse Fourier transform, we obtain the following expression as
\begin{equation}
\mathbf{J}^r (\mathbf{r}, t) = \frac{1}{4 \pi \varepsilon_i} \int  ( \phi_{th}^2  - \phi^2 ) \dot{\phi} \frac{\mathbf{r} - \mathbf{r}'}{ | \mathbf{r} - \mathbf{r}' |^3 } d^3 r' .
\end{equation}   \label{membranecurrentreal}
This equality means that the reactive membrane current density $\mathbf{J}^r ( \mathbf{r}, t)$ is just the Coulomb field of which magnitude is proportional to $( \phi_{th}^2  - \phi^2 ) \dot{\phi}$. This is a direct consequence of the construction of $\mathbf{j}^r_k$ as the longitudinal wave in the direction of wave vector $\mathbf{k}$ in reciprocal space. This mechanism becomes more obvious when the membrane current density $\mathbf{j}^m_k$ is expressed as the sum of $\mathbf{j}^c_k$ and $\mathbf{j}^r_k$ such as
\begin{equation}
\mathbf{j}_k^m =  -  i \sigma_i \mathbf{k} \frac{\lambda}{1 + \lambda}  (   \phi_k - \frac{1}{k^2} \dot{\phi}_k ) + i \varepsilon_i \frac{\mathbf{k} }{k^2} ( \phi_{th}^2  - \phi_k^2 ) \dot{\phi}_k + \mathbf{j}^{c \perp}_k.  \label{currentdensitychoice}
\end{equation}
Drawn from equation (\ref{currentdensitychoice}), Figure \ref{fig:mcd} demonstrates that the membrane current density $\mathbf{j}_k^m$ increases almost quadratically as the membrane potential $\phi_k$ increases. Note this phenomenon is almost universal for every $\dot{\phi}_k$ and $k$ ignoring its diverse magnitude. The reason that $\mathbf{j}_k^m$ is not exactly zero at $\phi_k = \phi_{th}$ is due to the scalar potential induced by the conducting membrane current density $\mathbf{j}_k^c$. Substituting the expression (\ref{currentdensitychoice}) of $\mathbf{j}^m_k$ into equation (\ref{bvp0}) yields
\begin{equation}
\ddot{\phi}_k +  \left [ \phi^2_k - \left ( \phi^2_{th} - k^2 +  \frac{\eta_i}{c^2}  \right ) \right ] \dot{\phi}_k + k^2 \eta_i \phi_k  = 0 ,\label{BvP1}
\end{equation}
where $\eta_i \equiv \sigma_i / \varepsilon_i $ and $\omega \equiv k c $.

The threshold potential $\phi_{th}$ is an arbitrary scalar quantity for the membrane current density $\mathbf{j}_k^m$ in real space solely depending on the type of the excitable media. In reciprocal space, $\phi_{th}$ can almost be randomly chosen dependent on the type of excitable media, but the following argument shows its first-order dependency on the wave number $k$. Suppose that $\phi_{th}$ is a constant or at most a function of $k^{\delta}$ where $\delta < 1$. Then, as the wave number $k$ is sufficiently large in equation (\ref{BvP1}), which roughly implies that the wave is highly fluctuate in times and can be interpreted as a motion in a shorter distance space, the equation for the oscillator converges to $\dot{\phi}_k + \eta_i \phi_k = 0$. Therefore, $\phi_k$ has the following analytic solution as $\phi_k = c_k \exp ( - \eta_i t)$ for an arbitrary constant $c_k$. But, this behavior of the solution contradicts the well known fact that the membrane potential $\phi_k$ is zero in the resting state independent of time $t$ and the wave number $k$. On the contrary, if $\phi_{th}$ is a function of $k$, then equation (\ref{BvP1}) for a sufficiently high frequency yields $\phi_k = 0$ satisfying the fundamental conditions of $\phi_k$, though the meaning of a high frequency in the excitation propagation remains largely unknown.

\subsection{Constructing BvP oscillator}

As a consequence, we may choose $\phi_{th}$ as $\phi_{th} (k) \equiv \sqrt{1 + k^2 - \eta_i / c^2}$. Then, the equation (\ref{BvP1}) reduces to a simpler expression as
\begin{equation}
\ddot{\phi}_k + \left ( \phi^2_k -1 \right ) \dot{\phi}_k + k^2 \eta_i \phi_k  = 0  \label{BvP2} .
\end{equation}
Equivalently by introducing the variable $\psi_k$ from the Li\'{e}nard's transformation such as $\psi_k =$ ${\dot{\phi}_k} / {c^2}$ $+ {\phi_k^3} / {3}$ $ - \left (  \phi^2_{th} - k^2 \right .$ $\left . +  \eta_i {k}/{c}  \right ) \phi_k$, the BvP oscillator in reciprocal space caused by $\mathbf{j}^m_k$ is given as: for $a,~b>0$,
\begin{align}
\dot{\phi}_k &= \psi_k + \left ( \phi^2_{th} - k^2 + \frac{\eta_i}{c^2} \right ) \phi_k - \frac{\phi_k^3}{3} , \label{BvP3} \\
\dot{\psi}_k &= - k^2 ( \eta_i \psi_k  - a + b \phi_k ).    \label{BvP4} 
\end{align}
Note the similarity between equations (\ref{BvP2}) and (\ref{BvPorig0}), or (\ref{BvP3}) - (\ref{BvP4}) and (\ref{BvPorg1}) - (\ref{BvPorg2}). Equation (\ref{BvP2}) has the additional component of $k$ partly because they lie in the different spaces, but the corresponding oscillators are in principle the same kind as the BvP oscillator of equation (\ref{BvPorig0}) because they have the same \textit{quadratic damping factor}. More analysis can be drawn from the vector potential that is also written as an oscillator in reciprocal space. With equations (\ref{MWk4}), (\ref{MWk6}), and (\ref{currentdensitychoice}), the dynamics of the vector potential $\mathbf{a}_k$ in reciprocal space is given by
\begin{align}
& \ddot{\mathbf{a}}_k^{\perp} + c^2 k^2 \mathbf{a}^{\perp}_k = c^2 \mu_i \mathbf{j}^{\perp}_k, \label{vectorpotentialdynamics1}\\
& \ddot{\mathbf{a}}_k^{\parallel} - c^2 k^2 \dot{\mathbf{a}}^{\parallel}_k + 2 c^2 k^2 \mathbf{a}^{\parallel}_k = c^2 \mu_i \mathbf{j}^{\parallel}_k .\label{vectorpotentialdynamics2}
\end{align}
The BvP oscillator does not change the dynamics of $\mathbf{a}^{\perp}_k$ which remains the same as a harmonic oscillator of classical electrodynamical waves. But it significantly changes the dynamics of $\mathbf{a}^{\parallel}_k$ crucial for the absorption and emission of the propagating charged particles. Comparing equation (\ref{vectorpotentialdynamics2}) with equation (\ref{Aosc}) immediately reveals that the equivalent term of $( c^2 b_{\alpha} + \dot{b}_{\alpha} )$ in reciprocal space is only substituted by $k \dot{\mathbf{a}}_k^{\parallel}$. Roughly stated, this means that the membrane potential and its current actually contribute to the oscillators representing $\mathbf{a}^{\parallel}$ as a \textit{damping factor}. This result is in accord with the BvP oscillator (\ref{BvPorig0}).

\section{Maxwell's equations with the BvP oscillator}

\subsection{Constructing the reaction function}

To obtain a set of Maxwell's equations equivalent to the FHN equations, the charge density difference $\rho$ only needs to be derived from the current density difference $\mathbf{j}_k$ (\ref{currentdensitychoice}) representing the BvP oscillator in reciprocal space. By substituting equation (\ref{currentdensitychoice}) into equation (\ref{MWk7}), the time derivative of the charge density difference is given by
\begin{equation}
\dot{\rho}_k = \dot{\rho}^c_k + \dot{\rho}^r_k = - \sigma_i k^2 \left ( \phi_k - \frac{1}{k^2} \dot{\phi}_k  \right )  + \varepsilon_i ( \phi^2_{th} - \phi^2_k ) \dot{\phi}_k .   \label{chargedensity1}
\end{equation}
The first term is clearly induced by the conducting current density $\mathbf{j}^c_k$ and is called the \textit{conducting membrane charge current}, denoted by $\dot{\rho}^c_k$. The second term is similarly induced by the reactive membrane current density $\mathbf{j}^r$ and is called  the \textit{reactive membrane charge current} denoted by $\dot{\rho}^r_k$. Then, equation (\ref{chargedensity1}) can be naturally decomposed into the two components of $\mathbf{j}^c_k$ and $\mathbf{j}^r_k$ such as $\dot{\rho}_k = \dot{\rho}^c_k + \dot{\rho}^r_k$, but we use the following decomposition for $\dot{\psi}_k$ and $\dot{\xi}_k$ such as $\dot{\rho}_k = (1/\varepsilon_i) \dot{\psi}_k + \dot{\xi}_k$ leading to the FHN equations for the BvP oscillator: For an arbitrary time-dependent function $f(t) \in \mathbb{R}$,
\begin{align*}
 \dot{\psi}_k&=  - \frac{\dot{\rho}^c_k }{\varepsilon_i}  - 2 \eta_i \dot{\phi_k}  - \eta_i \frac{\phi_k^3}{3} + f (t) \eta_i \phi_k  ,  \\
 \dot{\xi}_k &=   \dot{\rho}^r_k - f (t) \eta_i \phi_k +  \eta_i \frac{\phi_k^3}{3} + 2 \sigma_i  \dot{\phi_k}   .
 \end{align*} 
Then the charge density difference $\rho = \rho (\mathbf{r},t)$ in real space is simply expressed as
\begin{align*}
\frac{\partial \rho }{\partial t} = \frac{\partial \psi }{\partial t} -  \eta_i \left ( f(t) \phi  +  \frac{\phi^3}{3} \right ) + \varepsilon_i f_1\frac{\partial \phi }{\partial t} - \frac{\varepsilon_i}{3} \frac{\partial^3 \phi  }{\partial t} ,
\end{align*}
where $f_1 (\mathbf{r} ) = (2 / \pi)^{-3/2} \int \left ( \phi^2_{th} + 2 \sigma_i \right ) e^{i \mathbf{k} \cdot \mathbf{r} } d^3 k $. If $\phi_{th}$ is a constant independent of $k$, then $f_1$ is just $\phi^2_{th} + 2 \sigma_i$. Nevertheless we consider the general function of $\phi_{th} (k)$, thus we maintain this general expression of $f_1 \in \mathbf{R}$ being independent of the time $t$. Since $f(t)$ is an arbitrary function and $\phi(t)$ can be best approximated as a polynomial, it is always possible to choose $f(t)$ such that $\int_0^t \phi (\mathbf{r}, t) ( f(t) + \phi (\mathbf{r}, t)^2/3 ) dt'$ is a constant independent of time. Let the value of this integration be $f_1 \in \mathbf{R}$. Suppose that $ \phi$ and $\psi$ are all zero at $t=0$. By considering $\phi$ and $\sigma_i$ as being time independent, the above equation is integrated with respect to time to obtain
\begin{equation}
 \rho ( \mathbf{r}, t) =  \varepsilon_i  \left [ \psi ( \mathbf{r}, t) + f_0 + f_1 \phi ( \mathbf{r}, t) - \frac{\phi^3 ( \mathbf{r}, t)}{3} \right ] .  \label{chargedensitychoice}
\end{equation}
Substituting equation (\ref{chargedensitychoice}) and the expression of $\psi$ in real space into equation (\ref{DR1}), we obtain the following diffusion-reaction system:
\begin{align}
\frac{\partial \phi}{\partial t} &= \nabla^2 \phi +   \psi + f_0+  f_1 \phi - \frac{ \phi^3 }{ 3},  \label{DRf1}\\
\frac{\partial \psi}{\partial t} & = - \eta_i \left (  \psi + f_0+  ( f_1 - f_0 )  \phi \right ) .  \label{DRf2}
\end{align}
Compare the above equations with the FHN equations (\ref{FHN1}) and (\ref{FHN2}). The reaction functions of the FHN equations vary from model to model, but we may say that the above equations share the same properties of the reaction function with those of the FHN equations because the function has the components of $\phi$, $\psi$, and $\phi^3$ representing the BvP properties of the reaction function. For example, if we let $f_0 = 0$ and $f_1 = -1$, then the reaction functions (\ref{DRf1}) - (\ref{DRf2}) are exactly the same as those of the original FHN equations \cite{Zykov2} \cite{FitzHugh1} \cite{Nagumo}. The diffusion-reaction system with the BVP oscillator, which displays the dynamics of the scalar potential $\phi$ without the other three components of the vector potential $\mathbf{A}$, is the subsystem of the Maxwell's equations (\ref{MWf1}) - (\ref{MWf4}). But, all the four equations of the Maxwell's equations (\ref{MWf1}) - (\ref{MWf4}) should be used to derive the diffusion-reaction system (\ref{DRf1}) - (\ref{DRf2}) for the excitation propagation in real space. This means that neither of the Maxwell's equations is redundant for the diffusion-reaction system.

The derivation of the anisotropic diffusion-reaction system with the BvP model can be obtained with a similar procedure. Applying the divergence operator and tensor product with the electric conductivity tensor $\boldsymbol{\sigma}$ to the expression of equation (\ref{MWk6}) in real space, we obtain the following equation:
\begin{equation*}
\nabla \cdot ( \boldsymbol{\sigma} \mathbf{E} ) = - \nabla \cdot \boldsymbol{\sigma} \nabla \phi - \frac{\partial ( \nabla \cdot ( \boldsymbol{\sigma} \mathbf{A}) ) }{\partial t} .
\end{equation*}
Then, the gauge choice and the Coulomb's law are modified as
\begin{equation*}
 \nabla \cdot \left ( \boldsymbol{\sigma} \mathbf{A} \right )= - \phi ,~~~ \nabla \cdot \left ( \boldsymbol{\sigma} \mathbf{E} \right ) = \frac{  \rho_{ani} ( \phi )}{\varepsilon_i} ,
\end{equation*}
where ${\rho}_{ani}$ is the new charge density depending both on $\mathbf{E}$ and $\boldsymbol{\sigma}$. This means that the conductivity tensor $\boldsymbol{\sigma}$, which is  a non-identity tensor due to the cylindrical shape of the cardiac fibre, can significantly increase or decrease the charge density ${\rho}_{ani}$ and consequently the potentials $\phi$ and $\mathbf{A}$. Deriving the charge density from the BvP model for the above equation is beyond the scope of this paper, so we can simply use the same reaction functions of equation (\ref{chargedensitychoice}) again to obtain
\begin{align*}
\frac{\partial \phi}{\partial t} &= \nabla \cdot \boldsymbol{\sigma} \nabla \phi +   \psi + f_0+  f_1 \phi - \frac{ \phi^3 }{ 3}, \\
\frac{\partial \psi}{\partial t} & = - \eta_i \left (  \psi + f_0+  ( f_1 - f_0 )  \phi \right ) . 
\end{align*}
What remains is to obtain the expression for the Maxwell's equations with $\rho^m$ and $\mathbf{J}^m$. With $\phi_{th} (k) = \sqrt{1 + k^2 - \eta_i / c^2}$ as before and by substituting $\rho^m$ from equation (\ref{chargedensity1}) and $\mathbf{J}^m$ from equation (\ref{currentdensitychoice}) into the Maxwell's equations (\ref{MW1}) - (\ref{MW4}), we finally obtain the following Maxwell's equations with the BvP oscillators:
\begin{align}
\nabla \cdot \mathbf{E} &=  \eta_i \int_0^t \nabla^2 \phi dt' + \Gamma_i \phi - \nabla^2 \phi - \frac{\phi^3}{3}  ,    \label{MWf1} \\
\nabla \cdot \mathbf{B} &= 0,  \label{MWf2} \\  
\nabla \times \mathbf{E} &= - \frac{\partial \mathbf{B} }{\partial t}, \label{MWf3}  \\
\frac{1}{\mu_i} \nabla \times \mathbf{B}  &= \varepsilon_i \frac{\partial \mathbf{E} }{\partial t} + \sigma_i \nabla \phi  -\varepsilon_i \frac{\partial \nabla \phi}{\partial t}  \\
&~~- \frac{\varepsilon_i}{4 \pi }\int  \left [ \Gamma_i \frac{\partial \phi}{\partial t} + \frac{\partial}{\partial t} \left ( \frac{\phi^3 }{3} \right )  \right ] \frac{\mathbf{r} - \mathbf{r}'}{| \mathbf{r} - \mathbf{r}' |^3 } d^3 r' ,\label{MWf4}
\end{align}
where we introduced the new variable $\Gamma_i \equiv 1 + \eta_i - \eta_i / c^2$.\\
\\
\textbf{Proposition 2}: The Maxwell's equations (\ref{MWf1}) - (\ref{MWf4}), which are derived from equations (\ref{MW1}) - (\ref{MW4}) with the new gauge (\ref{DRgauge}) and the choice of the membrane current density (\ref{currentdensitychoice}), are equivalent to the diffusion-reaction system with the BvP oscillator for the membrane potential $\phi$. \\
\\
\textbf{Proof}: See sections II - IV. $\square$\\
\\
The Maxwell's equations (\ref{MWf1}) - (\ref{MWf4}) are now completely described in the macroscopic domain $\Pi$ of physical space similar to those of classical electromagnetic waves, but have two unique properties: (i) The first property is the presence of the additional point charge and the current density that are not directly induced by conductivity. In the bidomain, these terms represent the membrane current and ion-pumped point charge, but for the classical electrodynamics in the physical domain, they are often regarded as the sources of charges. Then a question arises on how we interpret these sources. (ii) The second property is the presence of a varying point charge $\chi_{\alpha}$ for the electromagnetic field. Depending on the activation of ion pumps, the magnitude of point charge changes or even the signs of it changes. This phenomenon may not exist in the real world since it is not likely to obey the conservation of charges. Then, what is the role of point charge $\chi_{\alpha}$ if it is not to perturb the conservation of total energy or momentum of the total system?

These non-classical components can be made into analogies of several physical phenomena. Leaving the role of $\chi_{\alpha}$ to quantum theory, the first unique property will be explained in the next subsection in the perspective of semiclassical theory of radiation \cite{Scully}.

\subsection{Interpretation by the semiclassical theory of radiation}

Let us consider the excitation propagation as the electromagnetic field generated by a traveling \textit{solid cation}. Let the \textit{radiation} of the excitation propagation mean the same as the classical electromagnetic waves such that the cell can be excited only by radiation, or the transverse electromagnetic waves excluding the weak Coulomb interaction between the cations. Then we soon realize that some energy should be \textit{radiated} by this cation to excite the media. Thus, the total energy of this cation should diminish as it travels. In other words, the energy loss occurs at every change of velocity of the cation, thus even if the states of motion of the particle is the same, the energy of it may be different. But, the energy of the cation remains the same all the time in a perfect homogeneous media. Otherwise, the excitation should depend on the distance from the initial source.

A similar argument is applied to the radiation of photon. If we assume that a photon is a solid particle, then light is radiated by an accelerating photon, thus a subsequent energy loss occurs. But, similar to the excitation propagation, the energy of the photon does not depend on the distance from the source in non-dissipative media. This contradiction gives birth to an idea of \textit{reaction} or \textit{self-force} to compensate for the energy loss. Consider a particle with mass $m$ traveling with acceleration $\dot{\mathbf{v}}$. Since the energy radiated per unit time by this particle is given by $(2/3) e^2 \dot{\mathbf{v}}/ c^3$, the external force $\mathbf{K}$ should be expressed with the self-force $\mathbf{K}_s$ for this energy loss such as
\begin{equation*}
\mathbf{K} + \mathbf{K}_s = m \dot{\mathbf{v}},
\end{equation*}
where $\mathbf{K}_s$ can be easily obtained as $( 2/3) (e^2/c^3) \ddot{\mathbf{v}}$ \cite{Heitler}. In the excitation propagation, the acceleration of the cation does not require the existence of the solid object of the cation, but can be deduced by the shape of the wavefront from the geometric relation between trajectory and wave front \cite{Chun1}. The source of the energy of $\mathbf{K}_s$ has remained largely unknown in a vacuum, but should respond immediately to the acceleration of the moving particle to maintain its total energy.

A similar idea has been adapted to the semiclassical theory in the \textit{self-consistent equations} \cite{Scully} \cite{Scully1972} . The only modification is the use of a dipole moment and polarization induced by the electromagnetic field. This dipole moment being induced by the field interacting with the atom yields the generation of polarization density $\mathbf{P}(\mathbf{R},t)$ which acts as a source in Maxwell's equations such that
\begin{equation}
\nabla \times ( \nabla \times \mathbf{E} (\mathbf{r}, t ) ) + \frac{1}{c^2} \frac{\partial  \mathbf{E} (\mathbf{r}, t)}{ \partial t^2 } = - \mu_i \frac{\partial^2 \mathbf{P(\mathbf{R},t)}}{\partial t^2} . \label{semiclassical}
\end{equation}
Comparing this equation with equations (\ref{MWf3}) and (\ref{MWf4}) confirms that the membrane current density is only substituted by the second derivative of polarization density with respect to time. Moreover, since polarization density can be expressed by the membrane current density, we may say that equation (\ref{semiclassical}) is actually in the same form as equations (\ref{MWf3}) and (\ref{MWf4}). Let us consider the equality relation that the polarization current $\dot{\mathbf{P}}$ is the same as the polarization current as the difference between the current density $\mathbf{J}$ and the magnetization current $ \mathbf{J}^{g}$ such as
\begin{equation*}
\dot{\mathbf{P}} = \mathbf{J} - \mathbf{J}^{g},~~~~~~~\mbox{where}~~\mathbf{J}^g = \nabla \times \mathbf{M},
\end{equation*}
where $\mathbf{M}$ is the magnetization density derived as $\sum_{\alpha} \int_0^1 u \chi_{\alpha} \mathbf{r}_{\alpha} \times \dot{\mathbf{r}}_{\alpha} \delta (\mathbf{r} - \mathbf{r}_{\alpha} ) d u$ for the discretized charge density. The membrane current density is just the portion of the current density except for the conducting current and the magnetic polarization such that
\begin{equation}
\mathbf{J}^m = \frac{\lambda}{1 + \lambda} \left ( \mathbf{J} - \mathbf{J}^g \right ).
\end{equation}

\section{Conservation of total charge, momentum and energy}
The Maxwell's equations (\ref{MWf1}) - (\ref{MWf4}) are now completely described and represent the diffusion-reaction system with the BvP oscillator in real space. But, it remains in question whether the system of these equations (\ref{MWf1}) - (\ref{MWf4}) is under conservational laws in views of charge, momentum, and energy.

\subsection{Conservation of total charge}

In principle, conservational laws fails if only one domain of the bidomain is considered, but the total number of point charges should remain constant in the bidomain $\Pi$ if no flux occurs at the boundaries. For simplicity, consider that only the cations are propagated along the velocity vector of the excitation propagation and the cations are transported through the membrane. For point charge $\chi_{\alpha}$ at $\mathbf{r}_{\alpha}$, let us consider the operator $m^+: \chi_{\alpha} \rightarrow \mathbf{I}^+$ and $m: \chi_{\alpha} \rightarrow \mathbf{I}^+$ returning the \textit{total number of the cation} in $\pi^i$ and $\pi^o$, respectively, such that
\begin{equation*}
m^+  \chi_{\alpha}  (\mathbf{r})= N_i,~~m  \chi_{\alpha}  = N_o ,
\end{equation*}
where $N_i$ and $N_o$ are the total number of the cations in $\pi^i$ or $\pi^o$, respectively. Let $Q$ be the electric charge of one cation both for $\pi^i$ and $\pi^o$. Then it is easy to verify that $Q ( m^+ - \lambda^{-1}m) \chi_{\alpha} = \chi_{\alpha}$. Moreover, let us introduce the operator $\tau^+$ which transports one cation from $\pi^o$ to $\pi^i$, called the \textit{membrane influx} operator. Similarly, the operator $\tau$ which transports one cation from $\pi^i$ to $\pi^o$, is called the \textit{membrane efflux} operator. Then it can be easily verified that
\begin{equation*}
m^+ ( \tau^+ \chi_{\alpha} ) = N_i + 1,~~ m ( \tau \chi_{\alpha} ) = N_i- 1  .
\end{equation*}
Let $\mathcal{P}$ be the ion pump operator which transports a number of the cations from $\pi^i$ to $\pi^o$ or vice versa. Note that $\mathcal{P}$ is no more than the combinations of $\tau^+$ and $\tau^-$ such that $\mathcal{P} =  ( \tau^+ )^{N_+} + ( \tau )^{N_-}$ for the number of the transport $N_+$ and $N_-$ of each operator $\tau^+$ and $\tau$, respectively. For example, after ion pumping, a new point charge $\chi_{\alpha}^{new}$ at $\mathbf{r}_{\alpha}$ can be obtained such that
\begin{align*}
\chi_{\alpha}^{new} =   Q ( N_i -  \lambda^{-1} N_o + N_+ -  \lambda^{-1} N_- ) .
\end{align*}
As we assume, $\chi_{\alpha} > 0$ for $N_i >N_o$, but the signs of $\chi_{\alpha}^{new}$ may vary such that
\begin{equation*}
\chi_{\alpha}^{new} = \left \{
\begin{array}{c}
   >0, ~~~~~\mbox{if}~~N_- -  \lambda N_+ <  \lambda N_i - N_o , \\
    <0, ~~~~~\mbox{if}~~N_- -  \lambda N_+ >  \lambda N_i -  N_o . \\
\end{array}
\right .
\end{equation*}
Moreover, $N_- - N_+$ is mainly determined by the BvP oscillator depending on  the membrane potential $\phi (\mathbf{r}_{\alpha}, t) $, thus we can say that $\chi_{\alpha}$ is also a function of the membrane potential $\phi$ such that $\chi_{\alpha} =\chi_{\alpha} ( \phi, \mathbf{r}, t  ) $. Then the conservation of total number of the cations and total charge is expressed as follows:\\
\\
\textbf{Proposition 3}: Suppose that there is no flux of charged particles at the boundaries of domain $\Pi$. Then, in the dynamical system for the Maxwell's equations (\ref{MWf1}) - (\ref{MWf4}), the total number of the cation for the finite number $N$ of point charge is conserved in $\Pi$ such that
\begin{equation}
\sum_{\alpha=1}^N (m^+ + m) \chi_{\alpha} (t) = \sum_{\alpha=1}^N (m^+ + m) \chi_{\alpha} (0), ~~~~\forall t >0 .   \label{TNconserved}
\end{equation}
Let $Q^i$ be the electric charge of the cation in $\pi^i$ and let $Q^o$ be the electric charge of it in $\pi^o$. If $Q^o$ is different from $- \lambda Q^i$, then the total charge $\sum_{\alpha} \chi_{\alpha}$ is not conserved such that
\begin{equation}
\sum_{\alpha=1}^N  \chi_{\alpha} (t) \neq  \sum_{\alpha=1}^N  \chi_{\alpha} (0),~~~~\mbox{any}~ t>0   . \label{TCconserved}
\end{equation}
\textbf{Proof}: The total number of point charges $N$ is fixed and the only changes can be made by the ion pump operator $\mathcal{P}$. Thus, it is enough to show that the above quantities are conserved by the operation of the membrane flux operator $\tau^+$ and $\tau$ because $\mathcal{P}$ is solely a function of $\tau^+$ and $\tau$. By applying the membrane influx operator $\tau^+$ to a number of the cations at $\mathbf{r}_{\alpha}$ of equation (\ref{TNconserved}), we obtain the conservation of the total number of the cations as
\begin{equation*}
 (m^+ + m) \tau^+ \chi_{\alpha}  = (N_i + 1 + N_o - 1 ) =  (m^+ + m) \chi_{\alpha} .
\end{equation*}
But note that the total difference of the number of the cation, i.e. $\sum_{\alpha} (m^+ - m) \chi_{\alpha} $, is not preserved as
\begin{equation*}
(m^+ - m) \tau^+ \chi_{\alpha} =   (m^+ - m) \chi_{\alpha} +2.
\end{equation*}
Similarly, by applying $\tau^+$ to the total charge of the cations at $\mathbf{r}_{\alpha}$ (\ref{TCconserved}) and by using $Q^o = - \lambda Q^i$, we obtain the conservation of total charge as
\begin{equation*}
\tau^+ \chi_{\alpha} = \left ( Q^i m^+ - \lambda^{-1} ( - \lambda  Q^i )  m \right ) \tau^+ \chi_{\alpha}  = \chi_{\alpha}.
\end{equation*}
If $Q^o \neq - \lambda Q^i$, then we can easily verify that the total charge is not conserved. For example, with the same electric charge $Q$ for one cation both in $\pi^i$ and $\pi^o$, total point charge $\sum_{\alpha} \chi_{\alpha}$ is not preserved as
\begin{equation*}
\tau^+  \chi_{\alpha} = \sum_{\alpha}^N Q (m^+ - \lambda^{-1} m) \tau ^+ \chi_{\alpha}   = \chi_{\alpha} - 1 - \lambda, 
\end{equation*}
though the magnitude is bounded due to the conservation of the total number of the cation. Similar arguments can be easily shown for the membrane efflux operator $\tau$ $\square$. \\
\\
Proposition 3 implies the different interpretation of ion pumps in the macroscopic domain $\Pi$. In the biological tissue, ion pumps simply transport a charged ion from $\pi^i$ to $\pi^o$ or vice versa. Thus, ion pumps initiate the change of locations of the ion while the electric charge of the ion remains unchanged. In $\Pi$, however, ion pumping only changes the sign and the electric charge of the ion, but does not change the location of it. Moreover, the equality condition ($Q^o = - \lambda Q^i$) leads to the conservation of point charge such that $\chi_{\alpha}$ never changes its signs and magnitude independent of ion pump $\mathcal{P}$. But, in the bidomain, it is natural to set $Q^o$ as the same sign and magnitude of $Q^i$, thus it is inevitable to violate the conservation of charge in $\Pi$, which causes many peculiar properties in the mechanism of the propagation of the biological waves different from physical waves. In the next sections, we will observe the effects of this varying point charge $\chi_{\alpha}$ on the Lagrangian and Hamiltonian of the Maxwells' equations (\ref{MWf1}) - (\ref{MWf4}).

\subsection{Conservation of total energy and momentum}
Before proceeding further, we will first verify that the Newton-Lorentz equation still holds for the Maxwell's equations (\ref{MWf1}) - (\ref{MWf4}) which are necessary for the proof of the conservation of total energy and momentum.\\
\\
\textbf{Lemma 1}: Let $\mathbf{v}_{\alpha}$ be the velocity of the particle indexed $\alpha$ which has mass $m_{\alpha}$ and point charge $\chi_{\alpha}$. Then, the Newton-Lorentz equation is valid for the Maxwell's equations (\ref{MWf1}) - (\ref{MWf4}) such that
\begin{equation}
{m}_{\alpha} \frac{d {\mathbf{v}}_{\alpha} }{d t} = \chi_{\alpha} \left [ \mathbf{E} + \mathbf{v}_{\alpha} \times \mathbf{B} \right ] .  \label{NewtonLorentz}
\end{equation}
\textbf{Proof}: Since the point charge lies microscopically either in $\pi^i$ or $\pi^o$, axiom 2 and 3 imply that it is sufficient to show that equation (\ref{NewtonLorentz}) expresses the Newton-Lorentz equation for each microscopic domain $\pi^i$ or $\pi^o$. If the point charge lies in $\pi^i$, then by equation (\ref{defpcandvel}), the above equation reduces to
\begin{equation*}
{m}_{\alpha} \frac{d {\mathbf{v}}_{\alpha}^i }{d t} = q^i_{\alpha} \left [ \mathbf{E}^i + \mathbf{v}_{\alpha} \times \mathbf{B}^i \right ] ,
\end{equation*}
which is just the Newton-Lorentz equation in the intercellular space $\pi^i$. On the other hand, if the point charge lies in $\pi^o$, then, by equation (\ref{defpcandvel}), equation (\ref{NewtonLorentz}) reduces to
\begin{equation*}
{m}_{\alpha} \frac{d {\mathbf{v}}_{\alpha}^o }{d t} =  q^o_{\alpha} \left [ \mathbf{E}^o + \mathbf{v}_{\alpha} \times \mathbf{B}^o \right ] , 
\end{equation*}
which is just the Newton-Lorentz equation in the interstitial space $\pi^o$ $\square$. \\
\\
Using the Newton-Lorentz equation (\ref{NewtonLorentz}), we obtain two propositions on the conservation of the total energy and the total momentum of the closed dynamical system. These conservational laws are actually the same as the classical Maxwell's equations with the Coulomb gauge \cite{CohenPA}. The conservation of total energy and momentum is the direct consequence of the Newton-Lorentz equation on the supposition of axiom 2 and 3, and the intact form of equations (\ref{MWf3}) and (\ref{MWf4}) resulting from the fact that the membrane current density is a point-wise current which only changes point charge $\chi_{\alpha}$ without adding charge current $\chi_{\alpha} \mathbf{v}_{\alpha}$. Thus, the proofs are similar and will be provided in Appendix I for interested readers. \\
\\
\textbf{Proposition 4}. Consider a closed domain $\Pi$ such that no flux occurs at the boundary. Then the energy of moving particles with mass $m_{\alpha}$ traveling in the electromagnetic field for the Maxwell's equations (\ref{MWf1}) - (\ref{MWf4}) is well defined as
\begin{equation}
\mathbf{U} = \sum_{\alpha} \frac{1}{2} m_{\alpha} \mathbf{v}^2_{\alpha} + \frac{\varepsilon_i}{2} \int \left [ \mathbf{E}^2 + c^2 \mathbf{B}^2  \right ] d^3 r   \label{totalenergy}
\end{equation}
and is conserved in $\Pi$ independent of time.\\
\\
\textbf{Proof}: With the Newton-Lorentz equation (\ref{NewtonLorentz}), see Appendix IA. \\
\\
\textbf{Proposition 5}. Consider a closed domain $\Pi$ such that no flux occurs at the boundary. Then the total momentum of moving particles with mass $m_{\alpha}$ traveling in the electromagnetic fields for the Maxwell's equations (\ref{MWf1}) - (\ref{MWf4}) is well defined as
\begin{equation}
\mathbf{P} = \sum_{\alpha} m_{\alpha} \mathbf{v}_{\alpha} + \varepsilon_i \int \left [ \mathbf{E} \times \mathbf{B}  \right ] d^3 r  \label{totalmomentum}
\end{equation}
and is also conserved in $\Pi$ independent of time.\\
\\
\textbf{Proof}: With the Newton-Lorentz equation (\ref{NewtonLorentz}), see Appendix IB.

\section{Lagrangian}

To study the effects of the time-varying point charge $\chi_{\alpha}$ and the membrane current density $\mathbf{J}^m$ on the classical or quantum mechanical paths of the cations in Maxwell's equations (\ref{MWf1}) - (\ref{MWf4}), we consider the most popularly-used Lagrangian $\mathcal{L}$, known as the \textit{standard Lagrangian}, for the system of the particles and the electromagnetic field \cite{LandauLifshitzV2} \cite{CohenQM}:
\begin{equation}
\mathcal{L} ( \mathbf{r} )= \sum_{\alpha} \frac{1}{2} m_{\alpha}  {\mathbf{v}}^2_{\alpha} + \int \mathcal{L} ( \mathbf{r} ) d^3 r ,  \label{Lag1}
\end{equation}
where $\mathcal{L} ( \mathbf{r} )$ is called the \textit{Lagrangian density} and is expressed as
\begin{equation}
\mathcal{L} ( \mathbf{r} )  =  \frac{\varepsilon_i}{2} \left [ \mathbf{E}^2( \mathbf{r} ) - c^2 \mathbf{B}^2 ( \mathbf{r} ) \right ] + \mathbf{J} ( \mathbf{r} ) \cdot \mathbf{A} ( \mathbf{r} ) - \rho ( \mathbf{r} ) \phi ( \mathbf{r} )  .  \label{Lagden}
\end{equation}
In equation (\ref{Lagden}), the first bracket represents the Lagrangian of the moving particles, $\mathbf{J} ( \mathbf{r} ) \cdot \mathbf{A} ( \mathbf{r} )$ the Lagrangian of the electromagnetic fields, and $\rho ( \mathbf{r} ) \phi ( \mathbf{r} )$ the interaction between the charge particles and the field. Note that the Lagrangian (\ref{Lag1}) only holds for $\pi^i$ and $\pi^o$, not for $\pi^i \cap \pi^o$ because the particles do not stay in the membrane as mentioned in axiom 3. In this section, we will study the difference between the standard Lagrangian of the Maxwell's equations (\ref{MWf1}) - (\ref{MWf4}) and that of the classical electrodynamical waves with the Coulomb gauge $\nabla \cdot \mathbf{A} = 0$ \cite{CohenPA}. This Lagrangian is known to be gauge invariance, thus the use of the new gauge $\nabla \cdot \mathbf{A} = - \phi$ does not change the Lagrangian, while $\mathbf{J}_m$ can significantly change it. We are particularly interested in the role of ion channels on the change of the trajectory, or equivalently the wavefront, of the excitation propagation, which is equivalently represented in the Maxwell's equations (\ref{MWf1}) - (\ref{MWf4}) as the dependency and sensitivity of the standard Lagrangian on $\chi_{\alpha}$. The Coulomb gauge is preferred over the Lorentz gauge because the excitation propagation is considered from the non-relativistic points of view in this paper. The validity of the standard Lagrangian in bidomain $\Pi$ can be easily verified by showing that the standard Lagrangian is valid for each $\pi^i$ and $\pi^o$, but this will not be shown here.

\subsection{Contribution of $\chi_{\alpha}$ and $J^m$ on the Lagrangian}

In order to study whether the Lagrangian is modified by the gauge choice and $\mathbf{J}_m$, the Lagrangian should be expressed as
\begin{equation*}
\mathcal{L} ( \mathbf{r} )= \sum_{\alpha} \frac{1}{2} m_{\alpha} {\mathbf{v}}^2_{\alpha} + \fint \mathcal{L} ( \mathbf{k} ) d^3 k ,  \label{Lag2}
\end{equation*}
where $\fint$ indicates the integration over the domain for $Re(\mathbf{k})>0$. Using the fact that all the fields are real, we can also express the Lagrangian density in reciprocal space as
\begin{align}
\mathcal{L} ( \mathbf{k} ) & = \varepsilon_i \left [ \left | \mathbf{e}_k  ( \mathbf{k} ) \right |^2 - c^2 \left | \mathbf{b}_k  ( \mathbf{k} ) \right |^2  \right ]   +  \mathbf{j}^*_k  ( \mathbf{k} ) \cdot \mathbf{a}_k  ( \mathbf{k} )  \nonumber  \\
& + \mathbf{j}_k  ( \mathbf{k} ) \cdot \mathbf{a}^*_k  ( \mathbf{k} )  - \rho^*_k  ( \mathbf{k} ) \phi_k  ( \mathbf{k} ) - \rho_k  ( \mathbf{k} ) \phi^*_k  ( \mathbf{k} )  .   \label{Lagden2}
\end{align}
Then, a lemma follows immediately.\\
\\
\textbf{Lemma 2}: The Lagrangian of the Maxwell's equations (\ref{MWf1}) - (\ref{MWf4}) can be expressed in reciprocal space as
\begin{align}
\mathcal{L} ( \mathbf{k} )  &= -  \frac{\rho_k \rho_k^* }{\varepsilon_i k^2 }  +  \varepsilon_i \left [ { \dot{\mathbf{a}}_k^{\perp *}} \cdot \dot{\mathbf{a}}^{\perp}_k - c^2 k^2   { {\mathbf{a}}_k^{\perp *}} \cdot {\mathbf{a}}^{\perp}_k \right ]  \\
& ~~~~+ \left [ \mathbf{j}^{\perp *}_k \cdot \mathbf{a}^{\perp}_k + \mathbf{j}^{\perp}_k \cdot { {\mathbf{a}}_k^{\perp *}} \right ]   , \label{stdLagrangian}
\end{align}
which is the same as that of the classical Maxwell's equations with the Coulomb gauge.\\
\\
\textbf{Proof}: See Appendix II.\\
\\
Let us pay our attention to the first term of the Lagrangian density in equation (\ref{stdLagrangian}), known as the \textit{Coulomb energy of a system of charges}. As mentioned in equation (\ref{chargedensity1}), charge density can be divided into two components: The charge density caused by conducting and the reactive membrane charge density such as $\rho = \rho^c + \rho^r$. Substituting this decomposition into the Coulomb energy yields
\begin{equation*}
 \frac{\rho_k \rho_k^* }{\varepsilon_i k^2 } = \frac{1}{\varepsilon_i k^2} \left [ \rho^c_k (\rho^c_k)^* + \rho^c_k (\rho^r_k)^* + \rho^r_k (\rho^c_k)^* + \rho^r_k (\rho^r_k)^* \right ] ,
\end{equation*}
which shows the effects of the reactive membrane charge density $\rho^r$ caused by $\mathbf{j}^r$ in the Lagrangian. Moreover, using the discrete expression of $\rho$ in reciprocal space such as $\rho_k ( \mathbf{k} ) = \sum_{\alpha} {\chi_{\alpha} ( t ) } {  (2 \pi )^{-3/2} } e^{- i \mathbf{k} \cdot \mathbf{r}_{\alpha} }$, the integration of this term is given by 
\begin{align}
\lefteqn{ \frac{1}{\varepsilon_i} \int \frac{\rho^*_k ( \mathbf{k} ) \rho_k ( \mathbf{k} ) }{k^2} d^3 k } \hspace{1cm} \nonumber \\
& = \frac{1}{ 8 \pi \varepsilon_i} \left [ \sum_{\alpha} \frac{ \chi^2_{\alpha} (t) }{ r (2 \pi)^{3/2} }  + \sum_{\alpha \neq \beta} \frac{\chi_{\alpha} (t) \chi_{\beta} (t) }{ | \mathbf{r}_{\alpha} - \mathbf{r}_{\beta} | }  \right ] .  \label{Coulchg}
\end{align}
The first term represents the Coulomb self energy of the particle $\alpha$ and the second term represents the Coulomb interaction between the particles $\alpha$ and $\beta$. Therefore, the changes of the Lagrangian due to the membrane current density $\mathbf{J}^m$ is \textit{implicitly} expressed in the Coulomb energy. The membrane current density only changes the magnitude and the signs of $\chi_{\alpha}$ without modifying the total number of them, thus the effect of the membrane current density is reflected in the \textit{qualitative} characteristics of $\chi_{\alpha}$. Nevertheless, the action of the above Coulomb potential can be regarded the same as that with a constant point charge as shown in the following lemma with a new definition:\\
\\
\textbf{Definition}: Suppose there exists a scalar function $F_{\alpha \beta}: \mathbf{R}^+ \rightarrow \mathbf{R}^+ $ such that the time integration of $\chi_{\alpha}\chi_{\beta}$ is equal to $ F_{\alpha \beta} (t)$ as
\begin{equation}
\int_{t_0}^t \chi_{\alpha} (t) \chi_{\beta} (t) dt' = F_{\alpha \beta} (t), ~~~~\forall~\alpha,~\beta  , \label{pointchargevariation}
\end{equation}
where the time $t_0$ is the minimum value of the latest time for the constant resting value of $\chi_{\alpha}$ and $\chi_{\beta}$. Then, the varying point charge $\chi_{\alpha}$ is called the \textit{time-integrable}. \\
\\
\textbf{Lemma 3}: Suppose that $\chi_{\alpha} (t)$ is time-integrable for all indexes $\alpha$. Then the action of the Coulomb energy of a system of charges (\ref{Coulchg}) with $\chi_{\alpha} (t)$ is the same as the action with a time-independent $\chi_{\alpha} (0)$. \\
\\
\textbf{Proof}: Differentiating equation (\ref{pointchargevariation}) with respect to $t$, we obtain
\begin{equation*}
\chi_{\alpha} (t) \chi_{\beta} (t) = \chi_{\alpha}  (t_0) \chi_{\beta} (t_0) + \frac{d F_{\alpha \beta}  (t) }{dt}.
\end{equation*}
By substituting the above equality into equation (\ref{Coulchg}) and by using the fact that the extremes of the action remain the same by the factor of $d F / dt$, we reach the conclusion $\square$. \\
\\
The condition (\ref{pointchargevariation}) is actually valid only if the action potential can be approximated as a polynomial. In fact, the action potential is sufficiently smooth, thus the polynomial approximation of the action potential is widely used explicitly or implicitly in most mathematical and computational modeling. Therefore, we may accept this condition naturally without more restrictions. As a consequence, the time-varying $\chi_{\alpha}$ of the Coulomb energy does not change the action, but the following proposition shows that it contributes to the change of the action by modifying the Lagrangian of the electrodynamic field.\\
\\
\textbf{Proposition 7}: Suppose that $\chi_{\alpha} (t)$ is time-integrable for all indexes $\alpha$. Then the standard Lagrangian (\ref{Lagfinal}) and (\ref{Lagdenfinal}) of the Maxwell's equations (\ref{MWf1}) - (\ref{MWf4}) is given by
\begin{align}
\lefteqn{ \mathcal{L} ( \mathbf{r} )= \sum_{\alpha} \frac{1}{2} m_{\alpha}  {\mathbf{v}}^2_{\alpha} - \frac{1}{ 8 \pi \varepsilon_i}  \sum_{\alpha} \frac{ \chi^2_{\alpha} (t ) }{ r (2 \pi)^{3/2} } } \hspace{1cm} \nonumber \\
 &  - \frac{1}{ 8 \pi \varepsilon_i} \sum_{\alpha \neq \beta} \frac{\chi_{\alpha} (t ) \chi_{\beta} (t ) }{ | \mathbf{r}_{\alpha} - \mathbf{r}_{\beta} | }  +  \int \mathcal{L} ( \mathbf{r} ) d^3 r ,  \label{Lagfinal}
\end{align}
where the Lagrangian density $\mathcal{L} ( \mathbf{r} )$ is
\begin{equation}
\mathcal{L} ( \mathbf{r} )  =  \frac{\varepsilon_i}{2} \left [ \left ( {\mathbf{E}}^{\perp} \right )^2  - c^2  \mathbf{B}^2  \right ] + \left ( \mathbf{J}^c \right )^{\perp} \cdot \mathbf{A}^{\perp} , \label{Lagdenfinal}
\end{equation}
which is independent of the reactive membrane current density $\mathbf{J}^r$, but depends on the time variation of point charge $\chi_{\alpha}$. Moreover, the action induced by $\chi_{\alpha} (t)$ with a constant velocity $\mathbf{v}_{\alpha} (0)$ is the same as that by constant point charge $\chi_{\alpha} (0)$ with a time-dependent velocity $\mathbf{v}'_{\alpha} (t)$.\\
\\
\textbf{Proof}: The derivation of the Lagrangian (\ref{Lagfinal}) with (\ref{Lagdenfinal}) is obtained directly from equation (\ref{stdLagrangian}) of Lemma 2 and equation (\ref{Coulchg}). The remaining task is to prove the independency of the Lagrangian density on $\mathbf{J}^r$. But, this is also a direct result from the choice of $\mathbf{j}^r_k$ (\ref{cdchoice2}) because $\mathbf{j}^r_k$ is only in the direction of $\mathbf{k}$ and consequently, $ \left ( \mathbf{j}^r_k  \right )^{\perp} (\mathbf{k}) $ = 0 or $\left ( \mathbf{J}^r  \right )^{\perp} (\mathbf{r}) =0$. Since the first term in the Lagrangian density remains constant independent of $\chi_{\alpha}$, we only need to study the second term $( \mathbf{J}^c  )^{\perp} \cdot \mathbf{A}^{\perp}$. Let us decompose $\chi_{\alpha}$ into two components as $\chi_{\alpha}^c$, or the point charge induced by the conducing charge density $\rho^c$, and $\chi_{\alpha}^r$, or the point charge induced by the reactive membrane charge density $\rho^r$, such as $\chi_{\alpha}= \chi_{\alpha}^c + \chi_{\alpha}^r$. Then, with the discrete expression of $\mathbf{J}^c = \sum_{\alpha} \chi_{\alpha}^c \mathbf{v}_{\alpha} \delta[ \mathbf{r} - \mathbf{r}_{\alpha} ]$, the action by the Lagrangian density for the electrodynamic field $S_2$ is given by 
\begin{equation*}
S_{2} = \int_{t_1}^{t_2}  \int \left ( \mathbf{J}^c \right )^{\perp} \cdot \mathbf{A}^{\perp} d^3 r d t = \int_{t_1}^{t_2}  \sum_{\alpha} \chi_{\alpha}^c (t) \left ( \mathbf{v}_{\alpha} \cdot \mathbf{A}^{\perp} \right ) d t .
\end{equation*}
Since $\chi_{\alpha}$ is time-integrable, it is easy to show that there exists a function $G_{\alpha} (t): \mathbf{R} \rightarrow \mathbf{R}$ such that 
\begin{equation}
G_{\alpha} (t) \equiv \int_{t_0}^t \frac{\chi_{\alpha}^c (t) }{\chi_{\alpha}^c (t_0)}  dt' ,   \label{defGalpha}
\end{equation}
where the time $t_0$ is the latest time for the constant resting value of $\chi_{\alpha}$. Then the above equation reduces to 
\begin{equation*}
S_{2} = \int_{t_1}^{t_2}  \sum_{\alpha} \chi_{\alpha}^c (0) \left ( \mathbf{v}'_{\alpha} (t) \cdot \mathbf{A}^{\perp} \right ) d t ,
\end{equation*}
where we introduced the new velocity $\mathbf{v}'_{\alpha} = ( d G_{\alpha} (t) / d t ) \mathbf{v}_{\alpha}$. The proof is done only by observing that the above equation is the action of the Lagrangian of electrodynamic field with a constant point charge $\chi_{\alpha} (0)$ with velocity $\mathbf{v}' (t)$ $\square$ .\\
\\
The following corollary also may show the practical interpretation of proposition 7.\\
\\
\textbf{Corollary}: Suppose that $\chi_{\alpha} (t)$ is time-integrable with $G_{\alpha}$ for all indexes $\alpha$. If $G_{\alpha}$ is the same for all indexes $\alpha$, then the Lagrangian (\ref{Lagfinal}) with (\ref{Lagdenfinal}) is the same as the Lagrangian of the classical electrodynamics in homogeneous media. On the other hand, if $G_{\alpha}$ is different for all indexes $\alpha$, the Lagrangian (\ref{Lagfinal}) with (\ref{Lagdenfinal}) corresponds to the Lagrangian of the classical electrodynamics in inhomogeneous media. \\
\\
Preposition 7 and corollary imply mostly two crucial characteristics of the excitation propagation: (i) The first characteristics is obviously that the operation of ion channels can be translated as the changes of material properties. The time variation of point charge is only induced by ion channels, but proposition 7 implies that this time dependency of point charge can be shifted to the time-dependent velocity that can be realized as the varying conductivity property of media. In the context of the original definition of geometry, any object to change the trajectory of the propagation, we may say that ion channels can be also regarded geometry, in addition to the shape of the domain and the conductivity property of media. (ii) The second characteristics is that the membrane current density can only change the Lagrangian of the electrodynamic field. In other words, this means that the membrane current does not modify the Lagrangian of the Coulomb energy, moving particles, or interaction between the particles and the fields. As we observe later from the Hamiltonian, the non-interference of the membrane current density especially to the interaction between the particles and the fields, gives birth to the simplest excitation system, the same as that of light propagation.

\subsection{Huygens' principle and the eikonal equation}
Proposition 7 and corollary 1 suggest that the trajectory of the excitation propagation is the same as the trajectory of light propagation in the homogeneous and isotropic media with \textit{normal} ion channels with a proper condition as mentioned as a supposition. But this fact could turn out to be of no surprise when we compare the fundamental mechanism of the excitation propagation with that of light propagation, known as the Huygens' principle saying  \cite{Huygens}:\\
\\
\textit{Each element of a wave-front may be regarded as the centre of a secondary disturbance which gives rise to spherical wavelets; and moreover, that the position of the wave-front at any later time is the envelope of all such wavelets.}\\
\\
However, no better description can be given than the above principle to the mechanism of the diffusion-reaction system, such as the classical FHN equations for the excitation propagation. If we replace \textit{secondary disturbance} and \textit{spherical wavelets} with \textit{reaction} and \textit{diffusion}, respectively, without losing its meaning, the above description of the propagation remains intact for the diffusion-reaction system. Consequently, without considering the velocity of the propagation, the trajectory and wavefront should remain same for both propagations.

If the two different systems share the same propagation mechanism, then their eikonal equation should be coincident. In geometric optics, the surface of light propagation is provided by the optical path $\mathcal{S}$ satisfying \cite{Born}
\begin{equation}
\left | \nabla \mathcal{S} \right |^2 = \left (  \frac{\partial \mathcal{S}}{\partial x}  \right )^2  + \left (  \frac{\partial \mathcal{S}}{\partial y}  \right )^2 + \left (  \frac{\partial \mathcal{S}}{\partial z}  \right )^2 = \sqrt{\varepsilon_i \mu_i}  .     \label{eikonal}
\end{equation}
This equation is derived for regions that are sufficiently far from the sources, or equivalent for a sufficiently large value of the wave number when the electrodynamic field is considered as a time-harmonic field. The Maxwell's equations (\ref{MWf1}) - (\ref{MWf4}) equivalent to the FHN equations (\ref{FHN1}) - (\ref{FHN2}) cannot be written without source terms because of the presence of ion channels almost everywhere. Thus, it is very difficult to prove mathematically that the eikonal equation of Maxwell's equations (\ref{MWf1}) - (\ref{MWf4}) or the FHN equations (\ref{FHN1}) - (\ref{FHN2}) can be written the same as equation (\ref{eikonal}). This is mostly because nor charge density $\rho$ nor the membrane current density $\mathbf{J}^m$ being divided by the wave number converges to zero even at a very high frequency. Even the computational study for the coincidence of the two eikonal equations are not trivial. Thus, in spite of strong inference from proposition 7 and corollary 1, we put it as a conjecture for validation in the later publications such that\\
\\
\textbf{Conjecture}: Suppose that $\chi_{\alpha} (t)$ is time-integrable for all indexes $\alpha$ and the media is homogeneous and isotropic with constant conductivity. Then the eikonal equation of the Maxwell's equations (\ref{MWf1}) - (\ref{MWf4}) is the same as the classical eikonal equation of optics (\ref{eikonal}).

\subsection{With external electromagnetic field}
We also can consider the effect of the external electromagnetic field on the excitation propagation in the heart, especially focusing on the changes of trajectory and velocity. Consider that the electric field $\mathbf{E}'_e$ and $\mathbf{B}'_e$ being measured in the vacuum are applied to the myocardial tissue. Let $\mathbf{A}'_e$ and $\phi'_e$ be the corresponding potentials in the vacuum. To be represented in the same myocardial medium, the field and potential measured in the vacuum should be expressed as those in the microscopic domain $\pi^i$ or $\pi^o$. Consider that the transformation of the field and potential between the vacuum and $\pi^i$ or $\pi^o$ can be simply performed by a linear projection operation $\mathcal{H}$ such that
\begin{align*}
\left . \mathcal{H} \mathbf{E}'_e \right |_{\pi^i} &= \mathbf{E}^i_e,~~\left . \mathcal{H}    \mathbf{E}'_e \right |_{\pi^o} = \mathbf{E}^o_e, \\
\left . \mathcal{H} \mathbf{B}'_e \right |_{\pi^i} &= \mathbf{B}^i_e,~~\left . \mathcal{H} \mathbf{B}'_e \right |_{\pi^o} = \mathbf{B}^o_e.
\end{align*}
Similar operations can be applied to the potentials to yield $\mathbf{A}_e^i(\mathbf{A}_e^o)$ and $\phi_e^i ( \phi_e^o)$. For example, if we consider the field as the consequential phenomena from the moving charge, then we may consider the operator $\mathcal{H}$ as the linear transformation caused by the proportional changes of the velocity of the moving charge from the vacuum to the bidomain or vice versa. But, according to axiom 1, the maximum velocity of the signal is the same in $\pi^i$ and $\pi^o$, thus the operator $\mathcal{H}$ should be the same operator for the field and potential in $\pi^i$ and $\pi^o$, but only differentiate according to the location of $q_{\alpha}$. The weighted difference of the external field and the external potential in $\Pi$ is therefore defined as
\begin{align*}
\mathbf{E}_e & \equiv \mathbf{E}^i_e - \lambda \mathbf{E}^o_e,~~~\mathbf{B}_e \equiv \mathbf{B}^i_e - \lambda \mathbf{B}^o_e, \\
\mathbf{A}_e & \equiv \mathbf{A}^i_e - \lambda \mathbf{A}^o_e,~~~\phi_e \equiv \phi^i_e - \lambda \phi^o_e .
\end{align*}
According to the above definitions, the strength of the external field or the external potential depends on the parameter $\lambda$ defined as $ \lambda = \sqrt{\varepsilon_i / \varepsilon_o} =\sqrt{\mu_o / \mu_i}$. For example, if $\lambda = 1$, $\mathbf{E}_e$ and $\mathbf{B}_e$ are always zero because we regard that the operator $\mathcal{H}$ performs the same for $\pi^i$ and $\pi^o$. Consequently $\mathbf{A}_e$ and $\phi_e$ are zero. Thus, there will be no effect of the external electromagnetic field. However, this does not reflect the real phenomena as shown in refs. \cite {Beck} \cite {Ferris} \cite {Zoll1} \cite {Zoll2}, but it is more reasonable to choose $\lambda$ different from $1.0$ for all the media. Then we can deduce that $\mathbf{E}_e$ and $\mathbf{B}_e$ are roughly proportional to $\mathbf{E}'_e$ and $\mathbf{B}'_e$, respectively, and similarly $\mathbf{A}_e$ and ${\phi}_e$ to $\mathbf{A}'_e$ and ${\phi}'_e$, respectively. When the external field is applied to $\Pi$, the new Lagrangian density is given by 
\begin{equation}
\mathcal{L} ( \mathbf{r} ) = \frac{\varepsilon_i}{2}  \left [ \left ( \mathbf{E}^{\perp} \right )^2 - c^2 \mathbf{B}^2  \right ] + \left ( \mathbf{J}^c \right )^{\perp} \cdot (\mathbf{A}^{\perp} + \mathbf{A}_e^{\perp}  ) - \rho \phi_e. \label{Lagwithexternalpotential}
\end{equation}
Or, by means of the standard procedure of the Power-Zienau-Woolley transformation \cite {Power, Woolley}, 
\begin{align}
\lefteqn{ \mathcal{L} ( \mathbf{r} ) = \frac{\varepsilon_i}{2} \left [ \left ( \mathbf{E}^{\perp} \right )^2  -c^2 \mathbf{B}^2  \right ] } \hspace{1cm} \nonumber \\
& +   \mathbf{M} \cdot ( \mathbf{B} + \mathbf{B}_e) +  \mathbf{P} \cdot ( \mathbf{E}^{\perp} + ( \mathbf{E}_e)^{\perp} ) - \rho \phi_e ,   \label{Lagdenpol}
\end{align}
where $\mathbf{P}  (\mathbf{r}) $ is the polarization density and $\mathbf{M}  (\mathbf{r}) $ is the magnetization density such as
\begin{align*}
\mathbf{P} (\mathbf{r}) &= \sum_{\alpha} \int_0^1 \chi_{\alpha} \mathbf{r}_{\alpha} \delta [ \mathbf{r} - p \mathbf{r}_{\alpha} ] dp,\\
\mathbf{M} ( \mathbf{r} ) &= \sum_{\alpha} \int_0^1 p \chi_{\alpha} \mathbf{r}_{\alpha} \times \dot{\mathbf{r}}_{\alpha} \delta [ \mathbf{r} - p \mathbf{r}_{\alpha} ] d p  .
\end{align*}
If the external field and potential is sufficiently large, $\mathbf{A}_e$ modifies the Lagrangian of the electrodynamic field and $\phi_e$ modifies the Lagrangian of the interaction between the particle and the field. Leaving the effects of $\phi_e$ on the Lagrangian of the interaction to the study of the Hamiltonian in the next section, the rest of this section focuses on the effect of $\mathbf{A}_e$ on the Lagrangian of the electrodynamics field. Consider the following proposition:\\
\\
\textbf{Proposition 8}: Suppose that $\chi_{\alpha}$ is time-integrable for all $\alpha$. Suppose that the perpendicular component to the wave vector of the external vector potential $\mathbf{A}_e$ during time $t_1$ to $t_2$ is non-trivial for $\chi_{\alpha}$ located at $\mathbf{r}_{\alpha}$. Then, applying the electric potential $\mathbf{A}_e$ to the bidomain $\Pi$ causes the changes of the propagational velocity of $\chi_{\alpha}$. Moreover, if $\mathbf{A}_e$ is in the opposite direction to $\mathbf{A}$, then there exists the critical magnitude of the electric potential $\mathbf{A}_e^*$ to stop the propagation of $\chi_{\alpha}$. \\
\\
\textbf{Proof}: For the Lagrangian density (\ref{Lagwithexternalpotential}), the action by the Lagrangian of the electrodynamic field is given by
\begin{align*}
S_{2} &= \int_{t_1}^{t_2}  \sum_{\alpha} \chi_{\alpha}^c (t)  \mathbf{v}_{\alpha} \cdot  ( \mathbf{A}^{\perp} + \mathbf{A}^{\perp}_e )  d t  \\
&= \int_{t_1}^{t_2}  \sum_{\alpha} \chi_{\alpha}^c (t)  \mathbf{v}'_{\alpha} \cdot  \mathbf{A}^{\perp}  d t  ,
\end{align*}
where $\mathbf{v}'_{\alpha} \equiv  \mathbf{v}_{\alpha} \cdot  ( 1 +  \mathbf{A}^{\perp}  \cdot \mathbf{A}^{\perp}_e )$. Note that this action is the same as the action without the external field, but with a different velocity. Thus, it can be deduced that applying $\mathbf{A}_e$ only leads to the changes of velocity of $\chi_{\alpha}$ without considering the interaction between the particle and the field. The existence of the critical $\mathbf{A}_e^*$ for the stopping of the propagation can be directly inferred from the existence of $\mathbf{A}_e^*$ to satisfy the equality $ \mathbf{A}^{\perp}  \cdot  \mathbf{A}^{\perp *}_e = - 1.0$, which implies that $\mathbf{A}^{\perp *}_e$ should be in the opposite direction to $\mathbf{A}^{\perp }$ $\square$.  \\
\\
If $\mathbf{A}^{\perp}_e$ is in the same direction as $\mathbf{A}^{\perp}$, then it is conjectured that the propagational velocity increases up to the maximum velocity $c$, defined as $c = \sqrt{\varepsilon_i \mu_i} = \sqrt{\varepsilon_o \mu_o}$. As posed in axiom 1, the propagational velocity is assumed to remain constant not exceeding $c$ independent of $\mathbf{A}^{\perp}_e$ that is greater than a certain magnitude, though what actually happens \textit{in vivo} is unknown.

\section{Hamiltonian}
The Hamiltonian of the Maxwell's equations (\ref{MWf1}) - (\ref{MWf4}), an operator associated with the total energy of the system, also may help us to enhance the understanding of (i) the excitation mechanism of the excitation propagation and (ii) the effect of the interaction between the intrinsic or external electrodynamic field and the cation for the excitation of the myocardial cell. Let us begin with the following lemma.\\
\\
\textbf{Lemma 4}: The Hamiltonian of the Maxwell's equations (\ref{MWf1}) - (\ref{MWf4}) is expressed as
\begin{align}
\lefteqn{\mathcal{H} = \sum_{\alpha} \frac{1}{2 m_{\alpha} } \left [ \mathbf{p}_{\alpha} - \chi_{\alpha} \mathbf{A}_{\alpha} ( \mathbf{r}_{\alpha} ) \right ]^2 }  \hspace{1cm} \\
& + \sum_{\alpha} \frac{1}{ 8 \pi \varepsilon_i ( 2 \pi )^3 } \frac{\chi^2_{\alpha}}{r} + \varepsilon_i \fint \mathcal{H} ( \mathbf{k} )d^3 k , \label{Hamilk}
\end{align}
where the Hamiltonian density $\mathcal{H}$ in reciprocal space is derived as
\begin{equation}
\mathcal{H} (\mathbf{k} ) = \left ( \dot{ \mathbf{a}}^{\perp}_k \right )^* \dot{\mathbf{a}}_k^{\perp}  + c^2 k^2 (\mathbf{a}^{\perp}_k)^* \cdot \mathbf{a}_k^{\perp}  .  \label{Hamildenk}
\end{equation}
\text{Proof}: See Appendix IIIA. \\
\\
The first term of the Hamiltonian (\ref{Hamilk}) represents the kinetic energy of the particles located at $\mathbf{r}_{\alpha}$ with momentum $\mathbf{p}_{\alpha} = ( \hbar / i ) \nabla_{\alpha}$ where the value of the constant $\hbar$ is unknown. Note that $\hbar$ is different from the Planck constant $6.62606957 \times 10^{-34} m^2 kg / s$ and should be defined such that the energy of the monochromatic wave of the Maxwell's equations  (\ref{MWf1}) - (\ref{MWf4}) with the frequency $\omega$ is expressed as $E = n \hbar \omega$ for an integer $n$. Intuitively, this modification is required because the traveling particle is the cation which carries a possibly larger energy and momentum than those of photon. The second term corresponds to the Coulomb energy, and the last term corresponds to the radiation energy of the transverse field, or the energy generated by the perpendicular component of the field to the wave vector $\mathbf{k}$. Similar to the Lagrangian, the only difference of the Hamiltonian (\ref{Hamilk}) to that of the Maxwell's equations with Coulomb gauge is the presence of the time-varying point charge $\chi_{\alpha} (t)$. To understand the impact of $\chi_{\alpha} (t)$ on the Hamiltonian, the Hamiltonian (\ref{Hamilk}) will be expressed with more distinguishable components by solving the bracket and using the normal variables from the second quantization \cite{CohenPA}.\\
\\
\textbf{Proposition 8}: The Hamiltonian of the Maxwell's equations (\ref{MWf1}) - (\ref{MWf4}) is given by 
\begin{equation}
\mathcal{H} = \mathcal{H}_{0}  + \mathcal{H}_R + \mathcal{H}_C + \mathcal{H}_{I1} + \mathcal{H}_{I2}  + \mathcal{H}_{I3} ,  \label{Hamiltonian}
\end{equation}
where
\begin{align}
\mathcal{H}_p & = \sum_{\alpha} \frac{\mathbf{p}_{\alpha}^2}{2 m_{\alpha} }   , \label{Hamil1} \\
\mathcal{H}_R & =  \hbar \omega  \left ( a^+ a + \frac{1}{2}  \right )  ,  \label{Hamil2} \\
\mathcal{H}_C & = \frac{1}{ 8 \pi \varepsilon_i} \left [ \sum_{\alpha} \frac{ \chi^2_{\alpha} (\phi,t ) }{ (2 \pi)^{3/2} r}  + \sum_{\alpha \neq \beta}  \frac{\chi_{\alpha} (\phi,t ) \chi_{\beta} (\phi,t ) }{ | \mathbf{r}_{\alpha} - \mathbf{r}_{\beta} | }  \right ] , \label{Hamil3}  \\
\mathcal{H}_{I1} & = - \sum_{\alpha} \frac{\chi_{\alpha}}{m_{\alpha}} \mathbf{p}_{\alpha} \cdot \mathbf{A} ( \mathbf{r}_{\alpha} ) , \label{Hamil4} \\
\mathcal{H}_{I2} & = - \sum_{\alpha} g_{\alpha} \frac{\chi_{\alpha}}{2 m_{\alpha} } \mathbf{S}_{\alpha} \cdot \mathbf{B} ( \mathbf{r}_{\alpha} ) , \label{Hamil5} \\
\mathcal{H}_{I3} & = \sum_{\alpha} \frac{\chi^2_{\alpha} }{2 m_{\alpha} } \mathbf{A}^2 ( \mathbf{r}_{\alpha} ) \label{Hamil6} ,
\end{align}
where $\mathbf{S}_{\alpha}$ is the spin of the particle $\alpha$ and $g_{\alpha}$ is the Land\'{e} factor. \\
\\
\textbf{Proof}: See Appendix IIIB. \\
\\
The \textit{particle Hamiltonian} $\mathcal{H}_p$ represents the kinetic energy of the particle with mass $m_{\alpha}$ and momentum $\mathbf{p}_{\alpha}= ( \hbar / i ) \nabla_{\alpha}$, independent of $\chi_{\alpha}$. The \textit{radiation field Hamiltonian} $\mathcal{H}_R$ depending on the operator $a^+$ and $a$, known as the \textit{creation operator} and \textit{annihilation operator} of the cation in the single mode, represents the energy of the radiation field with frequency $\omega$. If the propagating cations have various kinds of ions with various polarizations, then equation (\ref{Hamil2}) should be written as the sum over all the mode $j$ for corresponding $\hbar_j$, but both for simplicity and reflecting reality, we suppose that all the propagational cations are the same kind with the same polarization. Note that $\mathcal{H}_p$ and $\mathcal{H}_R$ are independent of $\chi_{\alpha} (t)$ and consequently independent of the reactive membrane current by ion channels. The \textit{Coulomb Hamiltonian} $\mathcal{H}_C$ is in the same form as that of the Lagrangian.

$\mathcal{H}_p$, $\mathcal{H}_R$ and $\mathcal{H}_C$ contain the dynamical variables of the particle \textit{or} the transverse field, but the other Hamiltonians, known as the \textit{interaction Hamiltonian}, contain both dynamic variables of the particle and the transverse field to indicated the interaction between them. The \textit{interaction Hamiltonian} $\mathcal{H}_I$ consists of three different components: (1) $\mathcal{H}_{I1}$ represents the energy caused by the momentum of the cation $\alpha$ in the direction of the potential $\mathbf{A}$. (2) $\mathcal{H}_{I2}$ represents the spin energy of the cation $\alpha$ caused by the magnetic field $\mathbf{B}$. (3) $\mathcal{H}_{I3}$ represents the kinetic energy of the oscillatory forced motion by $\mathbf{A}$. Note that $\mathcal{H}_C$ and $\mathcal{H}_I$ are all dependent on $\chi_{\alpha} (t)$, thus, consequently, dependent on the reactive membrane current density by ion channels.

In the next subsections, the Hamiltonian $\mathcal{H}$ will be divided into two components: One is the unperturbed $\mathcal{H}_0$ and the perturbed $\mathcal{H}_I$ where $\mathcal{H}_0 = \mathcal{H} - \mathcal{H}_I$. The motivations for this decomposition are well described in ref. \cite{CohenPA}, but will be described here in brief. The unperturbed $\mathcal{H}_0$ contains the radiation field Hamiltonian $\mathcal{H}_R$ and the particle Hamiltonian $\mathcal{H}_p$, thus the quantum state or energy state of $\mathcal{H}_0$, representing the free particle in the radiation field, remains constant during the propagation. On the other hand, the quantum state of the perturbed Hamiltonian $\mathcal{H}_I$ changes according to the interaction between the radiational field and the myocardial cell. The Coulomb Hamiltonian is the only undetermined component, but will be assorted as the unperturbed Hamiltonian to provide the lowest quantum number for the resting state.

\begin{figure}[h]
\centering
\vbox{
\includegraphics[height=4cm, width=4cm] {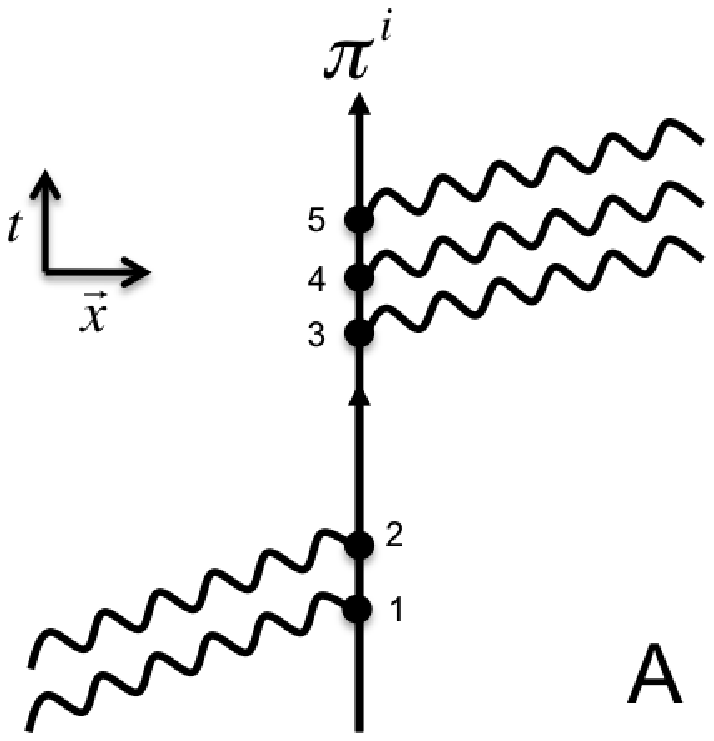}  \includegraphics[height=4cm, width=4cm] {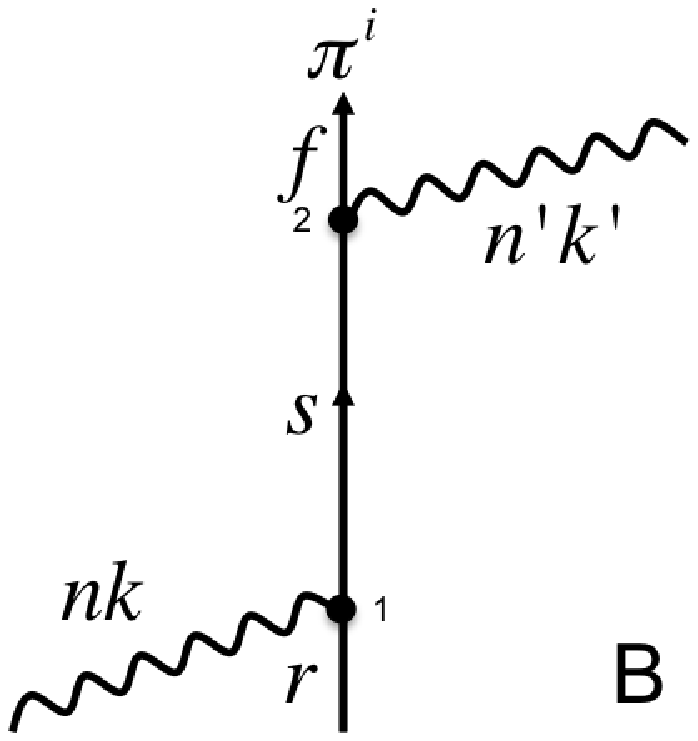}  }
\caption{Feynman diagram of the cardiac excitation without the membrane current. $\pi^i$ is stationary. $\mathbf{k}$ and $\mathbf{k}'$ are the wave vectors. $n$ and $n'$ are the number of the incident and emitting cations, respectively. }
\label {fig:Feymandiag0}
\end{figure}

\subsection{Transition amplitude without the membrane current}
In order to understand how the Hamiltonian is related to the cardiac excitation propagation, the transition amplitudes for the excitation mechanism of the myocardial cells will be derived from Feynman diagram \cite{QEDFeynman}. Let us state the following actions of the interaction between the cations and the myocardial cells reflecting the real biological phenomena:\\
\\
(ACTION i)     The cation travels in the space $\pi^i$ and time $t>0$. \\
(ACTION ii)    The cardiac cell in $\pi^i$ is stationary for all the time $t>0$. \\
(ACTION iii)   The point charge in $\pi^o$ fluctuates in the space $\pi^o$ and exchanges the cations with the myocardial cells. \\
(ACTION iv)   The cardiac cell absorbs and emits the cations. \\
\\
Note that these actions are similar to the interaction between electrons and photons in quantum optics where (ACTION iii) is analogous to the interaction between protons and electrons, but are strikingly different in (ACTION ii) because electrons also travels in space. (ACTION i) and (ACTION ii) are obvious. (ACTION iii) is also clear since the exchanges are induced by ion channels. The expression of (ACTION iv) could be less familiar, but is equivalent to other popular terminologies such as the safety factor (SF) which measures the ratio between the inward axial current $(I_{in})$ with the capacity current $(I_c)$ and the outward axial current $(I_{out})$ of a myocardial cell defined as \cite{Kleber}
\begin{equation}
\mathrm{SF} \equiv \left . \left ( \int I_c dt + \int I_{out} \right ) \right / \int I_{in} d t,
\end{equation}
where each current is integrated over the time interval only when $\rho$ is positive. Successful conduction thus means $\mathrm{SF}>1$.

In Feynman's diagram where the vertical line is the time and the horizontal line is the space, (ACTION ii) means that the cardiac cell is represented by a vertical line which marches in time forwardly from bottom to top. Let the wavy lines represent the motion of the cation traveling in $\pi^i$. Then, (ACTION i) and (ACTION iv) correspond to multiple wavy lines in Figure \ref{fig:Feymandiag0}A if the motion of each cation is represented by a single wavy line. Since the time marches from bottom to top in the diagram, two bottom wavy lines represent two cations traveling from other places and being absorbed by a myocardial cell. Accordingly, the points $1$ and $2$ correspond to the annihilation of a cation denoted by $a$ term in equation (\ref{Hamil2}). On the other hand, three upper wavy lines represent  three cations being emitted by the same myocardial cell and traveling to other places. Thus, the points $3,4,5$ correspond to the emission of a cation denoted by $a^+$ term in equation (\ref{Hamil2}). Without considering the capacity current, we can say that the safety factor for this myocardial cell is $1.5$.

This diagram can be also expressed by the transition amplitude which is crucial to the understanding of the quantum interaction. Let $ \left | \psi^i \right \rangle$ and $ \left | \psi^o \right \rangle$ be the state vector representing the energy spectral of $\pi^i$ and $\pi^o$, denoted by the quantum number $\{ n^i \}$ and $\{ n^o \}$, respectively. Let us define the \textit{relative state vector} $ \left | \psi \right \rangle$ is defined as the superpositions of these two states such that $\left | \psi \right \rangle \equiv  \left | \psi^i \right \rangle - \lambda \left |  \psi^o \right \rangle $ where the weight factor $\lambda$ is obtained from the normal variable $\boldsymbol{\alpha}$ for the operators $a$ and $a^+$. In quantum number, it is equivalent to $ n^i - \lambda n^o$. Then, the relative state vector $\left | \psi \right \rangle$ can be categorized as follows:\\
\\
(STATE i)~Resting state:~  $\left | \psi \right \rangle = \frac{1}{2}. $ \\
(STATE ii)~Excited state:~ $ \left | \psi \right \rangle > n_0,~~~\mbox{for a fixed }n_0 \gg 1.$ \\
(STATE iii) Refractory state: ~~$ \left | \psi \right \rangle \le 0. $ \\
\\
The resting state (STATE i), equivalent to the vacuum state in electrodynamics, is the direct consequence of equation (\ref{Hamil2}) with $a a^+ = n = 0$. For (STATE ii), $n_0$ is related to the threshold of the membrane potential, but is fixed and constant in homogeneous media. The refractory state (STATE iii) looks like an unnatural phenomenon, but is in fact a natural one, even from the point of view of physics, and will be explained in detail in the next subsection. Let $ \left | \psi_i \right \rangle $ and $ \left | \psi_f \right \rangle$ be the initial and final state vector. Then the transition amplitude is given by $\langle \psi_f | U(t_f, t_i) | \psi_i \rangle $ for the evolution operator $U(t_f, t_i)$ which transforms the initial state $ \left | \psi_i \right \rangle $ at $t_i$ into the final state $ \left | \psi_f \right \rangle $ at $t_f$. For example, $\exp [ - i E ( t_f - t_i ) / \hbar ]$, the solution of the Schr\"{o}dinger equation $i {\partial \psi} / {\partial t} =E \psi$, represents the free evolution of the energy state $E$ from time $t_i$ to $t_f$.

For the sake of simplicity, multiple absorption or emission cations will be represented by a single wavy line, but with different energy, as shown in Figure \ref{fig:Feymandiag0}B. But, this requires the following axiom: \\
\\
\textbf{Axiom 4}: The propagating cations are identical and indistinguishable, obeying the Bose-Einstein statistics. \\
\\
This axiom is obvious because the cations with different polarization are either rare in biological tissue or make no difference in functionality, especially in generating the membrane potential. Let $\mathbf{k}$ and $\mathbf{k}'$ be the wave vector of an incident and emitting cation, respectively. Let $n$ and $n'$ be the number of the incident cations and emitting cations, respectively. The polarization of each cations is disregarded. Let $\omega$ and $\omega'$ be its corresponding angular frequency such that $\omega = c | \mathbf{k} |$ and $\omega' = c | \mathbf{k} |'$ for the speed $c$ of the signal. Let $t_0$ be the resting phase, $t_1$ be the time when the cations are absorbed, $t_2$ be the time when the cations are emitted and $t_3$ be the time when it is back to the resting phase. The letters $r$ (resting state), $s$ (excited state), and $f$ (refractory state) next to the solid line indicate the quantum number for each procedure. Then Figure \ref{fig:Feymandiag0}B is expressed by the following evolution operator:
\begin{align}
&\exp \left [ - \frac{i}{\hbar}  ( E_f + \hbar n' \omega' ) ( t_3 - t_2 ) \right ]  \langle f | \mathcal{H}_I | s \rangle  \nonumber \\
&\times \exp \left [ - \frac{i}{\hbar}  E_s ( t_2- t_1 ) \right ]  \langle s | \mathcal{H}_I | r \rangle \label{diagram1} \\
& \times \exp \left [ - \frac{i}{\hbar}  ( E_r +  \hbar n \omega ) ( t_1 - t_0 ) \right ] .   \nonumber
\end{align}
The last component, $\exp \left [ - {i}/{\hbar}  ( E_a + \hbar n \omega ) ( t_1 - t_0 ) \right ]$, indicates the unitary transformation with respect to the energy level $E_r + \hbar n \omega$ from $t_0$ to $t_1$. The fourth component, $ \langle s | \mathcal{H}_I | r \rangle$, corresponds to the change of the states from $r$ to $s$ by the interaction Hamiltonian $\mathcal{H}_I$, due to the absorption of the cation at time $t_1$, in the Schr\"{o}dinger representation followed by the unitary transformation with respect to the energy level $E_s$ from $t_1$ to $t_2$. The second component, $\langle f | \mathcal{H}_I | s \rangle$, similarly describes the change of the states from $s$ to $f$ again by $\mathcal{H}_I$, due to the emission of the cation at time $t_2$, followed by the unitary transformation $\exp \left [ - {i}/{\hbar}  ( E_f + \hbar n' \omega' ) ( t_3 - t_2 ) \right ]$ with respect to the energy level $E_f + \hbar n' \omega'$.

The total energy of the incident cations $\hbar n \omega$ and the total energy $E_r$ of the myocardial cell depend on the microscopic \textit{coherence} of the incident cations which could be a reflection of the macroscopic \textit{geometry} of the neighboring cells (This will be discussed in part II of this series of papers). Since the energy $E_s$ is the sole parameter for the excited states, our only concern will be on the interacting Hamiltonian $\mathcal{H}_I$ changing the quantum state $r$ into the quantum state $s$. Thus, it is no surprise to find that each component of $\mathcal{H}_I$ is a function of $\chi_{\alpha}$ corresponding to the action of ion channels for the induction of the membrane current density. For example, if $\chi_{\alpha}$ is constant, the total amount of energy $ \mathcal{H}_I $ remains constant and consequently $\langle s | \mathcal{H}_I | r \rangle$ is close to zero if $s \gg r$. Thus, the only possible way to change the quantum states from $r$ to $s$ could be achieved by modifying the spin energy $\mathcal{H}_{I2}$ such that the perturbation Hamiltonian is only restricted to $ \mathcal{H}_{I1} +\mathcal{H}_{I3} $ to yield that $\langle s | \left ( \mathcal{H}_{I1} +\mathcal{H}_{I3} \right ) | r \rangle$ is not trivial even if $s \gg r$. But, if $\chi_{\alpha}$ varies according to the influx of ion channels, then the total energy of $\mathcal{H}_I$ varies as well, thus the transition amplitude $\langle s | \mathcal{H}_I | r \rangle$ is not trivial for any $s$ and $r$. We can reach the similar argument if we suppose that ion channels only response to the membrane potential and equivalently to $\mathcal{H}_{I3}$, not to the momentum $(\mathcal{H}_{I1})$ or the spin energy $(\mathcal{H}_{I2})$. In summary,\\
\\
\textbf{Lemma 4}: The interaction Hamiltonians (\ref{Hamil4}) - (\ref{Hamil6}) correspond to the activation of ion channels beyond the threshold. Moreover, if ion channels are activated by the (static) membrane potential, then the Hamiltonian $\mathcal{H}_{I3}$ (\ref{Hamil6}) is only involved for the excited states. \\
\\
Note that the Hamiltonian $\mathcal{H}_{I3}$ (\ref{Hamil6}) is proportional to $\chi_{\alpha}^2$, thus its energy changes quadratically as the membrane current occurs.

\begin{figure}[h]
\centering
\vbox{  \includegraphics[height=4cm, width=4cm] {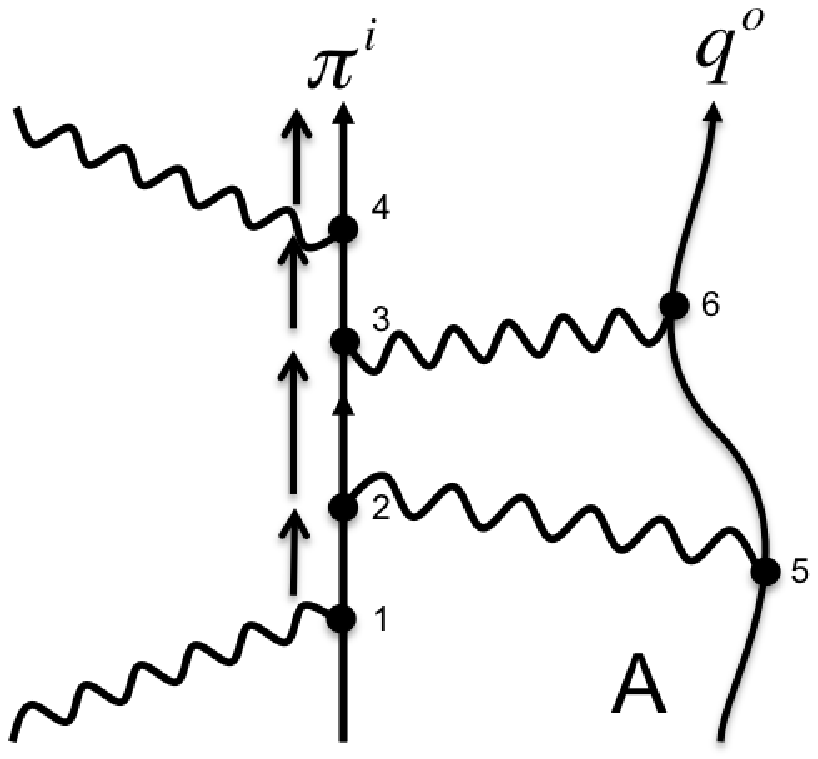}  \includegraphics[height=4cm, width=4cm] {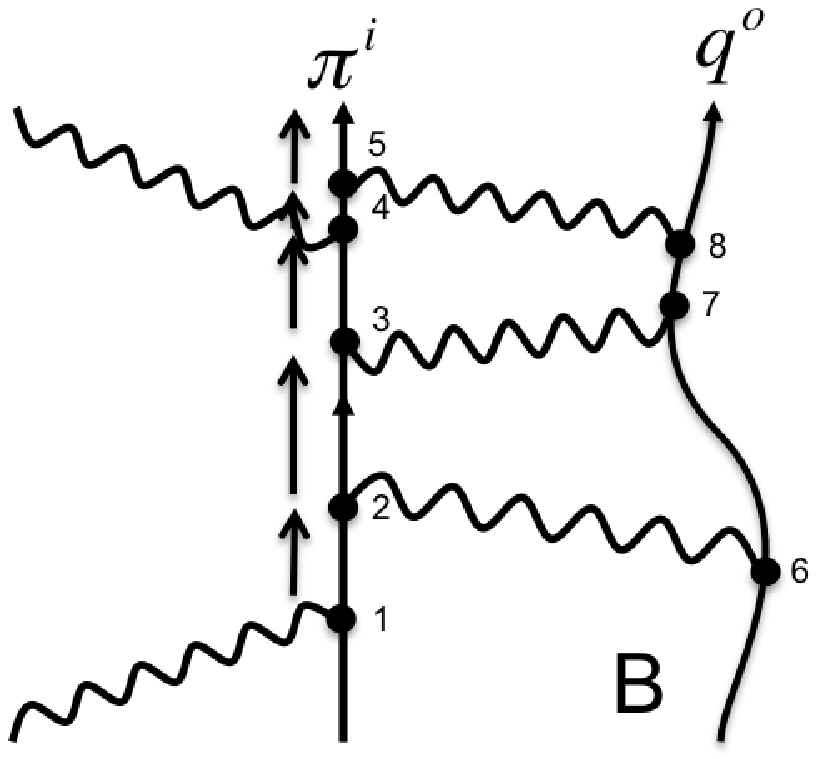}   \includegraphics[height=4cm, width=4cm] {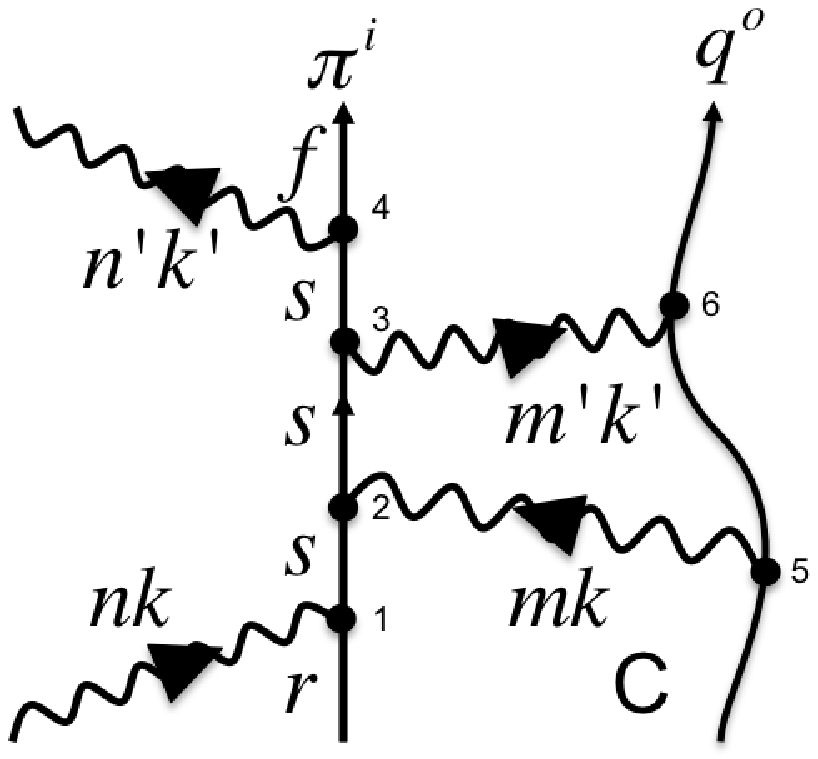}  }
\caption{Feynman diagram of the cardiac excitation with the membrane current. $\pi^i$ is stationary and $q^o$ fluctuates. $\mathbf{k}$ and $\mathbf{k}'$ are the wave vectors. $n$ and $n'$ are the number of the incident and emitting cations, respectively, while $m$ and $m'$ are the number of the influx and efflux cations, respectively. }
\label {fig:Feymandiag1}
\end{figure}

\subsection{Transition amplitude with the membrane current}

Figure \ref{fig:Feymandiag0} and the transition amplitude (\ref{diagram1}) display the simplest form of interactions between the cations and the myocardial cells, but in order to reflect more detail of interaction for more crucial behavior, another factor will be included in Feynman's diagram; point charge $q^o$ in $\pi^o$. Similar results may be obtained with $\pi^o$, but the use of $q^o$ seems to better reflect the complex mechanism of the cardiac excitation. According to (ACTION iii), the lines for $q^o$ are slightly wavy as shown in Figure (\ref{fig:Feymandiag1}) since $q^o$ fluctuates in $\pi^o$. The biggest advantage of introducing $q^o$ in the diagram is the strategical representation of the membrane current.

Let $t_1$ be the time when the incident cations are absorbed, $t_2$ be the time when the influx membrane current occurs, $t_3$ be the time when the efflux membrane current occurs, and $t_4$ be the time when cations are emitted. The wavy line between $2$ and $5$ indicate the influx membrane current because point $2$ occurs later than point $5$. Similarly, the wavy line between $3$ and $6$ indicates the efflux membrane current because point $6$ occurs later than point $3$. The left arrow besides the line of $\pi^o$ indicate that the time flows forward. Contrary to the depolarization phase at time $2$, the repolarization phase involves both of influx and efflux membrane current. Figure \ref{fig:Feymandiag1}B illustrates this fact. But, if we consider each absorption and emission by means of energy and momentum, Figure \ref{fig:Feymandiag1}A and \ref{fig:Feymandiag1}B can be displayed by the same plot Figure \ref{fig:Feymandiag1}C.

Let $n$ and $n'$ be the number of incident cations, respectively. Let $m$ and $m'$ be the number of the cations through the influx and efflux membrane currents, respectively. Suppose that the wave vector $\mathbf{k}$ and $\mathbf{k}'$ are the same for the incident cations in $\pi^i$ and the cations through the membrane. Also, we suppose that the number $n$ of the incident cations is sufficiently large such that the absorbed cations induce a sufficiently large electric potential in $\pi^i$ to create a larger membrane potential than the voltage threshold $(\phi_{th})$. The most critical step is to set up the quantum number between time $1$ and $4$ when an active membrane current occurs. For simplicity we let this period share the same quantum number $s$, like a plateau resulting from the equivalence between influx and efflux, but all the influx occurs prior to the efflux. Then, the evolution operator for Figure  \ref{fig:Feymandiag1}C will be obtained as
\begin{align*}
& \exp \left [ - \frac{i}{\hbar}  ( E_f + \hbar' m' \omega ' ) ( t_5 - t_4 ) \right ]  \langle f  | \mathcal{H}_I | s \rangle  \prod_{\ell=1}^{m'}  \\
&\times \exp \left [ - \frac{i}{\hbar}  ( E_f + \hbar ( n  +  \ell ) \omega'  ) \Delta t' \right ]   \\
 &\times \prod_{j=1}^m  \exp \left [ - \frac{i}{\hbar}  ( E_r + \hbar ( n  +  j ) \omega  )  \Delta t  \right ]   \langle s  | \mathcal{H}_I | r \rangle \\
 &\times \exp \left [ - \frac{i}{\hbar}  ( E_r + \hbar n \omega ) ( t_1 - t_0 ) \right ],   
\end{align*}
where $ \Delta t' $= ${  ( 2 t_4 - t_2 - t_ 3  ) }/ {2 m' }$ and $\Delta t $= ${  ( t_2 + t_ 3 - 2 t_1 ) }/{2 m }$. The evolution operator for the excited state without any perturbation Hamiltonian implies that the excitation state is considered the \textit{free evolution of various discrete energy levels of the myocardial cell induced by the membrane current}. This modeling could be an excessive simplification of various ion movements through ion channels during the cardiac excitation, but this better characterizes the critical properties of the excitation.

\subsection{Refractory period in QED}

\begin{figure}[h]
\centering
\vbox{
\includegraphics[height=4cm, width=4cm] {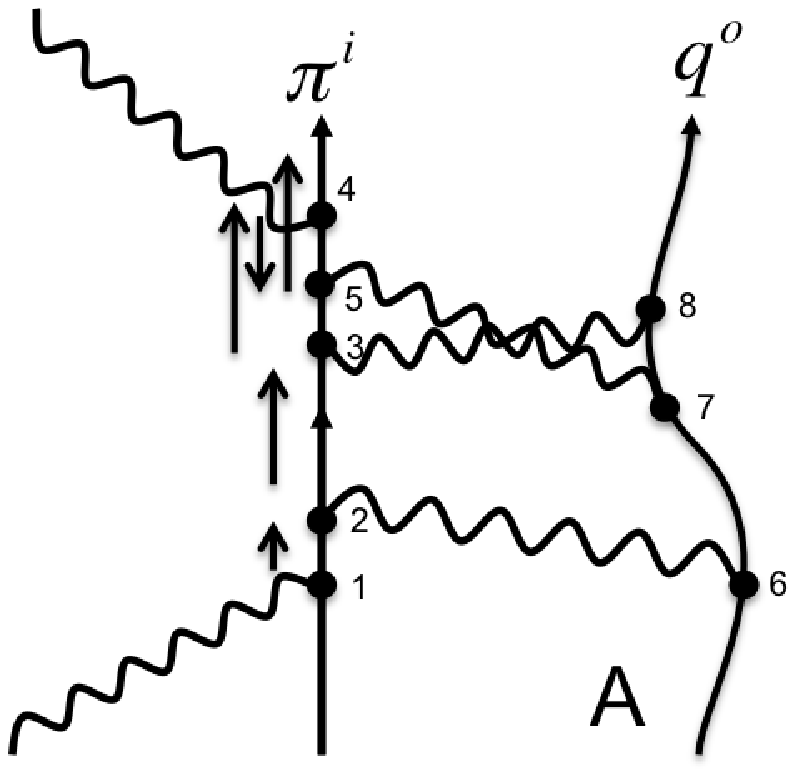}   \includegraphics[height=4cm, width=4cm] {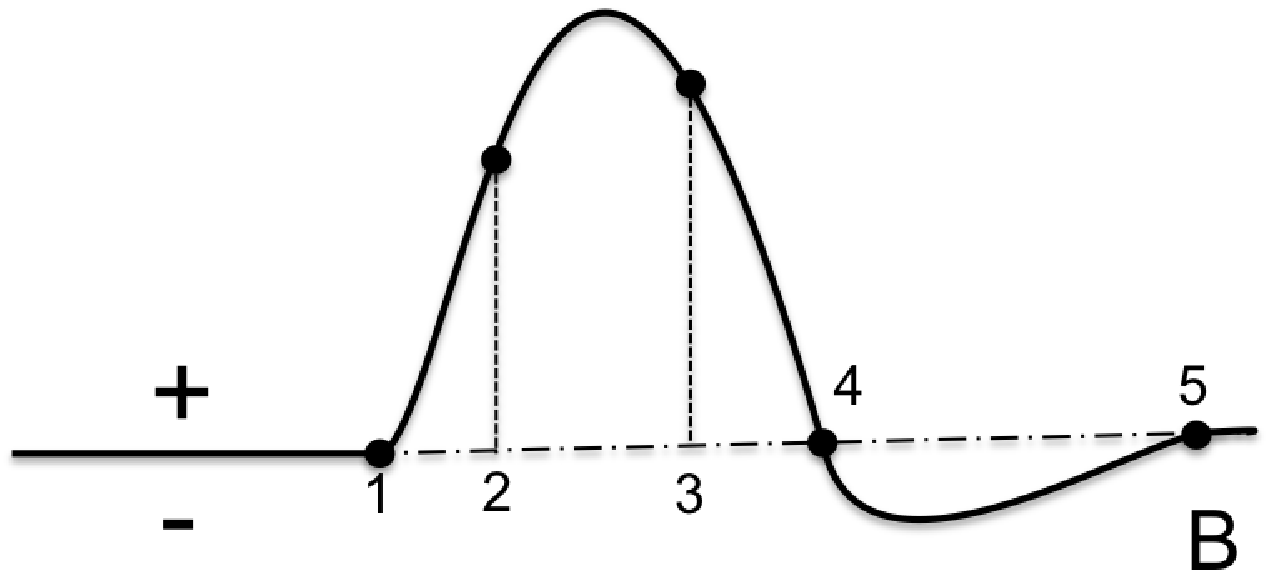}    }
\caption{Feynman diagram to represent the refractory period (A) and its correspondence to the action potential (B).}
\label {fig:Feymandiag2}
\end{figure}

Another important application of Feynman diagram and the transition amplitude is the representation of the refractory period in the perspective of quantum electrodynamics. The refractory period, indicated as the membrane potential below the resting potential as shown in Figure \ref{fig:Feymandiag2}B, is one of the unique features of the action potential and has been regarded as the possible causes of many unexplained nonlinear phenomena in cardiac electrophysiology. In a region during the refectory period shortly after the excitation propagates, the region becomes inactive to any excitation (\textit{absolution refractory period}) or requires more excitation than normal (\textit{relative refractory period}). This inactivation is biologically caused by the inactivation of a voltage-gated sodium channel and the slowly closing potassium channel \cite{Purves}, but the refractory period will be described only by quantum electrodynamical concepts. This means that the refractory period can be represented by Feynman's diagram without introducing ion channels. The goal of this description is to reveal the functionality of ion channels to generate the refractory region and its easier mathematical expression for important nonlinear phenomena, such as atrial reentry, caused by the refractory region.

In Figure \ref{fig:Feymandiag2}A, the influx of the membrane current occurs at time $5$ later than the emission of the propagation cations at time $4$. But, if we consider this happens \textit{before} the time $4$, then everything looks similar except the \textit{time travels backwards} between times $4$ and $5$. Since the events during times $3$ - $5$ happen almost continuously, the order of these events may change. Then the sequence of events occurs as follows: At time 3 when the efflux membrane current occurs, it marches to time $4$ when the emission of the propagation cations occurs. Then, suddenly time travels backward to reach time 5 when the influx membrane current occurs and proceeds to the final time to restore the resting potential. The backward traveling in time looks impossible, but is a very natural phenomenon which has been beautifully recognized as \textit{positrons} by Feynman \cite{Positron}.

The concept of the positron has been devised to explain the wave traveling backward in time to be annihilated to yield photons. This positron has often been observed in the laboratory, which reveals the same as the electron, but is attracted to normal electrons \cite{QEDStrange}. This is possible because electrons can have both positive and negative charges; the positron has only the positive charge, contrary to the negative charge of the normal electrons. This phenomenon is also explained by negative energy states created by scattering in a potential, equivalent to Dirac's Hole theory \cite{Dirac1931} that the vacuum is the sea of the negative energy states except one \textit{hole} that is occupied by positively-charged particles. But, the positron can be annihilated if it collides with an electron, emitting photons as a result. This is why the positron is known as the \textit{anti-particle}.

However, this annihilation by collision is not likely to occur in the process of the cardiac excitation because the myocardial cells are stationary and are separated by the membrane. Thus, the negative energy state exists relatively for a long period to account for the refractory period. Let us relate each time in Figure \ref{fig:Feymandiag2}A to each phase of the action potential in Figure \ref{fig:Feymandiag2}B. Time $1$ for the incident cations is obviously related to the initiation of the depolarization phase. Time $2$ corresponds to the rapid increase of the membrane potential above the threshold. Time $3$ corresponds to the beginning of the repolarization phase. The membrane potential continues to decrease until it reaches the resting potential again at time $4$, but the emission of the propagation cations into $\pi^i$ results in the lower membrane potential than the resting potential. However, the influx membrane current restores the membrane current up to the resting potential at time $5$. These relations between points in the diagram and the phase of the action potential again reveal that the backward time marching from time $4$ to time $5$ corresponds to the refractory period of which energy state can be considered negative as (STATE iii) when the energy state of the resting potential is $1/2$ close to zero as (STATE i).

The consequence of this mechanism clearly explains why the action potential cannot propagate into a region under the refractory period, or namely the \textit{refractory region}. It's natural to say that the resting state is in a positive energy state equivalent to a \textit{negatively-charged electron}. According to the insights from Feynman's diagram, we may regard that the myocardial cell changes into a negative energy state equivalent to \textit{positively-charged electron} during the refractory period. Since the cation is also positively charged, the positively-charged cation cannot propagate easily into the positively-charged myocardial cell, contrary to the normal absorption of the cations by the negatively-charged myocardial cell. This repulsion of the cations due to the same signs of charges as the myocardial cell leads to the non-excitability property of the refractory region. Note the similarities with the Dirac's Hole theory asserting the existence of a hole consisting of the positively-charged electrons in the sea of negatively-charge electrons. Thus, independent of the restoring frequency of the muscle fiber, the myocardial cell cannot be excited during the refractory period. This leads to the following lemma:\\
\\
\textbf{Lemma 4}: The refractory region corresponds to the Dirac's Hole filled with positively-charged electrons in the perspective of QED. \\
\\
However, the positively-charged electrons do not mean that the myocardial cell is filled with more positively-charged ions. What actually happens is to the contrary. The refractory period has more negative membrane potential than the resting potential, thus the intercellular space $\pi^i$ is likely to be filled with a smaller number of positive ions than that of the resting state. Instead, a lower membrane potential that the resting potential should be interpreted as a \textit{negative energy state} and correspondingly a cell in the refractory period as an \textit{anti-(excitable)-cell}. The term \textit{anti} makes more sense when one particle and anti-particle collides to be annihilated to yield new particles, but in the myocardial system, the cells do not move and there are no chance of collision between a cell and an anti-cell, thus the use of \textit{anti} may be not appropriate.

\subsection{Negative energy state in quantum operators}
In this subsection, the relative state vector $\left | \psi \right \rangle$ will be revisited here to better understand the meaning of the negative-state states for the refractory period. The influx membrane flux operator ($\tau^+$) and the efflux membrane flux operator ($\tau$) that are introduced in Section V will be used here again, but will be defined more rigorously. Similar to the annihilation $a$ and creation $a^+$ operator \cite{FeynSM}, suppose that the membrane flux operators $\tau$ and $\tau^+$ satisfy the canonical commutation relations;
\begin{equation*}
[\tau, ~\tau] = 0,~~~[ \tau^+,~\tau^+] = 0,~~~ [ \tau, \tau^+ ] = 1,
\end{equation*}
where $1$ means the identity in the space. Let us construct the eigenstate of $\tau^+ \tau$ as follows: First let the resting state $| 1/2 \rangle$ represent the relative quantum state $1/2$ such that $n^i - \lambda n^o$ is $1/2$. The state $| 1/2 \rangle$ does not mean that $n^i = n^o = 0$, thus $| 1/2 \rangle$ is not the ground state either for $\pi^i$ and $\pi^o$. Applying $\tau$ and $\tau^+$ to this resting state $| 1/2 \rangle$ yields
\begin{equation*}
\left | \frac{3}{2} + \lambda \right \rangle =  \tau^+  \left | \frac{1}{2} \right \rangle , ~~~\left | - \left ( \frac{1}{2} + \lambda \right ) \right \rangle =  \tau \left | \frac{1}{2} \right \rangle  .
\end{equation*}
Note that the relative quantum state is no longer an integer because $\lambda$ is generally not an integer. The quantum state $| 3/2 + \lambda \rangle$ simply means that the quantum state of $\pi^i$ is larger than that of $\pi^o$ by $3/2 + \lambda$. Similarly, the quantum state $| - 1/2 + \lambda \rangle$ simply means that the quantum state of $\pi^o$ is larger than that of $\pi^i$ by $1/2 + \lambda$. Thus, the positive or negative quantum states are well defined with the relative quantum states. In general, the relative eigenstates $| n \rangle $ and $|-n \rangle $ are defined as
\begin{align}
\left | \frac{2n + 1}{2} + n  \lambda  \right \rangle = \frac{1}{\sqrt{n}} { \tau^+ }^n \left | \frac{1}{2} \right \rangle , \\
\left | - \frac{2n - 1}{2} - n \lambda  \right \rangle = \sqrt{n}  \tau^n \left | \frac{1}{2} \right \rangle  ,  
\end{align}
and subsequently,
\begin{equation*}
 \tau^+ \tau \left | \frac{2n + 1}{2} + n  \lambda  \right \rangle  =  \left ( \frac{2n + 1}{2} + n  \lambda \right ) \left | \frac{2n + 1}{2} + n  \lambda  \right \rangle .
 \end{equation*}
On the other hand, because the annihilation $(a)$ and creation $(a^+)$ operator are applied to $\pi^i$ only, the relative eigenstate $| n \rangle $ and $|-n \rangle $ are derived as
\begin{equation*}
\left | n + \frac{1}{2} \right \rangle = \frac{1}{\sqrt{n}} { a ^+ }^n \left | \frac{1}{2} \right \rangle , ~~~\left | - n + \frac{1}{2} \right \rangle = \sqrt{n}  a^n \left | \frac{1}{2} \right \rangle  . \label{nstate2}
\end{equation*}
Then, by using above equations, the final quantum state can be also expressed at $t=5$ in Figure \ref{fig:Feymandiag2}A such that
\begin{align}
\lefteqn{ \left | \frac{1}{2} + n + m ( 1 + \lambda ) - m' ( 1 + \lambda ) - n' \right \rangle } \hspace{1cm} \nonumber \\
& = \frac{\sqrt{n' m' }}{\sqrt{n m }}   a ^{n'}   \tau^{m'}  { \tau^+  }^{m} {  a^+ }^n   \left | \frac{1}{2} \right \rangle . \label{fstate}
\end{align}
Simple calculus reveals that the negative energy state occurs if $   n' - n   > ( 1 + \lambda) ( m - m ' )$. Suppose that $n'/n$ is fixed as the constant safety factor. Then the ratio between $m$ and $m'$ can lead to the negative energy state for the following cases: (i) If $m'$ is larger than $m$, which means the efflux membrane current is larger than the influx membrane current. (ii) If $m$ is larger than $m'$, but its difference $m-m'$ is smaller than $(n' - n ) / ( 1 + \lambda)$. The case (i) is obvious since both $n - n'$ and $m- m'$ are all negative, but the case (ii) is worthy of being noticed because, in order to prevent negative energy states such as the refractory period, the influx membrane current should be \textit{sufficiently} larger than the efflux membrane current. But, if the number $m$ is relatively close to the number $m'$ to maintain the propagation from the conservation of the total number of cations as shown in proposition 3, then the negative energy state is likely to occur in the cardiac excitation propagation.

\subsection{With the external electromagnetic field}
As the similar studies of the changes of the Lagrangian with the external electromagnetic field, we also investigate the changes of the Hamiltonian with the external electromagnetic field. Let the magnetic field $\mathbf{B}_e$ be the external magnetic field and let $\mathbf{A}_e$ and $\phi_e$ be the external electromagnetic potentials which are constructed in the same way as in Section IV.C. Then, the Hamiltonian of the Maxwell's equations (\ref{MWf1}) - (\ref{MWf4}) with the external field is expressed as
\begin{equation}
\mathcal{H}^e = \mathcal{H}_{0}^e  + \mathcal{H}_R + \mathcal{H}_C + \mathcal{H}_{I1}^e + \mathcal{H}_{I2}^e + \mathcal{H}_{I3} + \mathcal{H}_{I4}^e,  \label{Hamilwithexternalpot}
\end{equation}
where $\mathcal{H}_R$, $\mathcal{H}_C$, and $\mathcal{H}_{I3}$ are the same as in the components of the Hamiltonian (\ref{Hamil2}), (\ref{Hamil3}), and (\ref{Hamil6}). Other Hamiltonians with the superscript $e$ are defined as follows:
\begin{align}
\mathcal{H}_p^e & = \sum_{\alpha} \frac{ ( \mathbf{p}_{\alpha}^e )^2 }{2 m_{\alpha} }   ,  \label{exHamil1} \\
\mathcal{H}_{I1}^e & = - \sum_{\alpha} \frac{\chi_{\alpha}}{m_{\alpha}} \mathbf{p}_{\alpha}^e \cdot \mathbf{A} ( \mathbf{r}_{\alpha} ) ,  \label{exHamil2} \\
\mathcal{H}_{I2}^e & = - \sum_{\alpha} g_{\alpha} \frac{\chi_{\alpha}}{2 m_{\alpha} } \mathbf{S}_{\alpha} \cdot ( \mathbf{B} + \mathbf{B}_e ) ( \mathbf{r}_{\alpha} ) , \label{exHamil3} \\
\mathcal{H}_{I4}^e &= \sum_{\alpha} \chi_{\alpha}  \phi_e (\mathbf{r}_{\alpha}, t) , \label{exHamil4}
\end{align}
where the momentum $\mathbf{p}^e$ is the new momentum affected by the external potential $\mathbf{A}_e$ defined as $\mathbf{p}^e_{\alpha} (\mathbf{r}_{\alpha}, t) = \mathbf{p}_{\alpha} - \chi_{\alpha} \mathbf{A}_e(\mathbf{r}_{\alpha}, t)$.

Observe that the external field significantly affects the particle Hamiltonian ($\mathcal{H}_p$), due to the additional particle Hamiltonian ($\mathcal{H}_p$) caused by the external field. On the contrary, the Hamiltonian for the transverse field $\mathcal{H}_R$ remains independent of the external field, which means that the trajectory of the cations remains unchanged even under the influence of the external field. However, dramatic changes occur for the interaction Hamiltonian $\mathcal{H}_I$. The kinetic energy of the oscillatory forced motion $\mathcal{H}_{I3}$ remains unchanged, but $\mathcal{H}_{I1}$ and $\mathcal{H}_{I2}$ are significantly changed because of the changes of the momentum $\mathbf{p}_{\alpha}$ and the additional effect of the external magnetic field $\mathbf{B}_e$, respectively. The effects of $\mathcal{H}_{I1}$ and $\mathcal{H}_{I2}$ on the membrane current may be negligible since the membrane current mostly is known to be sensitive only to the membrane potential, not to the momentum or the spin energy.

On the other hand, the new addition of $\mathcal{H}_{I4}$ to the interaction Hamiltonian dramatically changes the mechanism of the cardiac excitation in the following ways: (i) $\mathcal{H}_{I4}$ depends on $\chi_{\alpha}$ being related to ion channels, but the impact of $\mathcal{H}_{I4}$ is applied to any myocardial cell with non-zero $\chi_{\alpha}$. Reflecting that the resting state has the quantum number $1/2$ and the existence of the sources represented by ion channels, we may presume that $\chi_{\alpha}$ could be negligible in the resting state, but is not zero almost everywhere independent of the phase of the excitation. Then, $\mathcal{H}_{I4}$ may cause the excitation of the myocardial cells independent of the excitation propagation. (ii) Secondly, $\mathcal{H}_{I4}$ is proportional to charge density $\phi_e$, thus even for the region where the charge density $\rho$ is small, the myocardial cell can be excited by a sufficiently large external scalar potential $\phi_e$. The presence of $\mathcal{H}_{I4}$ may provide explanations on the effects of the external electric currents to terminate fibrillations. This is because the excitation of all the myocardial cells by a huge external field can trigger them into the resting state at the same time shortly after the electric shock, as a better condition for the normal propagation from a natural initiator such as the sinoatrial node.

\section{Conclusions and discussions}

The strength of the proposed QED theory for the cardiac excitation propagation lies in the fact that it provides analytical explanations on many electrophysiological phenomena which have been unexplained by previously-developed theories. This is mainly because the governing equations are Maxwell's equations under conservational laws. A simple expression of the Lagrangian provides many insights such as which factors are critical for the changes of the propagation and what the role of ion channels is in the action potential. Also, the Hamiltonian simplifies the excitation mechanism eligible for simpler mathematical analysis. It is also encouraging to see the clinically-supported explanations on the effects of the electromagnetic field generated by the cardiac excitation or the external electrodynamic fields.

The validation of this theory is mostly self-sufficient, especially by using the following proofs: (i) The derivation of a set of Maxwell's equations equivalent to the diffusion-reaction system. (ii) For the Lagrangian, the trajectory of the diffusion-reaction system is shown to be the same as the electrodynamic wave when $\chi_{\alpha}$ changes normally everywhere, which can be also deduced easily because they share the same propagational mechanism. But the validation for the theories by the Hamiltonian seems to be only possible by future experimental studies. The validation for the effects of the electromagnetic fields also seems to be supported by showing the consistency between clinical observations and what the theory explains.

However, there are also some drawbacks as mentioned in the Introduction. (i) The first is related to the propagating cations. If the propagating particle is more than one kind, then calculations become too complicated or becomes impossible. Also, the concept of the cations could be concrete, but at the same time could be abstract. (ii) The biggest drawback is the simplification of ion channels. The second quantization of Maxwell's equations does not show the existence of numerous ion channels for several charged ions. Consequently, the use of a corresponding quantum operator may remain restricted without describing in detail the complex dynamics of the real phenomena. One possible way is to consider the different charged ions as different modes. But, the negatively-charged ions are not relevant to this case and various values of the Planck constant $\hbar$ may only lead to much more complicated analysis which may be beyond our understanding. (iii) The last is the difficulty of using the Maxwell's equations (\ref{MWf1}) - (\ref{MWf4}) for computational simulations because the expressions for charge density and current density are too complicated. But, this can be easily solved by using diffusion-reaction equations as usual, but using, additionally, Maxwell's equations (\ref{MWf1}) - (\ref{MWf4}) to obtain $\mathbf{E}$, $\mathbf{B}$, and $\mathbf{A}$ from $\phi$.

In the future publication as a continuing effort of developing the QED theory for the cardiac excitation propagation, another quantum optical concept known as \textit{coherence} will be introduced in order to understand some important problems such as (i) when conduction fails (ii) what is the role of geometry in conduction failure (iii) how conduction failure can be prevented in the perspective of optical coherence, etc.

\section{Appendix I: Proofs in section IV}

\subsection{Proof of Proposition 4}
The total energy (\ref{totalenergy}) is well defined in the macroscopic domain $\Pi$, since for the particle in each space, the above equation will reduce to the classical energy for $\pi^i$ or $\pi^o$; if the particle lies in $\pi^i$, then equation (\ref{totalenergy}) with equation (\ref{defpcandvel}) reduces to
\begin{equation*}
\mathbf{U} = \sum_{\alpha} \frac{1}{2} m_{\alpha} \left ( \mathbf{v}^o_{\alpha} \right )^2 + \frac{\varepsilon_i}{2} \int \left [ ( \mathbf{E}^i)^2 + c^2 ( \mathbf{B}^i)^2  \right ] d^3 r  ,
\end{equation*}
and if the particle lies in $\pi^o$, then it reduces to
\begin{align*}
\mathbf{U} &= \sum_{\alpha} \frac{1}{2} m_{\alpha} \left (\mathbf{v}^o_{\alpha}  \right )^2+ \frac{\varepsilon_i}{2} \int \left [ ( \sqrt{\varepsilon_o / \varepsilon_i}\mathbf{E}^o)^2 + c^2 ( \sqrt{\mu_i / \mu_o} \mathbf{B}^o)^2  \right ] d^3 r   \\
&=  \sum_{\alpha} \frac{1}{2} m_{\alpha} \left ( \mathbf{v}^o_{\alpha} \right )^2 + \frac{\varepsilon_o}{2} \int \left [ ( \mathbf{E}^o)^2 + c^2 (\mathbf{B}^o)^2  \right ] d^3 r .
\end{align*}
Thus, the total energy (\ref{totalenergy}) is well defined in $\Pi$. The differentiation of the above equation with respect to the time $t$ yields
\begin{equation*}
\frac{\partial \mathbf{U}}{\partial t} = \sum_{\alpha} m_{\alpha} \mathbf{v}_{\alpha} \cdot \frac{ d \mathbf{v}_{\alpha}}{dt} + \varepsilon_i \int \left [ \mathbf{E} \cdot   \frac{\mathbf{\partial E}}{dt} +  c^2 \mathbf{B} \cdot  \frac{\partial \mathbf{B} }{dt}  \right ] d^3 r  .
\end{equation*}
By substituting the Maxwell's equations (\ref{MWf3}) and (\ref{MWf4}) and the Newton-Lorentz equation (\ref{NewtonLorentz}), we obtain
\begin{align*}
\frac{\partial \mathbf{U}}{\partial t} &= \sum_{\alpha} \mathbf{v}_{\alpha} \cdot (\chi_{\alpha} \mathbf{E} (\mathbf{r}_{\alpha},t)  ) - \int \mathbf{E} \cdot \mathbf{J} d^3 r \\
& + \varepsilon_i c^2 \int \left [ \mathbf{E} \cdot ( \nabla \times \mathbf{B} ) - \mathbf{B} \cdot ( \nabla \times \mathbf{E} ) \right ]  d^3 r.
\end{align*}
Due to the discrete expression of $\mathbf{J}$ as shown in equation (\ref{discreterhoandj}), we notice that the first two terms cancel out. The integrand in the last integration can be simplified as $\nabla \cdot (\mathbf{E} \times \mathbf{B} )$. Thus, the above equation reduces to 
\begin{equation*}
\frac{\partial \mathbf{U}}{\partial t} = \varepsilon_i c^2 \int \nabla \cdot (\mathbf{E} \times \mathbf{B} )  d^3 r  = \varepsilon_i c^2 \int_S ( \mathbf{E} \times \mathbf{B} )\cdot \mathbf{n} d S ,
\end{equation*}
where the last equation is obtained by the divergence theorem. Since no flux of the electromagnetic fields occurs at the boundaries in the closed system, the right hand side is zero, thus ${\partial \mathbf{U}} / {\partial t} = 0$ $\square$.

\subsection{Proof of Proposition 5}
The momentum (\ref{totalmomentum}) is also well defined in $\Pi$, since for the particle in each microscopic domain, equation (\ref{totalmomentum}) will reduce to the classical momentum for $\pi^i$ or $\pi^o$; if the particle lies in $\pi^i$, then equation (\ref{totalmomentum}) with equation (\ref{defpcandvel}) reduces to
\begin{equation*}
\mathbf{P} = \sum_{\alpha} m_{\alpha} \mathbf{v}_{\alpha}^i + \varepsilon_i \int \left [ \mathbf{E}^i \times \mathbf{B}^i  \right ] d^3 r  ,
\end{equation*}
and if the particle lies in $\pi^o$, then it reduces to
\begin{align*}
\mathbf{P} &= \sum_{\alpha} m_{\alpha}^o \mathbf{v}_{\alpha} + \varepsilon_i \int \left [ (- \sqrt{\varepsilon_o / \varepsilon_i } \mathbf{E}^o ) \times (- \sqrt{\mu_i / \mu_o } \mathbf{B}^o ) \right ] d^3 r \\
& =  \sum_{\alpha} m_{\alpha} \mathbf{v}_{\alpha}^o + \varepsilon_o \int \left [ \mathbf{E}^o  \times  \mathbf{B}^o \right ] d^3 r .
\end{align*}
Thus, the total momentum (\ref{totalmomentum}) is well defined in $\Pi$. After differentiating the above equation with respect to $t$, let us substitute again Maxwell's equations (\ref{MWf3}) and (\ref{MWf4}) and the Newton-Lorentz equation (\ref{NewtonLorentz}) in equation (\ref{totalmomentum}) to obtain
\begin{align}
\lefteqn{\frac{\partial \mathbf{P}}{\partial t} = \sum_{\alpha} \chi_{\alpha} \mathbf{E} (\mathbf{r}_{\alpha},t) + \chi_{\alpha} \mathbf{v}_{\alpha} \times \mathbf{B} (\mathbf{r}_{\alpha},t) - \int \mathbf{J} \times \mathbf{B} d^3 r} \hspace{0.2cm} \nonumber \\
& + \varepsilon_i \int \left [ c^2 ( \nabla \times \mathbf{B} ) \times \mathbf{B} - \mathbf{E} \times ( \nabla \times \mathbf{E} )  \right ] d^3 r  .    \label{proof20}
\end{align}
Substituting equation (\ref{discreterhoandj}) into the above equation will cancel out the second and the third terms. For the integration term, we use
\begin{equation*}
\mathbf{V} \times ( \nabla \times \mathbf{V} ) = \frac{1}{2} \nabla ( \mathbf{V}^2) - \sum_j \mathbf{e}_j \nabla \cdot (V_j \mathbf{V} ) + \mathbf{V} ( \nabla \cdot \mathbf{V} ) ,
\end{equation*}
where $\mathbf{e}_j$ is the directional vector of the Cartesian coordinates, then for the first part of the integration we obtain 
\begin{equation}
\int  ( \nabla \times \mathbf{B} ) \times \mathbf{B} d^3 r = 0, \label{proof21}
\end{equation}
where the first and the second terms on the right hand side are zero because there is no flux of $\mathbf{B}$ across the boundaries in the closed system and the third term is zero because of equation (\ref{MWf2}). Similarly, for the second part of the integration, we obtain
\begin{align}
\lefteqn{ \int \mathbf{E} \times  ( \nabla \times \mathbf{E} ) d^3 r = - \int_S \mathbf{E}^2 \cdot \mathbf{n} dS } \hspace{0.2cm} \nonumber \\
 &- \int_S B_j \mathbf{E} \cdot \mathbf{n} dS  - \int \mathbf{E} ( \nabla \cdot \mathbf{E} ) d^3 r  = \sum_{\alpha} \chi_{\alpha} \mathbf{E} (\mathbf{r}_{\alpha},t) ,  \label{proof22}
\end{align}
where the first and second terms on the right hand side are zero because there is no flux of $\mathbf{E}$ across the boundaries in the closed system and the third term is obtained by the use of equation (\ref{MWf1}). Finally, substituting equalities (\ref{proof21}) and (\ref{proof22}) into equation (\ref{proof20}) yields $\partial \mathbf{P} / \partial t = 0 $ $\square$.

\section{Appendix II: Proof in section V}

\subsection{Proof of Lemma 2}
Substituting equation (\ref{MWk5}) and (\ref{MWk6}) into equation (\ref{Lagden2}) yields
\begin{align*}
\lefteqn{ \mathcal{L} ( \mathbf{k} )  = \varepsilon_i \left [ \left | \dot{\mathbf{a}}_k ( \mathbf{k} )+ i \mathbf{k} \phi_k ( \mathbf{k} ) \right |^2 - c^2 \left | \mathbf{k} \times \mathbf{a}_k ( \mathbf{k} ) \right |^2  \right ]  } \hspace{0.1cm} \\
& + \left [ \mathbf{j}^*_k  ( \mathbf{k} ) \cdot \mathbf{a}_k  ( \mathbf{k} ) + \mathbf{j}_k  ( \mathbf{k} ) \cdot \mathbf{a}^*_k  ( \mathbf{k} )  - \rho^*_k  ( \mathbf{k} ) \phi_k  ( \mathbf{k} ) - \rho_k  ( \mathbf{k} ) \phi^*_k  ( \mathbf{k} ) \right ] .
\end{align*}
To eliminate the scalar potential $\phi_k$ in this equation, we substitute the equality $\phi_k = ({1}/{k^2})  ( i k \dot{a}^{\parallel}_k + ({\rho_k} / {\varepsilon_i} ) )$ being obtained from equations (\ref{MWk1}) and (\ref{MWk6}). Then, we obtain
\begin{align}
 \mathcal{L} ( \mathbf{k} )  &= -  \frac{\rho_k \rho_k^* }{\varepsilon_i k^2 }   +  \varepsilon_i \left [ { \dot{\mathbf{a}}_k^{\perp *}} \cdot \dot{\mathbf{a}}^{\perp}_k - c^2 k^2  { {\mathbf{a}}_k^{\perp *}} \cdot {\mathbf{a}}^{\perp}_k \right ]   \nonumber \\
 &~~~~~~~~  + \frac{i}{k} \frac{d}{dt} \left [ \rho a^{\parallel *}_k  - \rho^* a^{\parallel}_k \right ] , \label{Lagdenap2}
\end{align}
where we used $\dot{\mathbf{a}}^{\perp}_k = \dot{\mathbf{a}}_ k - (\mathbf{k}/k) \dot{{a}}^{\parallel}_k $ and used the new variables ${a}^{\parallel}_k = \boldsymbol{\kappa} \cdot \mathbf{a}_k$ and ${j}^{\parallel}_k = \boldsymbol{\kappa} \cdot \mathbf{j}_k$. With the conservation of charge (\ref{conservcharge}), or equivalently $\dot{\rho} = - i k j^{\parallel}$ in the $k$-space, the last total time derivative is obtained from the following equality:
\begin{equation*}
j_k^{\parallel *} a_k^{\parallel} +  j_k^{\parallel} a_k^{\parallel *}  +  \frac{i}{k}  \left ( \rho_k \dot{a}^{\parallel *}_k - \rho_k^* \dot{a}^{\parallel}_k \right )  =  \frac{i}{k} \frac{d}{dt} \left [ \rho a^{\parallel *}_k  - \rho^* a^{\parallel}_k \right ].
\end{equation*}
Then the only difference of Lagrangian with $\mathcal{L} ( \mathbf{k} ) $ (\ref{Lagdenap2}) to Lagrangian with $\mathcal{L} ( \mathbf{k} ) $ (\ref{stdLagrangian}) is the above total derivative term with respect to time, thus we only need to show that this term can be subtracted without changing the extremes of the action integral from the Lagrangian of our system. Let $S_0$ be the action integral for the classical Maxwell's equations to the Lagrangian $\mathcal{L}_0$. Then the action integral $\mathcal{S}$ for the Lagrangian density $\mathcal{L}$ (\ref{Lagdenap2}) is expressed as
\begin{equation*}
\mathcal{S} = \int_{t_1}^{t_2} \mathcal{L} = S_0 + \frac{1}{ 4 \pi} \int_{t_1}^{t_2}  \left [ \frac{d}{dt} \int \frac{ \rho(\mathbf{r},t ) \phi (\mathbf{r}', t ) }{ | \mathbf{r} - \mathbf{r}' | } d^3 r \right ] dt ,
\end{equation*}
thus,
\begin{equation*}
\mathcal{S} = \mathcal{S}_0 + \frac{1}{ 4 \pi} \frac{1}{ | \mathbf{r} - \mathbf{r}' | } \left [  \rho(\mathbf{r},t_1 ) \phi (\mathbf{r}', t_1 )  - \rho(\mathbf{r}, t_2 ) \phi (\mathbf{r}', t_2 )    \right ]  .
\end{equation*}
Observe that $S$ and $S_0$ are only different in terms of a constant, thus have the same extreme values. Therefore, the total derivative term with respect to $t$ is redundant and can be subtracted from the Lagrangian density in equation (\ref{Lagdenap2}) without changing its extremes. $\square$

\section{Appendix III: Proof in section VI}

\subsection{Proof of Lemma 4}
Consider the conjugate momentum $\mathbf{p}_{\alpha}$ for the particle $\alpha$ and the conjugate momenta $\boldsymbol{\xi}$ for $\dot{\mathbf{A}}^{\perp}$ defined as
\begin{equation}
\mathbf{p}_{\alpha} ( \mathbf{r}_{\alpha} )  \equiv \frac{\partial \mathcal{L}}{\partial \dot{\mathbf{r}}_{\alpha} },~~~~~\boldsymbol{\xi} ( \mathbf{k} ) \equiv  \left ( \frac{\partial \mathcal{L}}{\partial \dot{{\mathbf{a}_k^{\perp}}} } \right )^*,  \label{conjmom}
\end{equation}
where the differentiation with respect to a vector $\mathbf{V}$ is just considered as a vector whose component is the differentiation with respect to each component $V^i$. Then, the Hamiltonian for the Lagrangian (\ref{Lagfinal}) and (\ref{Lagdenfinal}) can be derived as \cite{CohenPA}
\begin{equation}
\mathcal{H} = \sum_{\alpha} \mathbf{p}_{\alpha} \cdot \dot{\mathbf{r}}_{\alpha}  + \fint  \left [ \boldsymbol{\xi}^* \cdot \dot{\mathbf{a}}_k^{\perp} + \boldsymbol{\xi} \cdot \left ( \dot{\mathbf{a}}_k^{\perp} \right )^* \right ] d^3 k  - \mathcal{L} .  \label{Hamil0}
\end{equation} 
Substituting the Lagrangian (\ref{Lagfinal}) and (\ref{Lagdenfinal}) with the conjugate momenta (\ref{conjmom}) into the Hamiltonian (\ref{Hamil0}), we obtain the Hamiltonian (\ref{Hamilk}) with (\ref{Hamildenk}) $\square$.

\subsection{Proof of Proposition 8}
Using the normal variable $\boldsymbol{\alpha}$ = $- i / ( 2 \mathcal{N} ( k )  )$ $ \left [ \mathbf{e}_k^{\perp} - c (\mathbf{k}/k ) \times \boldsymbol{b}_k  \right ]$ which satisfies
\begin{equation}
\dot{\boldsymbol{\alpha} } (\mathbf{k}) + i \omega  \boldsymbol{\alpha} (\mathbf{k})  = \frac{i}{2 \varepsilon_i }{ \mathcal{N} } \mathbf{j}^{\perp}_k  , \label{normalvar}
\end{equation}
where $\omega = c k$ and $\mathcal{N}$ is the normalization coefficient, normally chosen as $\mathcal{N} (k) = \sqrt{ \hbar \omega / 2 \varepsilon_i}$, we can express $\mathbf{e}_k^{\perp}$ and $\mathbf{b}_k$ with $\boldsymbol{\alpha}$ and its conjugate $\boldsymbol{\alpha}^*$, thus the Hamiltonian density is expressed as
\begin{equation}
\mathcal{H} (\mathbf{k}) = \mathcal{N}^2 \left [ \boldsymbol{\alpha}^* (\mathbf{k}) \cdot \boldsymbol{\alpha}  (\mathbf{k})+   \boldsymbol{\alpha} (\mathbf{k}) \cdot \boldsymbol{\alpha}^* (\mathbf{k}) \right ]  .
\end{equation}
In fact, this is the same as the classical Maxwell's equations with the Coulomb gauge because the gauge choice and the reactive membrane current density (\ref{cdchoice2}) does not affect the Maxwell's equations (\ref{MWk3}) and (\ref{MWk4}) and consequently the oscillator of the normal variable $\boldsymbol{\alpha}$ (\ref{normalvar}) remains the same. For $\mathbf{j}^{\perp}_k=0$, the equation (\ref{normalvar}) boils down to a Schr\"{o}dinger equation for the wave function $\boldsymbol{\alpha}$ such as $i \hbar \dot{\boldsymbol{\alpha}} ( \mathbf{k}, t) = \hbar \omega \boldsymbol{\alpha} ( \mathbf{k}, t) $ as a Schr\"{o}dinger's form for the equation of motion \cite{Dirac}. This similarity often leads to the substitution of the normal variable $\boldsymbol{\alpha}$ and its conjugate $\boldsymbol{\alpha}^*$ with the annihilation operator $a_i$ and $a^+_i$, respectively \cite{Loudon}. As a consequence, by using $\left [  a, a^+ \right ] = a a ^+ - a^+ a = 1 $ and by introducing the spin magnetic moment, we have the following Hamiltonian (\ref{Hamilk}) and (\ref{Hamildenk}) $\square$.

\bibliographystyle{plain}    

\bibliography{Article}

\end{document}